\PassOptionsToPackage{pdfpagelabels=false}{hyperref} 
\documentclass[twocolumn]{aastex62}

\usepackage{hyperref}

\usepackage{graphicx}
\usepackage{epsfig}
\usepackage{hyperref}
\usepackage{natbib}

\usepackage{newtxtext,newtxmath}
\usepackage{graphicx}	
\usepackage{amsmath}	
\usepackage{amssymb}	
\usepackage{lipsum}
\usepackage{float}

\newcommand{\kms}{km~s$^{-1}$ }
\newcommand{\kmsp}{km~s$^{ -1}$}
\newcommand{\oi}{\ion{O}{1}~}

\newcommand{\nv}{\ion{N}{5}}

\newcommand{\siii}{\ion{Si}{2}~}
\newcommand{\siiii}{\ion{Si}{3}~}

\newcommand{\siiifs}{\ion{Si}{2}$^\ast$~}
\newcommand{\siiv}{\ion{Si}{4}~}
\newcommand{\heii}{\ion{He}{2}}

\newcommand{\ov}{\ion{O}{5}}

\newcommand{\ciii}{\ion{C}{3}}
\newcommand{\nvp}{\ion{N}{5}}

\newcommand{\siiiip}{\ion{Si}{3}}

\newcommand{\civp}{\ion{C}{4}}
\newcommand{\civ}{\ion{C}{4}~}
\newcommand{\siivp}{\ion{Si}{4}}

\newcommand{\zism}{Z$_\mathrm{neb}$}

\newcommand{\zs}{Z$_\ast$}

\newcommand{\frat}{F$_{900}$/F$_{1500}$}
\newcommand{\fratf}{F$_{525}$/F$_{1500}$}
\newcommand{\fratt}{F$_{300}$/F$_{1500}$}

\newcommand{\ebv}{E(B-V)}

\newcommand{\megasaura}{M\textsc{eg}a\textsc{S}a\textsc{ura}}
\newcommand{\bpass}{\textsc{bpass}}
\newcommand{\starburst}{\textsc{starburst99}}
\newcommand{\megasauralong}{The Magellan Evolution of Galaxies Spectroscopic and Ultraviolet Reference Atlas}

\begin{document}
\shorttitle{Properties and ionizing continua of extragalactic massive stars populations}
\shortauthors{Chisholm et al.}
\title{Constraining the metallicities, ages, star formation histories, \\ and ionizing continua of extragalactic massive star populations\footnote{
Based on observations made with the NASA/ESA Hubble Space Telescope,
obtained from the Data Archive at the Space Telescope Science Institute, which
is operated by the Association of Universities for Research in Astronomy, Inc.,
under NASA contract NAS 5-26555.}}

\email{jochisho@ucsc.edu}
\author{J. Chisholm}
\affil{University of California--Santa Cruz, 1156 High Street, Santa Cruz, CA, 95064, USA}
\author{J.R. Rigby}
\affil{Observational Cosmology Lab, NASA Goddard Space Flight Center, 8800 Greenbelt Rd., Greenbelt, MD 20771, USA}
\author{M. Bayliss}
\affil{MIT Kavli Institute for Astrophysics and Space Research, 77 Massachusetts Ave., Cambridge, MA 02139, USA}
\author{D.A. Berg}
\affil{Department of Astronomy, The Ohio State University, 140 W. 18th Avenue, Columbus, OH 43202, USA}
\author{H. Dahle}
\affil{Institute of Theoretical Astrophysics, University of Oslo, P.O. Box 1029, Blindern, NO-0315 Oslo, Norway}
\author{M. Gladders}
\affil{Kavli Institute for Cosmological Physics, University of Chicago, 5640 South Ellis Ave., Chicago, IL 60637, USA}
\author{K. Sharon}
\affil{Department of Astronomy, University of Michigan, 500 Church St., Ann Arbor, MI 48109, USA}

\begin{abstract}
We infer the properties of massive star populations using the far-ultraviolet stellar continua of 61 star-forming galaxies: 42 at low-redshift observed with \textit{HST} and 19 at $z\sim2$ from the \megasaura\ sample. We fit each stellar continuum with a linear combination of up to 50 single age and single metallicity \starburst\ models. From these fits, we derive light-weighted ages and metallicities, which agree with stellar wind and photospheric spectral features, and infer the spectral shapes and strengths of the ionizing continua. Inferred light-weighted stellar metallicities span 0.05--1.5~Z$_\odot$ and are similar to the measured nebular metallicities. We quantify the ionizing continua using the ratio of the ionizing flux at 900\AA\ to the non-ionizing flux at 1500\AA\ and demonstrate the evolution of this ratio with stellar age and metallicity using theoretical single burst models. These single burst models only match the inferred ionizing continua of half of the sample, while the other half are described by a mixture of stellar ages. Mixed age populations produce stronger and harder ionizing spectra than continuous star formation histories, but, contrary to previous studies that assume constant star formation, have similar stellar and nebular metallicities. Stellar population age and metallicity affect the far-UV continua in different and distinguishable ways; assuming a constant star formation history diminishes the diagnostic power. Finally, we provide simple prescriptions to determine the ionizing photon production efficiency ($\xi_{\rm ion}$) from the stellar population properties. $\xi_{\rm ion}$ inferred from the observed star-forming galaxies has a range of log($\xi_{\rm ion}) =24.4-25.7$~Hz~erg$^{-1}$ that depends on the stellar population age, metallicity, star formation history, and contributions from binary star evolution. These stellar population properties must be observationally determined to accurately determine the number of ionizing photons generated by massive stars.
\end{abstract}
\keywords{binaries: general -- dark ages, reionization, first stars -- galaxies: abundances -- galaxies: starburst -- stars: abundances -- stars: massive}

\section{Introduction}

O-stars, which have masses >15~M$_\odot$ and lifetimes <10~Myr, are the the only main-sequence stars hot enough to generate a significant number of ionizing photons ($\lambda < 912$\AA). These photons ionize hydrogen in the interstellar medium, powering the nebular emission lines that diagnose the physical state \citep{stromgren, seyfert43, baldwin1991} and chemical evolution \citep{tinsley80} of star-forming galaxies. The emission lines trace the most recent star formation and measure the rate at which stars form \citep{kennicutt98, kennicutt2012}. These observations describe how galaxies build up their stellar mass \citep{brinchmann2004, noeske07, elbaz07} and how star formation evolves with cosmic time \citep{madau99, madau14}. Massive stars impact more than just their host galaxies: the ionizing photons produced by the earliest stars may have been sufficient to reionize the universe \citep{ouchi09,robertson13, robertson15, finkelstein19}. Ionizing photons from massive stars generate the fundamental observables which describe the formation and evolution of star-forming galaxies. As such, determining how stars produce ionizing photons is fundamental to understanding galaxy formation and evolution.
 
Stellar ionizing photons are challenging to directly observe because neutral hydrogen within galaxies efficiently absorbs ionizing photons. Nearly all inferences about the flux and spectral shape of the stellar ionizing continua have been made either from emission lines that have been reprocessed through nebular gas adjacent to massive stars, or from the technique of stellar population synthesis. Stellar population synthesis constructs a model stellar spectrum by first determining a hypothetical stellar population (with a given age, composition, and star  formation  history) and then creating a theoretical spectrum of that stellar population using model stellar atmospheres. The stellar age, metallicity and star formation history are inferred by constructing models with a range of these parameters and using statistical methods to determine which population values best match the observed spectrum. Large libraries of rest-frame optical spectra, from surveys such as the Sloan Digital Sky Survey \citep{sdss}, have revolutionized population synthesis at optical wavelengths \citep{bruzual, maraston05, conroy13}.

Stellar population synthesis of the most massive stars can, in principle, constrain the ionizing continua of massive stars. However, massive stars have largely featureless optical spectra that do not change appreciably with stellar metallicity or age. Therefore, optical stellar population synthesis has a temporal resolution on the order of 10-100~Myr when B-stars, with significant Balmer absorption features, begin to appear in optical spectra. In contrast, the O-stars that produce the majority of the ionizing photons have much shorter lifetimes of 2-10~Myr.

The ideal wavelength range to capture the rapid temporal evolution of massive stars is the rest-frame far-ultraviolet (FUV).  The FUV contains spectral features of massive stars, namely stellar wind lines \citep[][]{walborn85, howarth, lamers, walborn02,pellerin02}, which have been observed in star-forming galaxies over most of cosmic time \citep{kinney, heckman98, pettini2002, leitherer11, steidel16, rigbyb}. The shape and strength of these spectral features strongly depend on both the age and metallicity of the stellar populations \citep{leitherer95, claus99, smith}, enabling FUV spectral synthesis to determine the population properties of massive stars.


Both stellar physics and stellar population properties dictate the production of ionizing photons. In spectral population synthesis, the stellar models amass the complicated underlying stellar physics of the individual stellar properties and evolution which leads to the observed stellar continuum. These vital stellar physical properties include the initial mass function  \citep[IMF; ][]{salpeter, kroupa, chabrier}, stellar rotation \citep{meynet2000, levesque, leitherer14}, and the stellar evolution tracks that may include interactions among binary stars \citep{geneva94, leitherer95, eldridge09, stanway16}. Ultimately, stellar spectral population synthesis is founded upon the individual stellar models.  The success or failure of the stellar synthesis relies upon the models properly incorporating the crucial stellar physics. 

Predominately, this paper focuses on using stellar models to constrain the stellar population properties such as age, metallicity, and star formation history. We then use these properties to infer their ionizing continua. More massive stars must be hotter to counteract their intense gravity and remain in hydrostatic equilibrium. Increased stellar temperatures produce bluer spectra and fully ionized stellar atmospheres. Both effects lead to the production of a copious amount of ionizing photons. These massive stars rapidly exhaust the hydrogen in their cores and have much shorter lifetimes than cooler stars.  Consequently, ionizing photons are only produced by the youngest and most massive stars. Further, since hydrogen is highly ionized in their photospheres, metals are the main opacity source of ionizing photons in massive stars. Thus, lower metallicity stars produce significantly more ionizing photons than stars of similar ages but higher metallicities. The stellar age and metallicity must be observationally constrained to determine the number of ionizing photons generated by massive stars. 

The 2--10~Myr lifetimes of the most massive stars are only 1--10\% of the dynamical timescales of galaxies. Thus, the relative proportion of massive stars depends on when stars were formed and how many stars formed at each epoch. This is referred to as the star formation history. A starburst galaxy is typically defined as recently forming a large fraction of the total stellar mass, thus, their star formation histories are typically assumed to be nearly a delta-function of a single burst \citep{kmcquinn}. Meanwhile, the entire disks of normal star-forming galaxies have more moderate, nearly constant, star formation histories that generate new massive stars at a nearly constant rate \citep{leitherer95}. A constant star formation history always has a component of young massive stars capable of producing ionizing photons that is diluted by the older population. Nature is unlikely to comply with these simplified star formation histories, and the true star formation histories are assuredly somewhere between these two extremes \citep{kmcquinn2}. To understand the relative strength of the youngest stellar populations and their role in producing ionizing photons, there must be an observationally motivated method to determine the star formation history.

In this paper, we perform FUV stellar population synthesis to constrain the age, metallicity, star formation history, and ionizing continua of extragalactic massive star populations. We fit the non-ionizing FUV continua of a sample of 61 low and moderate redshift star-forming galaxies as a linear combination of single age, fully theoretical stellar continuum models. We infer the light-weighted ages, metallicities, and the ionizing continua of the massive star populations from these fits. We compare the stellar and nebular metallicities (\autoref{metal}) and explore the inferred ionizing continua of the stellar populations (\autoref{ion}).  The star formation histories are derived by comparing the inferred stellar continuum fits to single burst models (\autoref{cont}). We test the observational differences between populations that contain binary stars (\autoref{comp}) and illustrate how the stellar continuum fits predict the total number of ionizing photons (\autoref{esc}).

Throughout this paper we follow the literature convention and assume that stellar solar metallicity is 0.02 \citep{claus99, claus2010, stanway18}. It is debated whether solar abundance is actually higher or lower than this value \citep{nieva, villante}, but we retain the 0.02 value used in stellar models because it determines the stellar evolution tracks and stellar wind profiles. We take the solar gas-phase metallicity to be 12+log(O/H) = 8.69 and Z$^{\rm neb}_\odot =0.0142$  \citep{asplund}. All spectra and flux densities are plotted and quoted in F$_\lambda$ units (erg~s$^{-1}$~cm$^{-2}$~\AA$^{-1}$). The equivalent widths of absorption lines are defined to be positive; emission lines are defined to be negative. 

\section{Data}
\subsection{Moderate redshift galaxies}

\subsubsection{\megasaura\ data}
Here we predominately display spectra of nineteen star-forming galaxies from project \megasaura: \megasauralong\ \citep{rigbya}. The extended \megasaura\ sample includes the brightest southern lensed galaxies found in the Red-sequence Cluster Survey \citep[RCS;][]{gladders05}, the Sloan Giant Arcs Survey \citep[SGAS;][]{bayliss11}, the South Pole Telescope \citep[SPT;][]{spt}, and the ESA {\it Planck} survey \citep{planck1, planck2}. These surveys found $z \sim 2$ star-forming galaxies behind massive foreground galaxy clusters. The mass of the foreground clusters magnifies, stretches, and amplifies the light from the background star-forming galaxies, enabling high signal-to-noise ratio (SNR) and moderate spectral resolution rest-frame FUV observations of $z\sim2$ galaxies with ground based telescopes. \megasaura\ is the ideal individual galaxy, FUV stellar spectral reference sample.  

\megasaura\ spectra were taken with the Magellan Echellette (MagE) Spectrograph \citep{Marshall:2008bs} on the Magellan telescopes. Thirteen of the nineteen spectra presented here were included in the original \megasaura\ data release \citep{rigbya}. Additionally, we include six galaxies from the upcoming expanded \megasaura\ sample (Rigby et al.\ in preparation): the Sunburst Arc \citep{dahle, rivera-thorsen17}, SPT~0142, SPT~0310, SPT~0356, PSZ~0441, and SPT~2325. We only include \megasaura\ spectra with SNR$>5$ per resolution element and without AGN signatures, which means that we exclude S1050+0017 (low SNR) and S2243$-$0935 (rest-frame optical AGN emission lines) from the original sample. The data reduction and full spectra of the original \megasaura\ spectra were presented in \citet{rigbya}.

The \megasaura\ galaxies span a redshift range $1.6<z<3.1$ (see \autoref{tab:megasaura}), with a rest-frame FUV spectral coverage of 1220--1950\AA\ for all nineteen galaxies at a median spectral resolution of $R = 3300$ (90~\kmsp, or 0.5\AA\ at 1500\AA) and SNR~$=21$ \citep{rigbya}. The spectra were corrected for Milky Way reddening in the observed frame using the \citet{cardelli} attenuation curve and the dust maps from \citet{green15}. We normalized the spectra to the median of the flux in the line-free region of 1267--1276\AA\ in the rest-frame. The \megasaura\ spectra contain all of the strong stellar features that constrain the stellar fits at a high SNR with a resolution similar to the stellar continuum models. Due to the superior combination of wavelength coverage and sensitivity, we use the \megasaura\ sample as our main sample instead of the HST/COS sample introduced below.

\subsubsection{Moderate-redshift stacked data}

While the individual \megasaura\ spectra have high SNR, many of the important stellar features are extremely weak. By averaging many observations together (often called \lq{}\lq{}stacking\rq{}\rq{}), the SNR increases by a factor of $\sqrt{N}$, where N is the number of spectra included in the stack. Consequently, a composite provides an average spectrum of an ensemble of galaxies at an extremely high SNR (see \autoref{stack}).

This stacking procedure has demonstrated the average FUV spectrum of galaxies at moderate redshifts \citep{shapley03, steidel16, rigbyb, steidel18}. We use two recent stacks: (1) the \megasaura\ lensed galaxies \citep{rigbyb} and (2) a stack of 30 star-forming, field galaxies at $z \sim 2.4$ from the Keck Baryonic Structure Survey \citep[KBSS;][]{steidel16}. The \megasaura\ stack has a peak SNR of 104 per spectral resolution with an average spectral resolution of R~$= 3300$ in the rest-frame wavelength range of 900--3000\AA. The \citet{steidel16} stack has a peak SNR of 38 per spectral resolution at a resolution of R~$=1400$ and rest-frame wavelength coverage between 1000--2200\AA.

\subsection{Low-redshift galaxies}
Our low-redshift sample consists of spectra from recent observations of 2 low-metallicity galaxies (PID: HST-GO-15099, PI: Chisholm) and the compilation from \citet{chisholm16} which are 40 local star-forming galaxies at $0.0007 < z < 0.1816$ with high SNR observations using the \textit{Cosmic Origins Spectrograph} \citep[COS;][]{cos} on the \textit{Hubble Space Telescope} (HST). The data were compiled from eight different HST programs, and we include the HST program IDs and references in \autoref{tab:cos}. The spectra were processed through CalCOS v2.20.1, reduced following the procedures in \citet{wakker2015}, binned by 20 pixels (0.2\AA, or 48~\kms at 1240\AA), and convolved to the resolution of the \starburst\ models (0.4\AA). We normalized the spectra near 1270\AA, similar to the \megasaura\ observations. These spectra are also corrected for Milky Way reddening in the observed frame. The COS observations are typically only made with one grating (G130M), such that the average rest-frame wavelength coverage is from 1150-1450\AA. This spectral regime contains many, but not all, of the stellar features that define the stellar age and metallicity (see \autoref{features}). As such, we display the \megasaura\ sample throughout this paper rather than the narrow HST/COS wavelength range.

\subsection{Host galaxy properties}
\label{properties}
The 61 galaxies studied here sample a wide range in host galaxy properties. \autoref{tab:megasaura} \& \ref{tab:cos} give literature nebular metallicity values, measured as 12+log(O/H) and referred to as \zism, which were determined using rest-frame optical emission lines. The \lq{}\lq{}gold-standard\rq{}\rq{} nebular metallicity method, the direct method, uses the temperature sensitive [\ion{O}{3}]~4363\AA\ emission line to determine the emission-line emissivities, which directly translates into oxygen abundances. However, [\ion{O}{3}] 4363\AA\ is a weak emission line that is challenging to observe in faint galaxies and is substantially weaker in higher metallicity regions. In the absence of [\ion{O}{3}] 4363\AA\ detections, calibration techniques have been developed using strong nebular emission lines to infer the ionization structure and nebular metallicity. These strong-line abundances are easily observed, however, the inferred absolute abundances from different calibration methods can be discrepant by as much as 0.7 dex \citep{kewley08}.

We have used direct metallicities whenever possible, however, the [\ion{O}{3}] 4363\AA\ emission line is faint and rarely observed at high-redshift; accordingly, the \megasaura\ \zism\ values are calculated using the \citet{pettini04} [\ion{N}{2}]/H$\alpha$ calibration \citep[table~2 of][]{rigbya}. Meanwhile, optical spectra of the entire low-redshift COS sample are not publicly available and the literature values have used different strong-line calibrations, precluding a uniform metallicity analysis. Of these, the O3N2 method from \citet{pettini04}, which uses the ([\ion{O}{3}]~5007/H$\beta$)/([\ion{N}{2}]~6583/H$\alpha$) ratio, is the most common empirical metallicity calibration used for our sample. We have also calculated 12+log(O/H) using the direct method for three galaxies in the sample following the methods of \citet{berg19}. Consequently, 12+log(O/H) (or \zism) is not uniformly calculated and may include systematic calibration uncertainties \citep{kewley08}. The low-redshift galaxies span a factor of 50 in \zism\, from 0.03--1.5~Z$_\odot$ (correspond to 12+log(O/H)$=7.22-8.87$).

All of the galaxies were selected as rest-frame UV-bright, star-forming galaxies such that the sample generally resides above the so-called star-forming main-sequence \citep[fig.~1 of][]{chisholm16}. Importantly, these rest-frame UV spectra probe a large, and varying, spatial scale. However, each spectrum samples a spatially \textit{unresolved} stellar population. In other words, each spectrum samples multiple young, UV-bright massive stars. At high redshifts, the physical scale depends on the lensing magnification, which varies from 2-200 \citep{sharon19}, such that the \megasaura\ spectra probe multiple  star-forming regions within the same galaxy \citep[see the analysis in ][]{bordoloi16}. Similarly, COS is a fixed circular aperture spectrograph with a 2\farcs5 diameter. The low-redshift sample spans a range of $z = 0.0007-0.1816$, which corresponds to the COS aperture covering a physical diameter of 50~pc to 10~kpc. Many different physical scales (multiple star-forming regions to entire galaxies) reside within the COS aperture.

\section{Stellar continuum modeling}
\subsection{Fitting procedure}
\label{procedure}
We fit the stellar continua of both the \megasaura\ and low-redshift samples by assuming that the observed spectra are combinations of multiple bursts of single age, single metallicity stellar populations. The light from these stellar populations then propagated through an ambient interstellar medium which attenuated the stellar continuum to produce the observed spectral shape. We fit this with a uniform dust screen model as:
\begin{equation}
    F_\text{obs}(\lambda) = 10^{-0.4\text{E(B-V)}k(\lambda)} \Sigma_i X_i M_i(\lambda) ,
    \label{eq:flux}
\end{equation}
where \ebv\ is the stellar attenuation parameter, $k(\lambda)$ is the reddening curve  from \citet{reddy16}, and $X_i$ is the linear coefficient multiplied by the $i$th single age stellar population model, $M_i$. Each $M_i$ corresponds to a single age and single metallicity (\zs) fully theoretical stellar continuum model (see \autoref{models}). Thus, $k(\lambda)$, E(B-V), $X_i$ and $M_i$ completely describe the shape and spectral features of the observed stellar continuum. 

We chose the \citet{reddy16} attenuation law because it is observationally defined down to 950\AA, closer to the ionizing continuum than other models \citep[e.g.][]{calzetti}. This is important because we are predominately interested in inferring the ionizing continua of massive stars. We tested the effect that changing the attenuation law has on the derived stellar properties and found it to most strongly affect the inferred E(B-V) values which were 0.01~mag redder, on average, using the \citet{calzetti} law versus the \citet{reddy16} law.

We fit the entire observable wavelength regime between 1220-2000\AA. We did not include wavelengths below 1220\AA\ due to strong Ly$\alpha$ features and the Ly$\alpha$ forest. The available rest-frame wavelength regime for each galaxy depends on the observational setup and the redshift of the galaxy. We masked out $\pm500$~\kms around strong ISM absorption and emission lines as well as absorption from foreground systems at lower redshifts (Rigby et al. in preparation). We optimized this masking velocity interval by studying the individual ISM features at our spectral resolution. One exception is the \civp~1550\AA\ region, where we only masked out from $-500$ to $+50$~\kms in order to include the crucial \civ P-Cygni emission (\autoref{civ}). Further, we manually masked out regions that are not stellar continuum features, such as abnormally large ISM absorption, Milky Way features, or sky-emission lines. We then fit for the $X_i$ and E(B-V) of each single age, single metallicity, fully theoretical stellar continuum model in \autoref{eq:flux} using \textsc{MPFIT} \citep{mpfit}. 

\subsection{Stellar models}
\label{models}

The theoretical stellar continuum models ($M_i(\lambda)$) are key to the spectral population synthesis. We used both single star models \citep{claus99, claus2010, leitherer14} and models that include binary evolution \citep{eldridge17, stanway18} to quantify the effect that binary evolution has on the ionizing and non-ionizing continua of massive stars (\autoref{comp}). We chose the stellar atmosphere models below because they are the most comparable to each other \citep{eldridge17} and have the most observationally motivated mass outflow rates \citep{claus2010}. For both models we assumed a standard Kroupa initial mass function \citep[IMF;][]{kroupa} with a broken power-law with a high-(low-) mass exponent of 2.3 (1.3) and a high-mass cut-off of 100~M$_\odot$. \citet{steidel16} demonstrated that the high-mass cut-off weakly impacted the spectral synthesis fits to FUV stellar continua by showing that the best-fit stellar continua did not change drastically using either a 300~M$_\odot$ or 100~M$_\odot$ cut-off (their fig.~7). 

Star light between 1200-2000\AA\ is dominated by young, massive O-stars. Consequently, we used fully theoretical stellar models with young ages corresponding to 1, 2, 3, 4, 5, 8, 10, 15, 20 and 40~Myr. At 1270\AA, a 20~Myr stellar population of a given initial mass is nearly two orders of magnitude fainter than a 1~Myr stellar population, while older populations are fainter still \citep{claus99, eldridge17}. Moreover, the UV stellar continua of older stellar populations evolve more slowly with time, such that there are small spectral differences between a 50 and 100~Myr stellar population \citep[fig. 5 in ][]{demello}. Finally, the effective temperature of B-star populations greater than 40~Myr drops below the 20,000~K threshold where high-resolution stellar templates are computed using the \textsc{WM-Basic} code \citep{claus2010}. Thus, the selected age range includes models of the most luminous O-stars whose spectral features vary rapidly with time at sufficient spectral resolution to resolve these important spectral features. 

We use the five \zs\ models that are available from the Geneva stellar atmospheres \citep{geneva94}: 0.05, 0.2, 0.4, 1.0, and 2.0~Z$_\odot$. Combined, each observed spectrum is fit with 50 fully theoretical stellar models and one free parameter for the dust attenuation, for a total of 51 total free parameters.

Other stellar physics impact the production of ionizing photons, such as rapid rotation \citep{levesque, leitherer14, choi17} and a varying (or stochastically populated) IMF \citep{leitherer95, rigby04, crowther07}. However, the stellar population synthesis routines only include two \zs\ \citep[0.14~Z$_\odot$ and 1~Z$_\odot$, but see the recent extension to 0.02~Z$_\odot$ from][]{groh}, which samples metallicity too coarsely for our fitting. Consequently, binary models are the only alternative stellar model that we discuss below.

\subsubsection{The fiducial case: \starburst\ single-star models}
\begin{figure*}
\begin{centering}
  \includegraphics[width = \textwidth]{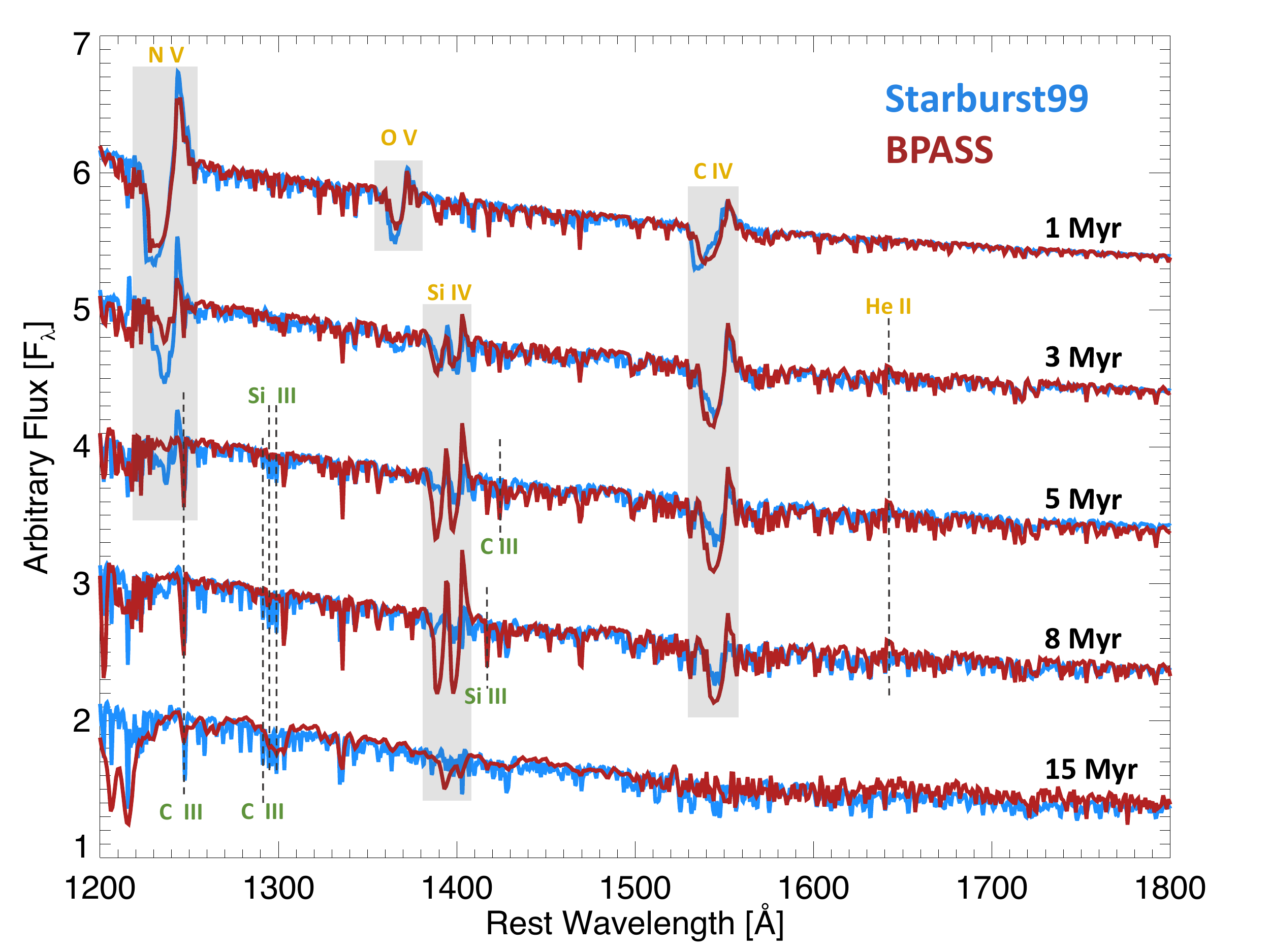}
\end{centering}
\caption{The fully theoretical, rest-frame FUV stellar continuum of five \starburst\ (blue) and \bpass\ (red) single burst models with a metallicity of 0.4~Z$_\odot$. Each spectrum shows a different age, with age decreasing from 1~Myr at the top to 15~Myr at the bottom (labeled on the right). Prominent stellar wind (orange) and photospheric (green) lines are labeled near the models where the features are strongest. The displayed models are a subset of the 50 stellar continuum models that determine the stellar population properties and demonstrate the spectral variations in the rest-frame FUV with stellar population age. }
\label{fig:mods}
\end{figure*}

We used the fully theoretical \starburst\ models with Geneva atmospheres that incorporate high-mass loss rates \citep{geneva94} as our fiducial model. These models have a spectral resolution of 0.4\AA, which match the spectral resolution of the \megasaura\ spectra. We convolved the models with a Gaussian to match the observed spectral resolution of each individual \megasaura\ spectrum, as measured from the optical sky emission lines \citep{rigbya}, and resampled the stellar models onto the wavelength grid of the observations. Similarly, we convolved the higher resolution HST/COS data to the 0.4\AA\ spectral resolution of the \starburst\ models from the spectral resolution measured from the Milky Way absorption lines \citep{chisholm16}. Each model spectrum was normalized to the median flux density between 1267--1276\AA. 

The \starburst\ stellar models were created using the \textsc{WM-Basic} method \citep{pauldrach01} and densely sample the high-mass portion of the Hertzsprung-Russell diagram up to temperatures of 20,000~K. \textsc{WM-Basic} does not calculate high-resolution models below these temperatures \citep{claus2010}. Consequently, we chose the ten stellar ages between 1--40~Myr listed above with stellar temperatures greater than 20,000~K. These models include Wolf-Rayet (WR) stars using the Potsdam Wolf-Rayet code  \citep[PoWR; ][]{sanders15}, but the evolutionary tracks predict that few if any WR stars are present in low-metallicity stellar population, such that the WR spectra are rarely incorporated into 0.2--0.4~Z$_\odot$ \starburst\ models \citep{leitherer18}.

\subsubsection{The binary evolution case: \bpass\ models}
\label{bpass}
We also used the Binary Population and Spectral Synthesis (\bpass) v2.2.1 models\footnote{https://flexiblelearning.auckland.ac.nz/bpass/9.html}, which include binary star evolution \citep[][]{stanway18}. \bpass\ models have a larger metallicity range, but, for consistency, we used the same five \zs\ available from the Geneva models. \bpass\ models use a custom set of O-star models created with \textsc{WM-Basic} at 1\AA\ resolution for O-stars with temperatures greater than 25,000~K \citep{eldridge17}. Temperatures less than this have the \textsc{basel}v3.1 and C3K models which have spectral resolution of 20\AA\ below 1500\AA\ \citep{westera, leborgne, conroy12, conroy14}.  Therefore,  \bpass\ models are lower resolution when ages are greater than 20~Myr for any metallicity and when ages are greater than 15~Myr for metallicities greater than 0.4~Z$_\odot$ (see \autoref{fig:mods}). This spectral resolution is too low to diagnose many of the narrow B-star features of older stellar populations and cannot be used to distinguish older stellar populations.  For this reason, we chose the \starburst\ models as our fiducial model. We return to this issue in \autoref{bpass_props}.

\subsection{The nebular continuum}
\begin{figure}
\includegraphics[width = 0.5\textwidth]{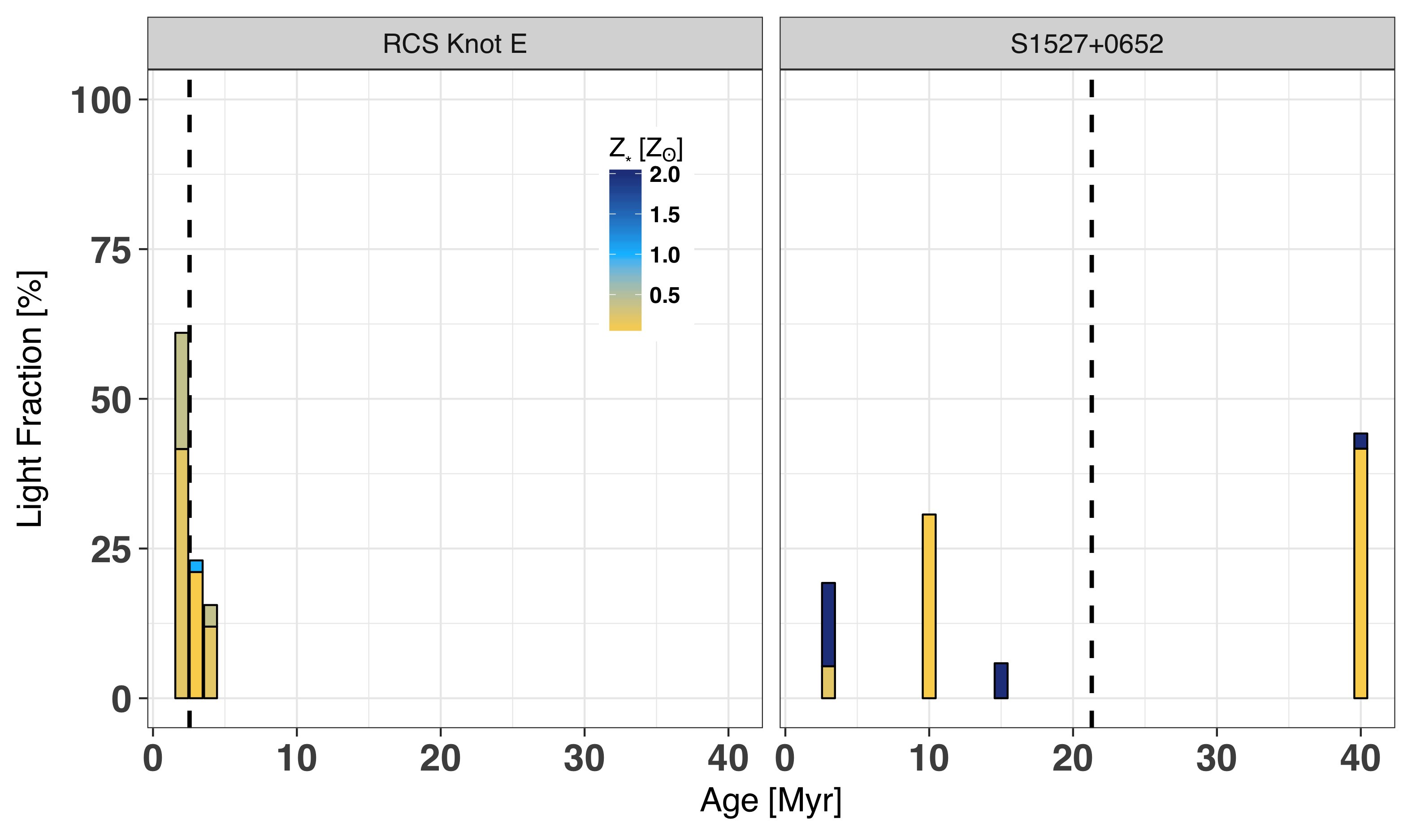}
    \caption{Comparison of the fitted light fractions for each stellar age used in \autoref{eq:flux} to determine the fit to the FUV spectra for two galaxies: RCS~Knot~E (left panel) and S1527+0652 (right panel). The total light-weighted ages are given as the dashed vertical lines. RCS~Knot~E has a very young light-weighted age, 2.5~Myr, with all of the light coming from three ages: 2, 3, and 4~Myr. S1527+0652 has an older light-weighted age and the fitted models have a mix of old and young populations. Each bar is color-coded by the stellar metallicity of the corresponding model. }
    \label{fig:light_frac}
\end{figure}

Young massive stars produce large amounts of ionizing photons which produce free-free, free-bound, and two-photon nebular continuum emission. The nebular continuum heavily contributes to the total continuum flux at young ages, low metallicities, and redder wavelengths \citep{steidel16, byler18}. For a stellar continuum metallicity of 0.05~Z$_\odot$ and a stellar age of 1~Myr, the nebular continuum is 25\% of the stellar continuum at 2000\AA. 

We created a nebular continuum for each age, metallicity, and stellar model by processing the stellar continuum models through \textsc{cloudy} v17.0 \citep{ferland}. We assumed that the gas-phase metallicity and stellar metallicity were the same (\autoref{metal}), an ionization parameter of log(U)~$ =-2.5$, and $n_\text{H} = 100$~cm$^{-3}$. We produced a nebular continuum for each stellar population, added the output nebular continua to the stellar models, and normalized by the flux between 1267--1276\AA. The inclusion of the nebular continuum produces redder stellar models than before, which has a pronounced impact on the fitted E(B-V) of young stellar populations. 

We tested the effect that different ionization parameters have on the fitted stellar ages and metallicities by also creating models with log(U)~$= -2$, $-$2.3, $-$2.7, and $-$3.0. We found that the reduced $\chi^2$ of the resultant fits do not change statistically for the different log(U) values. Consequently, we adopted a midrange log(U)~=~$-$2.5 for all galaxies.

\subsection{Stellar population parameters derived from the fits}
\label{parameters}

The observed stellar continua are fit by statistically determining the linear multiplicative coefficient, $X_i$, for the 50 single age, single metallicity stellar models ($M_i$; \autoref{eq:flux}). These linear coefficients can take any value greater than or equal to 0 and \textsc{MPFIT} determines the linear combination that best fits the observed stellar continuum. In practice, the code typically assigns $X_i = 0$ values to most of the stellar models, and only gives power to a small subset of the models (on average 6 models have light fractions >0\%). \autoref{fig:light_frac} shows the distribution of the light at a given age for two observed stellar populations fitted with the multiple population method. All of the light from RCS~Knot~E comes from a very young, moderate metallicity stellar population (left panel); while the stellar light from S1527+0652 is broadly distributed across age and metallicity (right panel).

We determined intrinsic stellar parameters from the fitted coefficients in \autoref{eq:flux}. The derived parameters are a weighted average of the total light at 1270\AA\ attributed to the individual stellar models ($X_i$). Thus, each property derived below is a \lq{}\lq{}light weighted\rq{}\rq{} property. First, the light fraction ($f_i$) that each model contributes to the total intrinsic flux at 1270\AA\ is defined as:
\begin{equation}
    f_i = \frac{X_i}{\Sigma_i X_i} .
\end{equation}
Secondly, the light-weighted age at 1270\AA\ is defined as:
\begin{equation}
    \text{Age} =  \frac{\Sigma_i X_i \text{Age}_i}{\Sigma_i X_i} .
    \label{eq:age}
\end{equation}
The light-weighted ages of RCS~Knot~E and S1527+0652 are indicated as vertical lines in the left panel of \autoref{fig:light_frac}. Finally, we computed the light-weighted stellar metallicity as:
\begin{equation}
    Z_s = \frac{\Sigma_i X_i Z_i}{\Sigma_i X_i} .
\end{equation}
 These three parameters describe the properties of the observed stellar populations. The uncertainties on these parameters were derived by varying the observed flux density at every wavelength by a random Gaussian kernel with width equal to the flux uncertainty at that wavelength. We then recalculated the ages and \zs, tabulated each value, and repeated the procedure 100 times.  The standard deviation of each age and \zs\ distribution is the uncertainty on the age and \zs, respectively. We include the $f_i$, \zs, and light-weighted ages in the electronic version of the Appendix. 

All of these stellar population properties (age and metallicity) are light-weighted at 1270\AA; they cannot be directly compared to similar properties derived at other wavelengths. Younger stars produce relatively more light at bluer wavelengths than older stars, biasing the light from young stars to bluer wavelengths.  The ages derived from full SED modeling using optical and near-infrared observations will inherently return older ages than we estimated because optical light comes from older stars. 

An important measure of the ionizing continuum is the ratio of the intrinsic flux density at 900\AA\ to the flux density at 1500\AA\ (\frat). Observations compare the observed \frat\ to the intrinsic \frat\ to determine the fraction of ionizing photons that escape galaxies \citep{steidel01}. We estimate the intrinsic \frat\ by extending the stellar population fits to bluer wavelengths than the observations and removing the contributions from dust attenuation (setting E(B-V)~=~0 in \autoref{eq:flux}). The high-resolution \starburst\ models are only defined at $>905$\AA, consequently, we created low-resolution \starburst\ models \citep[20\AA\ resolution;][]{claus99} using the same model parameters and fitted light-fractions as the high-resolution models. We then measured the median model flux density between 895--906\AA, for F$_{900}$, and 1495--1506\AA, for F$_{1500}$.

Ionizing photons with higher energies create high-ionization gas (e.g. O$^{++}$). We also inferred the stellar flux density between 510--540\AA\ (F$_{525}$) and between 280--320\AA\ (F$_{300}$) to determine how many high-energy photons a given stellar populations produces. F$_{525}$ (with photon energies of 24~eV) probes photons that singly ionize oxygen, but do not ionize helium. Meanwhile, F$_{300}$ (photon energies of 41~eV) probes photons that doubly ionize oxygen and singly ionize helium, but do not doubly ionize helium. These wavelengths were carefully chosen to probe the peak of the stellar SEDs, while avoiding contributions from strong stellar absorption and emission features (see \autoref{fig:mods}).

Finally, all derived parameters are either flux density ratios (e.g., \frat) or derived from normalized spectra (e.g., stellar age). This means that the stellar population parameters are independent of the intrinsic luminosity, which depends on the magnification from gravitational lensing.



\section{Relating spectral features and inferred stellar population properties}
\label{features}

\begin{table}
\caption{Prominent stellar features}
\begin{tabular}{lcc}
Line & \bpass\ Age & \starburst\ Age \\
& [Myr] & [Myr]\\
\hline
& Stellar wind lines & \\
\hline
\nv\ 1240\AA\ & 1$-$5 & 1$-$10 \\
\ion{O}{5} 1371\AA\ & 1$-$2 & 1$-$4 \\
\siiv 1400\AA\ & 3$-$15 & 3$-$5 \\
\civ 1550\AA\ & 1$-$15 & 1$-$10 \\
\heii\ 1640\AA\ & 4$-$20 & 3$-$4 \\
\hline
& Photospheric lines & \\
\hline
\ciii\ 1247.4\AA\ & 4$-$15 & 5$-$40 \\
\siiii\ 1294.5\AA\ & -- & 5$-$40 \\
\ciii\ 1296.3\AA\ & -- & -- \\
\siiii\ 1296.7\AA\ & -- & 5$-$40 \\
\siiii\ 1298.9\AA\ & -- & 5$-$40 \\
\ion{Fe}{5} 1346$-$1365\AA\ & 1$-$15 & 1$-$20 \\
\ion{Fe}{5} 1427$-$1430\AA\ & 1$-$15 & 1$-$20 \\
\ion{S}{5} 1501.8\AA & 1$-$8 & 1$-$40\\ 
\ion{Fe}{4} 1526$-$1534\AA\ & 3$-$10 & 3$-$40 \\
\siii 1533\AA\ & -- & 8$-$40 \\
\ion{Fe}{3} 1923$-$1966\AA\ & 10$-$40 & 10$-$40 \\
\end{tabular}
\tablecomments{The first column gives the feature and wavelength. The second and third columns give the age range that the features are found in a 0.4~Z$_\odot$ \bpass\ and \starburst\ model, respectively. }
\label{tab:lines}
\end{table}

Often times the ages of stellar populations are deduced using the broadband UV through IR SED shapes. While the SED shape provides important age information, it is often degenerate with metallicity and dust attenuation. By contrast, fitting the stellar spectral features with theoretical stellar templates simultaneously determines the age, metallicity, and dust attenuation of the stellar populations. Consequently, the light-weighted ages and metallicities derived above are driven by spectral features which are less degenerate than the spectral shape alone (see \autoref{fig:mods} and \autoref{tab:lines}).

The two main types of FUV stellar features are strong, broad stellar wind P-Cygni features and weak stellar photospheric absorption lines. Both types of features can be contaminated by neighboring ISM absorption, and require high SNR and moderate spectral resolutions to resolve. In the following two subsections, we discuss both types of spectral features individually, and illustrate how the individual features relate to the inferred stellar population ages and metallicities. The purpose of these subsections is not to advocate for determining the stellar ages and metallicities using single features, but rather to demonstrate that the stellar properties inferred from the full spectral fits are entirely consistent with the trends of stellar spectral features.

\subsection{Stellar wind features}
The most notable stellar features in the FUV are the broad blueshifted absorption and redshifted emission profiles (called P-Cygni profiles; orange labels in \autoref{fig:mods}). These P-Cygni profiles arise from strong winds that are radiatively driven off of stellar photospheres \citep{castor75, lamers}. The terminal velocity, ionization structure, and mass-outflow rates sensitively depend on the stellar luminosity \citep{castor75, lamers93, lamers95, leitherer95, puls96, kudritzki00}. The terminal velocity describes the maximal velocity extent of the absorption component, while the mass-loss rate determines the depth of the profile. In turn, these establish the shape of the P-Cygni absorption and emission.

Whether a given ion is observed as a P-Cygni profile in the wind depends on the ionization structure of the stellar wind. The peak ionization stages for stellar winds typically are the \ion{C}{5}, \ion{N}{4}, \ion{O}{4}, and \ion{Si}{5} states \citep{lamers}, none of which have resonant transitions in the rest-frame FUV. Alternatively, the presence of adjacent ionization stages with P-Cygni profiles (\ion{N}{5}, \ion{C}{4}, or \ion{Si}{4}) provides information on the stellar temperature of the most luminous stars and, by inference, the stellar population age. The age of a stellar population can be inferred from the strength of the observed P-Cygni transitions: \ov\ and \nv\ have the highest ionization states and are strongest in stars with lifetimes of 2-3~Myr,  while \civ and \siiv are lower ionization and peak in stars with lifetimes near 5~Myr. 

Stellar temperature, or age, is not the sole determinant of the stellar wind profiles. Metals in the photospheres of hot stars absorb continuum photons which is what accelerates the gas off the stellar surface. Consequently, the stellar metallicity determines both the acceleration and mass outflow rate of the stellar wind \citep{lamers, vink02}. The terminal velocities and mass-loss rates of O-stars with $>0.2$~Z$_\odot$ have been empirically determined to scale as Z$_\ast^{0.13}$ and Z$_\ast^{0.69}$, respectively, \citep{leitherer92, vink02}. Lower metallicity stellar winds of resolved individual stars have not been observed, consequentially the mass outflow rate and terminal velocity relations have been extrapolated to lower metallicities. These relations illustrate how stellar wind P-Cygni absorption profiles scale with stellar metallicity.

In the next five sub-sections we walk through the individual P-Cygni lines in the FUV. Each subsection explores the theoretical and observed wind features and their relationship to the inferred stellar population age and metallicity. In \autoref{young} we conclude that the \nv\ P-Cygni and \heii\ emission are strong in very young stellar populations, while the shape of the \civ P-Cygni profile changes both with stellar age and metallicity (see \autoref{fig:age}). Conversely, the \siiv of our sample is dominated by interstellar absorption, and the \ov\ line is not observed. Collectively, the stellar wind profiles mimic the inferred stellar ages and metallicities.

\subsubsection{The \ion{C}{4} P-Cygni feature}
\label{civ}
\begin{figure}
\begin{centering}
  \includegraphics[width = 0.5\textwidth]{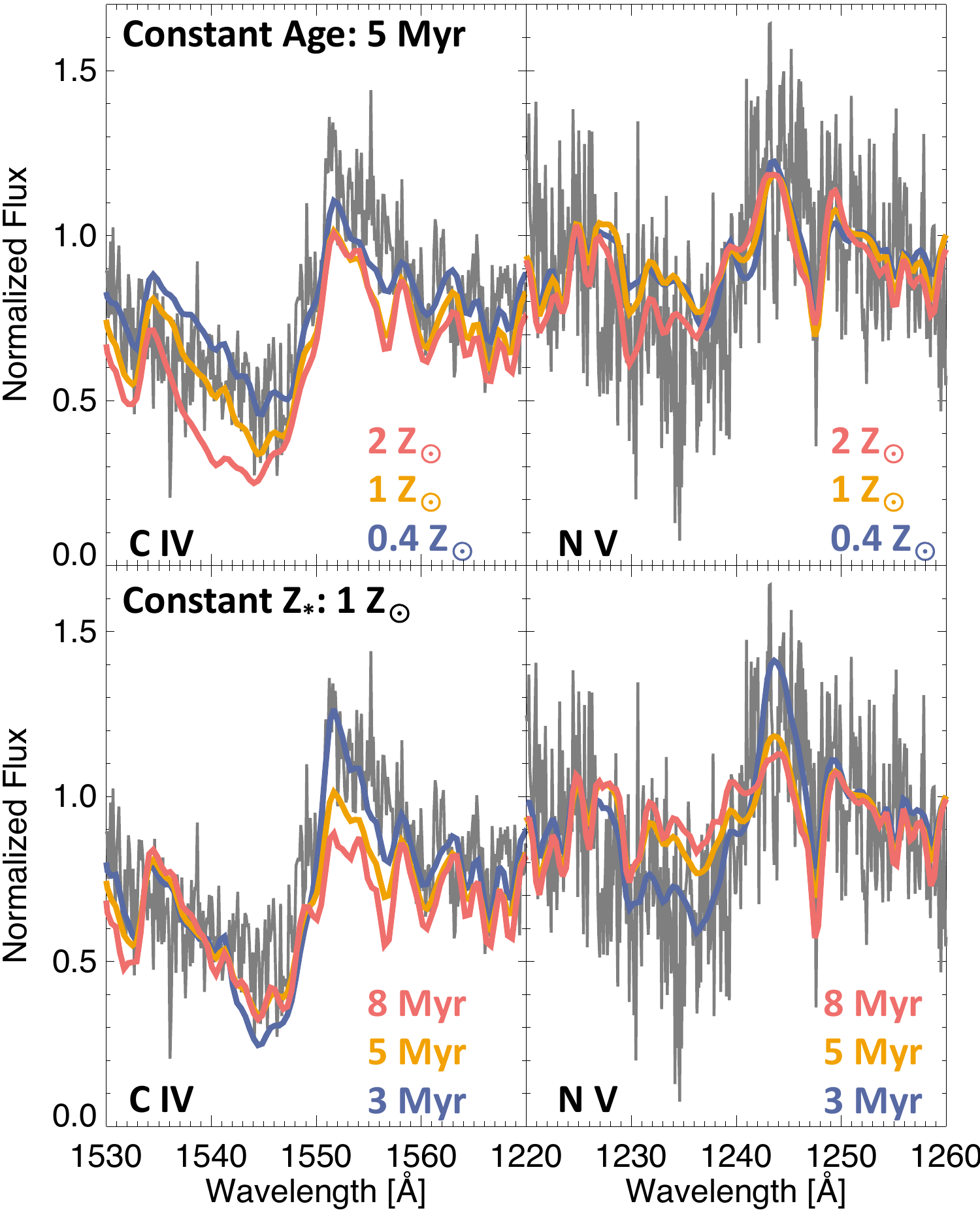}
\end{centering}
\caption{The dependence  of  the \civp~1550\AA\ (left panels) and \nvp~1240\AA\ (right panels) stellar wind profiles on stellar age and metallicity. The observed stellar wind profile from S0033+0242 is shown in gray in each panel. In the upper panels we overlay three \starburst\ theoretical models with a constant 5~Myr age but with a varying stellar metallicity (\zs) of 2 (red), 1 (gold), and 0.4~Z$_\odot$ (blue). The bottom two panels show three \starburst\ models with a constant \zs~=~1~Z$_\odot$ but with varying stellar age of 8 (red), 5 (gold) and 3~Myr (blue). \zs\ does not change the \nv\ profile (upper right panel), but the emission profile strongly peaks for ages <5~Myr (lower right panel). \zs\ strongly impacts the \civ absorption (wavelengths less than 1550\AA), but weakly impacts the \civ emission (upper left panel). Conversely, the stellar age strongly impacts the \civ emission (wavelengths greater than 1550\AA) but weakly impacts the \civ absorption (lower left panel). By eye, the \civ and \nv\ profiles of S0033+0242 are best-fit between the blue and gold profiles in each panel. This is consistent with the light-weighted age and metallicity of $5\pm0.5$~Myr and $0.84\pm0.04$~Z$_\odot$. } 
\label{fig:age}
\end{figure}

\begin{figure}
\begin{centering}
  \includegraphics[width = .5\textwidth]{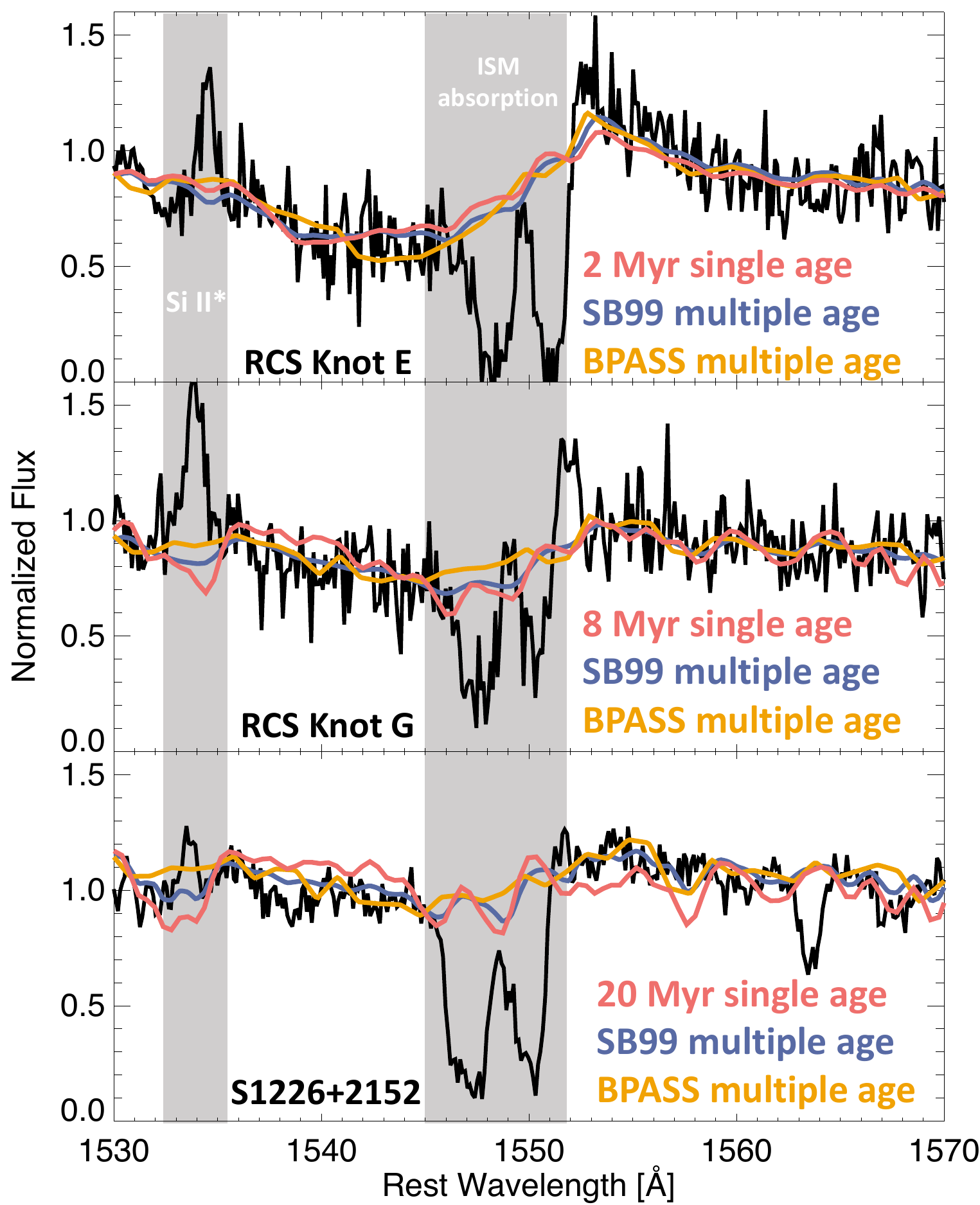}
\end{centering}
\caption{The dependence of the \civ wind profile on stellar age. Each panel shows an observed galaxy spectra (black line) ordered by descending fitted light-weighted ages of $2.5\pm0.1$, $11\pm1.8$, and $26\pm1.3$~Myr. The fitted metallicities are similar ($0.29\pm0.04$~Z$_\odot$). The colored curves are theoretical models. A single age, 0.2~Z$_\odot$ \starburst\ theoretical model with an age nearest to the fitted light-weighted age is overplotted in red. The multiple age, multiple metallicity \bpass\ (blue) and \starburst\ (gold) fits are also shown. The single age models describe the overall shape of the younger \civ profiles (e.g. RCS Knot E and G), while the multiple age fits are required to capture the full details of the older profiles (e.g. S1226+2152). Gray regions correspond to interstellar \civ absorption and \siiifs emission.}
\label{fig:age_mods}
\end{figure}

\begin{figure}
\begin{centering}
  \includegraphics[width = .5\textwidth]{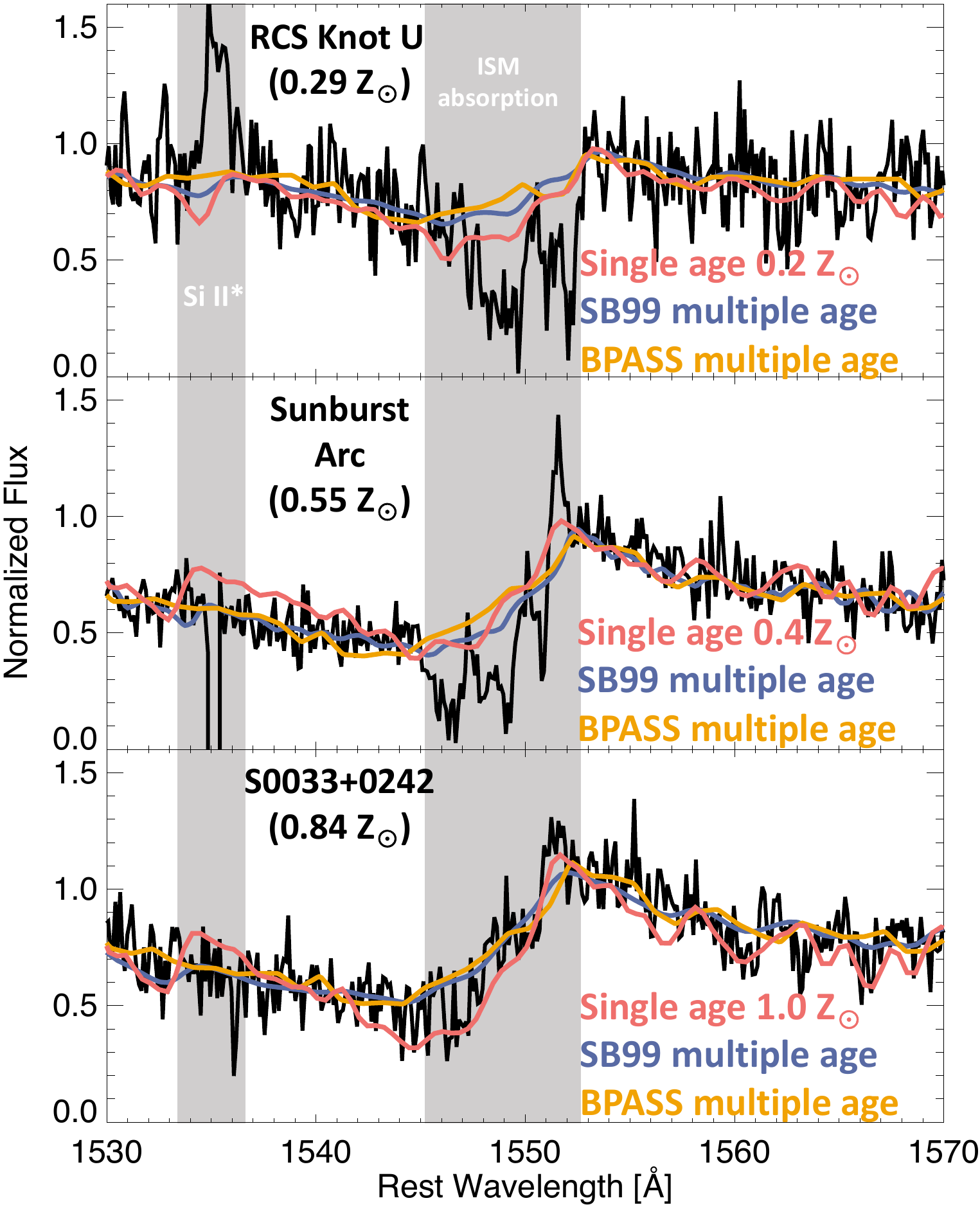}
\end{centering}
\caption{The dependence of the \civ stellar wind profile on stellar metallicity (\zs) at nearly constant stellar age ($4\pm1$~Myr). The three panels show the observed spectra (in black) ordered by descending fitted \zs\ of $0.29\pm0.07$, $0.55\pm0.04$, and $0.84\pm0.04$~Z$_\odot$. The \civ stellar absorption increases with increasing \zs. The colored curves correspond to various models. A 5~Myr \starburst\ model is shown in red in each panel with \zs\ closest to the inferred light-weighted \zs. The multiple age, multiple metallicity \bpass\ (blue) and \starburst\ (gold) fits are also shown. At these young ages, the single age models describe the overall shape of the \civ lines. Gray regions correspond to interstellar \civ absorption and \siiifs emission.}
\label{fig:age_mods_z}
\end{figure}

The \civ feature is strong, broad, and has a P-Cygni profile for all of the \megasaura\ galaxies (there is not \civ coverage for many of the low-redshift COS spectra). Further, the \civ P-Cygni profile is sufficiently broad that stellar and interstellar components can easily be separated with moderate spectral resolution (see shaded regions in \autoref{fig:age_mods} and \autoref{fig:age_mods_z}). The \civ profile probes stellar outflows from 1-10~Myr populations and the absorption component distinctly varies with \zs\ (\autoref{fig:age}). This makes it an ideal diagnostic of stellar age and metallicity.

Different portions of the \civ profile depend on either \zs\ or stellar age (left panels of \autoref{fig:age}). At a constant 5~Myr age (upper left panel), the \civ absorption of the \starburst\ models deepen and broaden as \zs\ increases from 0.4~Z$_\odot$ (blue line) to 1~Z$_\odot$ (gold line) to 2~Z$_\odot$ (pink line), because the \civ mass-loss rate strongly increases with \zs. Conversely, the models predict that the \civ emission is nearly independent of \zs. Meanwhile, at a constant 1~Z$_\odot$ metallicity (lower left panel), the modeled \civ absorption is independent of stellar age, but the emission strongly peaks at ages <5~Myr (blue line). In summary, the \civ absorption varies with \zs\ and the \civ emission varies with stellar age.

The temporal evolution of \civ found in the models is clearly corroborated by the light-weighted ages. \autoref{fig:age_mods} shows three  \megasaura\ spectra ordered in descending light-weighted stellar age and each have similar light-weighted metallicities ($0.29\pm0.06$~Z$_\odot$). Overplotted in red is a single age \starburst\ model with a population age nearest to the inferred light-weighted age. The \civ emission is strongest in the youngest populations and adequately matches the triangular wind emission from the RCS~Knot~E spectrum. Meanwhile, RCS~Knot~G, a different region from the same galaxy, has weaker P-Cygni emission (although note the narrow nebular \civ emission), similar to an 8~Myr single age population. Finally, S1226+2152 has an even older inferred light-weighted age; the \civ P-Cygni feature has nearly disappeared and has been replaced by a flat \civ feature with narrow, interstellar absorption. The derived light-weighted ages of each spectrum are consistent with the by-eye P-Cygni variation: Knot~E, Knot~G, and S1226+2152 have \starburst\ light-weighted ages of 2.5, 11, and 26~Myr, respectively.  

While the single age models reflect the \civ profiles of the youngest populations, they do not describe all of the galaxy-to-galaxy P-Cygni variations in the older populations. The \civ absorption from RCS Knot~G is deeper at bluer velocities than the 8~Myr single age model, while S1226+2152 has a small amount of redshifted emission. The observed stellar populations are not single age populations, but the multiple age fits capture the different populations (blue and gold lines in \autoref{fig:age_mods} are the \starburst\ and \bpass\ fits). While a single age population largely describes the \civ emission from the youngest populations, older populations require a mix of both young and old stars to match the observations.
\begin{figure}
  \includegraphics[width = 0.5\textwidth]{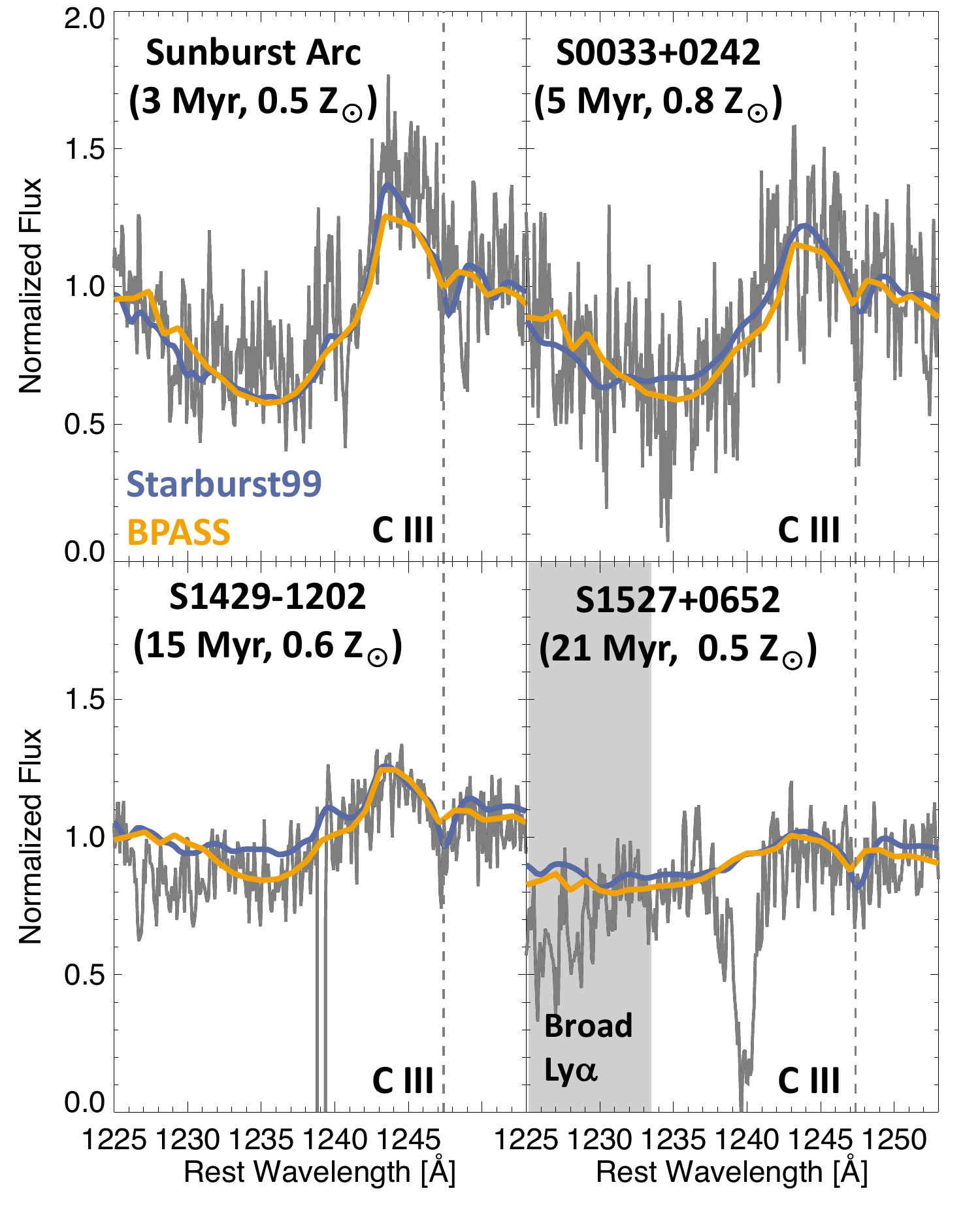}
\caption{ The observed \nv~1240\AA\ stellar wind profiles of four galaxies in gray. The panels are arranged from youngest stellar age in the upper left to oldest in the lower right. The  \nv\ wind profile transitions from a strong P-Cygni profile in the upper left to a nearly flat continuum in the lower right as the light-weighted age declines. The four stellar populations have similar \zs\ but different light-weighted ages, emphasizing that the \nv\ line chiefly depends on stellar age (see the right panel of \autoref{fig:age}). The \starburst\ (blue) and \bpass\ (gold) fits are included in each panel. The dashed gray line denotes the \ciii~1247\AA\ photospheric line.}
\label{fig:nv}
\end{figure}

Metallicity has a similarly strong impact on the shape of the observed \civ profile, but this time on the absorption portion of the P-Cygni profile. The upper left panel of \autoref{fig:age} shows that \zs\ mostly impacts the depth and width of the absorption component. This is further illustrated in \autoref{fig:age_mods_z} which shows three \megasaura\ spectra with nearly constant stellar age (5, 3, and 5~Myr) but with increasing stellar metallicity. At 0.3~Z$_\odot$ (top panel), the \civ absorption only reaches a depth of 0.6 in normalized flux units. Increasing the metallicity to 0.6~Z$_\odot$ (middle panel) creates a pronounced, broad P-Cygni absorption profile that reaches 0.5 in normalized flux units. Finally, by 0.8~Z$_\odot$ (bottom panel) the stellar wind dominates the \civ spectral region  and the absorption reaches 0.4 in normalized flux units. The stellar emission does not strongly change with \zs, as the stellar emission peaks near $1.05\pm0.05$ for all of these spectra. Both the stellar models and the observed spectra indicate that the stellar metallicity strongly shapes the absorption component of the \civ P-Cygni profile.

To summarize: the \civ profile strongly varies with the inferred light-weighted stellar population properties. The \civ P-Cygni absorption depends on the \zs\ and the emission depends on the stellar age (\autoref{fig:age}). The multiple age and multiple metallicity fits to the \civ P-Cygni profiles mimic changes in the inferred stellar age (\autoref{fig:age_mods}) and metallicity (\autoref{fig:age_mods_z}).

\subsubsection{The \ion{N}{5} P-Cygni feature}
\label{nv}

The \nvp~1240\AA\ stellar wind profile largely depends only on stellar age. The age dependence arises because the dominant ionization state of an O-star wind is N$^{+++}$, but as the stellar temperature increases with decreasing \zs, N$^{+++}$ gas is heated into the N$^{4+}$ state. This heating produces relatively more gas in the N$^{4+}$ ionization state for low \zs\ winds than higher \zs\ winds and nearly balances the decreasing metallicity \citep{kudritzki98, lamers, claus2010}. The models in the upper right panel of \autoref{fig:age} show the negligible \nv\ variation with \zs, while strong absorption and emission only occurs at ages young enough to produce \nv\ stellar winds  (bottom right panel of \autoref{fig:age}). \nv\ only arises from very young (<5~Myr) stellar populations. 

A strong \nv\ profile is detected in most FUV spectra in \autoref{fig:nv}. The exceptions are the oldest stellar populations which do not show P-Cygni profiles in either the \civ or the \nv\ ionization state (e.g. S1527+0652). Weak \civ and a non-detection of \nv\ indicates that there is not currently a young (<8~Myr) stellar population in these older galaxies (see the discussion in \autoref{cont}). In contrast, the normalized flux of the \nv\ profile from the Sunburst Arc varies by a factor of 2.5 from the absorption depth to the emission peak, illustrating the strong \nv\ P-Cygni profiles in populations with very young light-weighted ages. This trough-to-peak ratio decreases with increasing inferred stellar population age in \autoref{fig:nv}: it is 2.2 for S0033+0242 with an age of 5~Myr and 1.4 for S1429-1202 with an age of 15~Myr.


\subsubsection{The \ion{Si}{4} P-Cygni feature}
\label{siiv}
\begin{figure}
  \includegraphics[width = 0.5\textwidth]{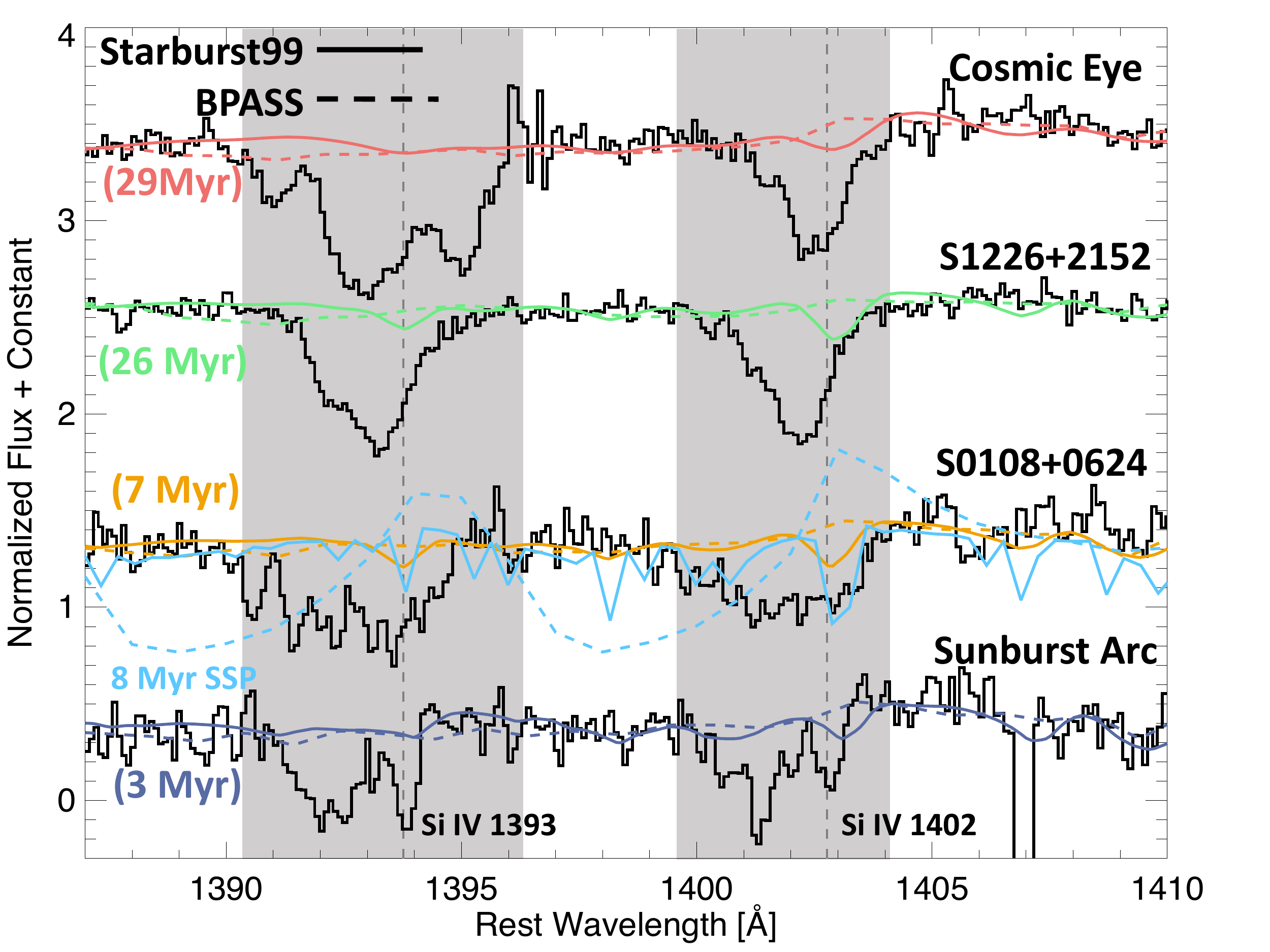}
\caption{ The \siiv~1400\AA\ stellar wind region for four galaxies ordered by ascending light-weighted age. The multiple age and multiple metallicity \starburst\ (solid line) and \bpass\ (dashed line) fits are overplotted for each galaxy. Areas of strong ISM absorption, which dominates the \siiv spectral regime, are shaded in gray. An 8~Myr single age model is included over the S0108+0624 spectra in light-blue to demonstrate that \bpass\ models predict strong P-Cygni profiles at these ages, while the \starburst\ models do not.  }
\label{fig:siiv}
\end{figure}

P-Cygni \siiv is generally observed in supergiants (evolved stars) or in metal-rich stars because the dominant ionization state in main-sequence stellar winds is the Si$^{4+}$ state. Thus, the Si$^{+++}$ ionization state is too low of a temperature to trace the bulk of the O-star wind \citep[the opposite behavior as \nv\ above;][]{walborn84}. As O-stars evolve into supergiants, their stellar photospheres expand and the stellar winds become denser, causing the dominant ionization state, Si$^{4+}$, to recombine into Si$^{+++}$. These denser winds produce prominent \siiv P-Cygni features in the spectra of evolved O-stars \citep{drew89, pauldrach90}. Similarly, the ionization structure also shifts towards lower ionization stages as larger stellar metallicities produce relatively more Si$^{+++}$ in their stellar winds.  Thus, the stellar models predict that \siiv P-Cygni profiles are only strong in stellar populations with a lower ionization structure, either from higher metallicity stars or an out-sized contribution of evolved stars.

Strong \siivp~1400\AA\ ISM absorption is seen in \autoref{fig:siiv}, but \siiv is rarely observed to have a stellar P-Cygni profile. The entirety of the observed \siiv absorption can be explained by narrow interstellar absorption that reaches nearly zero flux. After accounting for interstellar absorption (gray regions in \autoref{fig:siiv}), the observed \siiv region is nearly featureless. The lack of strong P-Cygni \siiv suggests that the stellar populations within our sample are either metal poor (see \autoref{metal}) or not dominated by evolved stars. We conclude that the observed \siiv regions are dominated by interstellar absorption and that \siiv does not strongly vary with the stellar population properties.

\subsubsection{The \ion{He}{2} emission feature}

\label{heii}
\begin{figure}
\includegraphics[width = 0.5\textwidth]{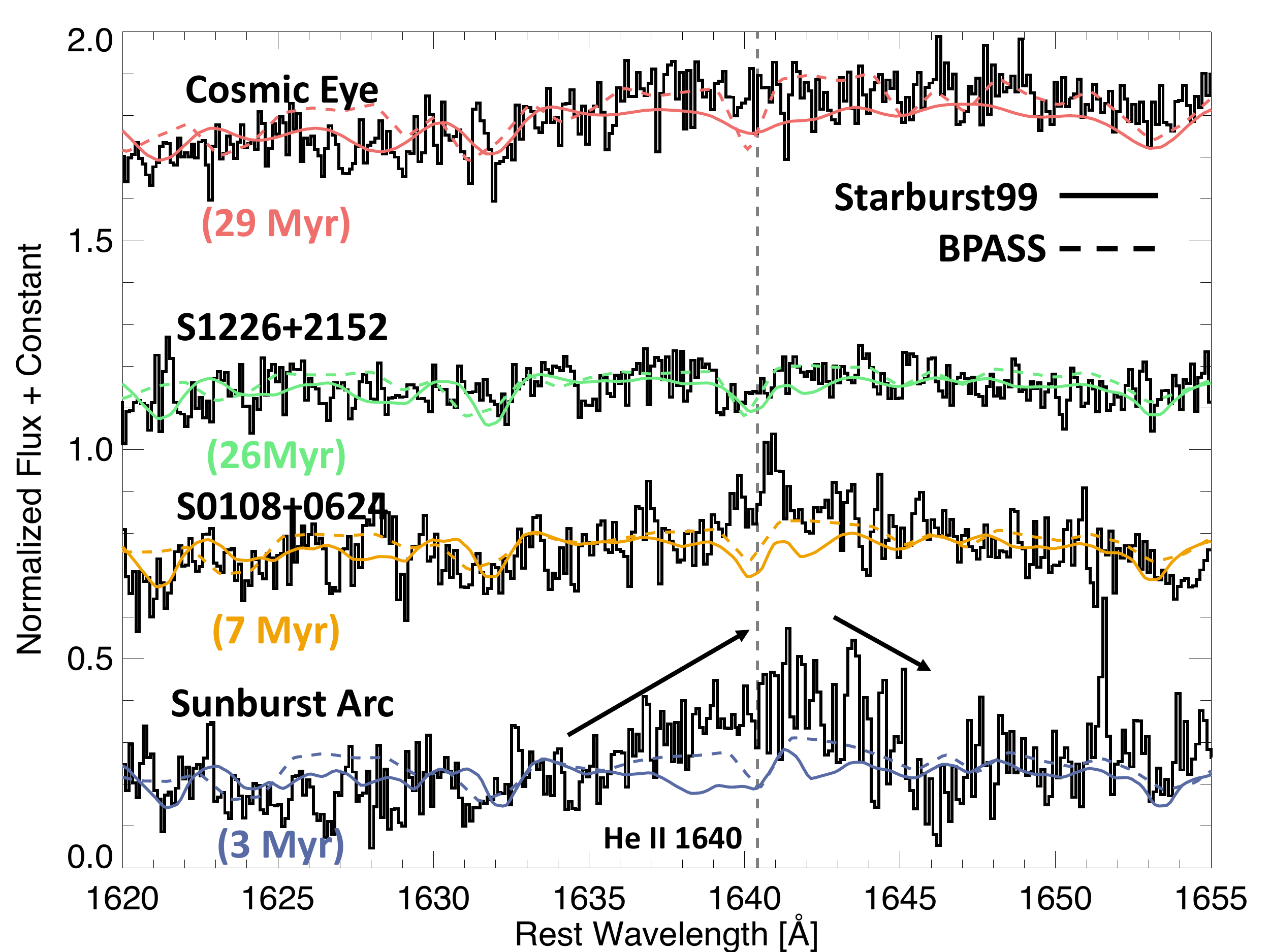}
\caption{The observed \heii~1640\AA\ region for four \megasaura\ spectra in black. Overplotted on each is the \starburst\ (solid line) and \bpass\ (dashed line) multi-age stellar population fit. Generally, the \bpass\ models fit the \heii\ region better than the \starburst\ models. However, neither model adequately fits J0108+0624 nor the Sunburst Arc (bottom two spectra). J0108+0624 has narrow emission (96~\kmsp) that is likely nebular in origin, while the Sunburst Arc has broad emission (379~\kmsp) that is likely stellar in origin. The arrows above the Sunburst Arc's spectrum emphasize the broad emission feature. The other young populations have similarly broad \heii\ emission.}
\label{fig:heii}
\end{figure}

\heii~1640\AA\ stellar emission arises from extremely hot evolved Wolf-Rayet stars. Wolf-Rayet stars are a short-lived supergiant phase associated with stars that have main-sequence lifetimes less than 5~Myr \citep{abbott87, crowther07}. Wolf-Rayet stars have broad \heii\ emission lines in both the optical and FUV that are strongest in higher metallicity stars \citep{schaerer98}.

\heii\ in \autoref{fig:heii} is not observed as a broad P-Cygni profile like \nv\ or \civp. For the oldest light-weighted populations, \heii\ is a weak absorption line (equivalent width of $0.13\pm0.04$\AA\ in S1226+2152). Conversely, the youngest populations (Sunburst Arc, RCS~Knot~E, S0033+0242, and RCS~Knot~U) have a broad triangular shaped \heii\ emission profile (see the arrows in \autoref{fig:heii}). The \heii\ emission resembles the redshifted triangular \civ emission profile but without the blueshifted absorption component that creates the P-Cygni profile. There is not \heii\ absorption because \heii~1640\AA\ is not a resonant transition (it is analogous to H$\alpha$), therefore the stellar wind is optically thin to the \heii\ recombination emission. The \heii\ emission equivalent width from the Sunburst Arc spectrum is $-$1.2\AA\ and it has a FWHM~=~379~\kmsp, consistent with a 3~Myr, 0.5~Z$_\odot$ Wolf-Rayet model \citep[][]{schaerer98}. The presence of a broad \heii\ emission profile strongly suggests that the spectrum is dominated by a $<5$~Myr stellar population. 

Finally, S0108+0624 is the only galaxy in the sample with statistically significant ($>3\sigma$), narrow \heii\ emission (equivalent width of $-0.33\pm0.08$\AA\ and FWHM~=~96~\kmsp, which is resolved by the 69~\kms spectral resolution). This narrow emission appears nebular in origin when compared to the broad Wolf-Rayet feature of the Sunburst Arc (compare the bottom and second-to-bottom spectra in \autoref{fig:heii}). This \heii\ emission is at the weak end of the range of \heii\ equivalent widths seen in local dwarf galaxies \citep[$-0.4$ to $-3.4$\AA;][]{berg16, berg19}.  We return to possible origins of the nebular \heii\ emission in \autoref{heii_comp}. 

The multiple age fits to the \heii\ region are fairly poor. This is especially true for galaxies with strong WR emission (e.g.\ the Sunburst Arc). The \bpass\ models fit the \heii\ region substantially better than the \starburst\ models do, although they still do not match the WR features of younger populations. While both stellar models include WR models, neither model appears to produce a sufficient number of WR stars to match the broad \heii\ emission observed in the youngest stellar populations \citep[e.g.,][]{leitherer18}.

Overall, we find broad Wolf-Rayet \heii\ emission in stellar populations with the youngest light-weighted ages and weak \heii\ absorption in older populations. 

\subsubsection{The \ion{O}{5} wind feature}
\label{ov}
With an ionization potential of 117~eV, \ov~1371\AA\ is the highest ionization line between 1200-2000\AA\ and traces the hottest outflowing phase from the most massive stars (up to 10$^{6}$~K). This feature is an unambiguous indicator of the youngest stars and disappears once the stellar population ages past 3~Myr (see \autoref{fig:mods} and \autoref{tab:lines}). 

While \ov\ is an ideal tracer of the most massive stars, it is not strongly observed in any of the \megasaura\ or COS spectra. This may indicate that there is not a significant extremely massive stellar population ($\sim$300~M$_\odot$), but more likely it indicates that the winds of the most massive stars are significantly clumpier--thus denser--than assumed in the model atmospheres. Denser, clumpier winds lead to lower ionization stellar winds and reduced \ov\ profiles \citep{bouret, claus2010}. 

\subsection{Photospheric absorption lines}
\label{photo}
\begin{figure}
\begin{centering}
  \includegraphics[width = 0.5\textwidth]{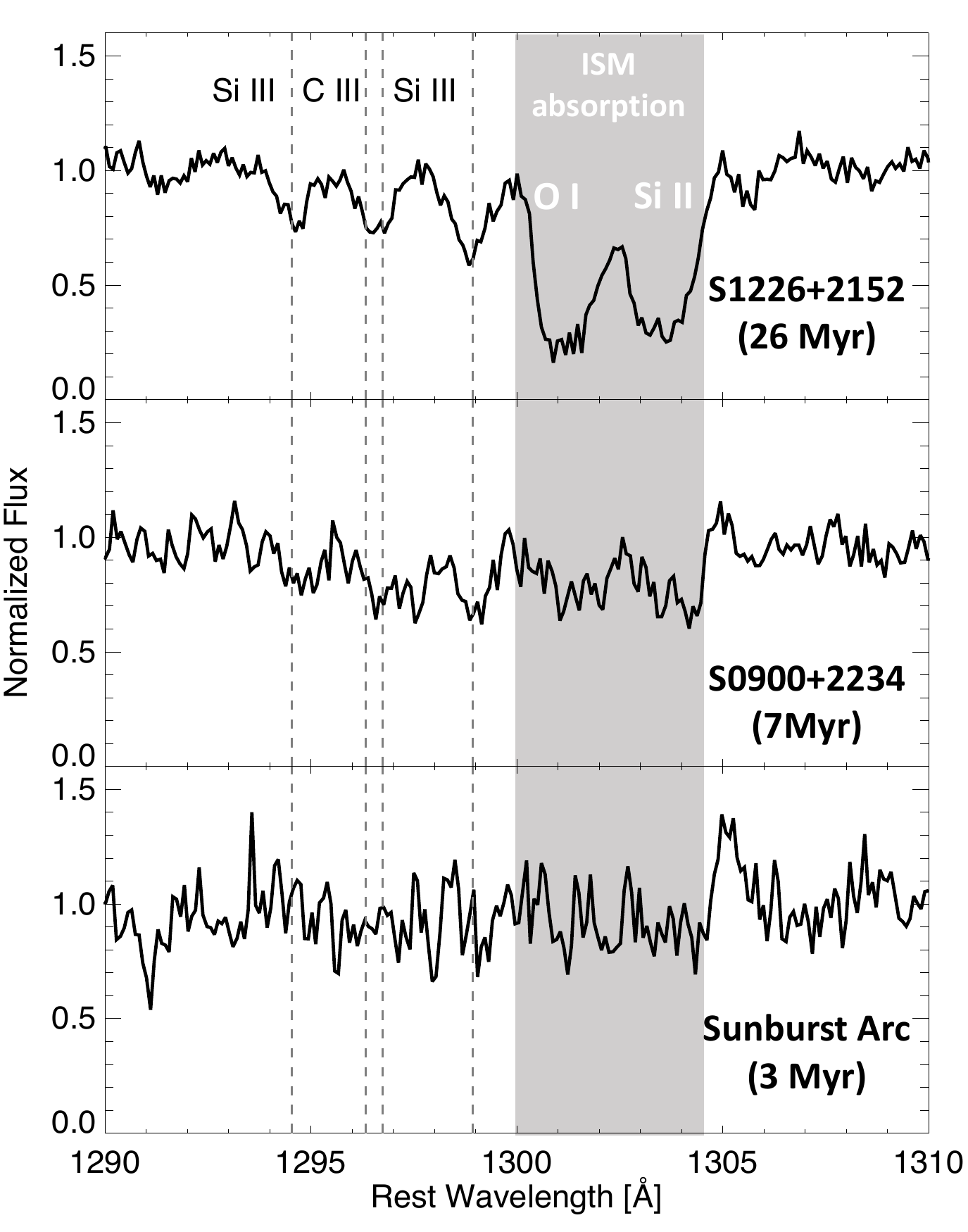}
\end{centering}
\caption{A set of photospheric absorption lines (1290--1300\AA) from three galaxies: S1226+2152 (top panel), S0900+2234 (middle panel), and the Sunburst
Arc (bottom panel). The four photospheric absorption lines are marked by dashed gray lines. These lines originate in the photospheres of B-type stars with ages greater than 8~Myr and strengthen with increasing light-weighted age (printed underneath each name). Gray regions indicate possible ISM absorption from \ion{O}{1}~1302\AA\ and \ion{Si}{2}~1304\AA. }
\label{fig:photo}
\end{figure}

The broad, pronounced stellar wind features, fully discussed in the previous section, arise as intense radiation fields accelerate gas off of stellar surfaces. However, gas within the stellar photospheres also absorbs stellar radiation. These photospheric absorption lines are narrower and weaker than the stellar wind lines because the photosphere is relatively static, but provide similarly robust indicators of stellar age and \zs\ \citep{demello}. Since photospheric gas is denser than interstellar gas, photospheric lines typically arise from excited states that are easily separated from ISM lines.

High-ionization photospheric lines, like \ion{S}{5}~1502\AA, are found in the spectra of young stars, but the rest-frame FUV has a host of O-star diagnostics, such as stellar wind lines, that are stronger features and diagnose the stellar population properties at a lower SNR. The photospheric lines excel in diagnosing older (>7Myr), B-star dominated populations  \citep{demello}, because the photospheric features are the only distinguishing features of B-stars in the FUV continuum. Previous authors have suggested that individual lines \citep[see \autoref{tab:lines};][]{demello} and bands of photospheric lines near 1417\AA\ and 1935-2020\AA\ identify B-star populations \citep{rix04}.

Our moderate-resolution data reveal that the \ciii\ and \siiii photospheric lines at 1247\AA\ (see \autoref{fig:nv}) and between 1295--1299\AA\ (specifically \ciii~1299\AA; see \autoref{fig:photo}) are the strongest due to their low excitation energies \citep[6~eV versus 10-30~eV for the features near 1420\AA;][]{leitherer11}. The 1290\AA\ lines blend with neighboring \oi and \siii ISM absorption features at low spectral resolution, however, at moderate resolution the \siiiip~1299\AA\ line is resolved. \autoref{fig:age} clearly shows that \siiii\ strengthens in stellar populations older than 5-10~Myr. The \siiiip~1299\AA\ equivalent width decreases from $0.40 \pm 0.02$\AA\ for S1226+2152 with an age of 26~Myr, to $0.29 \pm 0.15$ for S0900+2234 with an age of 7~Myr, to undetected ($0.07\pm0.05$\AA) for the Sunburst Arc at 3~Myr. Similarly, the other three \megasaura\ galaxies with ages less than 6~Myr (RCS~Knot E with an age of 2~Myr, S0033+0242  with an age of 5~Myr, and RCS~Knot U  with an age of 5~Myr) have undetected \siiiip~1299\AA\ (equivalent widths of $0.12\pm 0.21$, $0.05\pm0.07$, and $0.08\pm0.17$\AA, respectively). Very young populations do not contain the \ciii~1299\AA\ photospheric feature.

There are multiple \textsc{F}e photospheric lines in the FUV that arise from the \ion{Fe}{3}, \ion{Fe}{4}, and \ion{Fe}{5} ionization states \citep[\autoref{tab:lines};][]{nemry91,demello, rix04}. These \textsc{F}e lines (specifically \ion{Fe}{5} in O-stars) blanket the spectral regions, overlap in wavelength, and form large-scale continuum features that change the shape of the stellar continuum. These \textsc{F}e features are included in the full spectral fitting, however, even at the high SNRs of the \megasaura\ sample, the \textsc{F}e lines are very weak.

The power and utility of the stellar photospheric lines is limited because the photospheric lines are significantly weaker than the stellar wind features, requiring extremely high SNR$ > 20$ observations. At the SNR and spectral resolution of our observations, the most prominent photospheric lines are near 1299\AA\ and these features indicate the presence of stars with ages $>$7~Myr.

\subsection{Summary: Individual spectral features are consistent with inferred stellar ages and metallicities}
\label{young}
\begin{figure}
  \includegraphics[width = 0.5\textwidth]{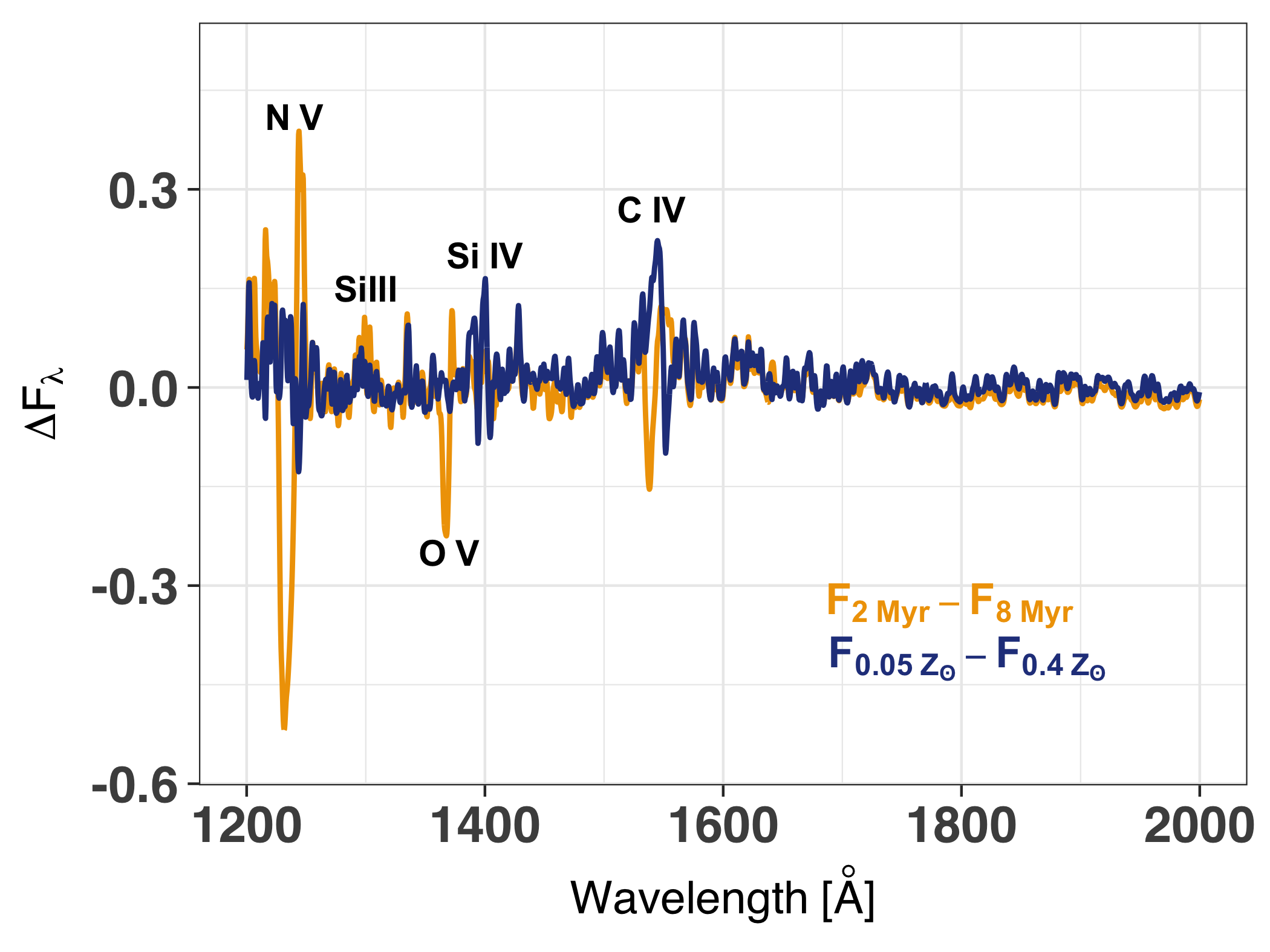}
\caption{The change in flux ($\Delta$F$_\lambda$)  incurred by increasing the age from 2~Myr to 8~Myr of a 0.2~Z$_\odot$ population (gold line) or increasing the stellar metallicity from 0.05~Z$_\odot$ to 0.4~Z$_\odot$ of a 5~Myr population (blue line). The strong spectral features discussed in the text are labeled, but significant continua changes occur outside of these spectral features in "featureless" regions (see \autoref{tab:df}). }
\label{fig:df}
\end{figure}

\begin{table}
\caption{The integrated $\sqrt{\Delta F_\lambda^2}$ incurred when changing stellar population properties}
\begin{tabular}{clcc}
Region & Spectral Feature within Region & $\Delta$Age & $\Delta$Z$_\odot$ \\
\hline
1225-1250 & \nv & 7.3 & 1.4 \\
1250-1350 & \ion{Si}{3}~1299\AA\ & 2.4  & 2.1 \\
1380-1410 & \siiv & 0.8 & 1.7 \\
1420-1520 & Featureless & 2.0 & 2.4 \\
1530-1560 & \civ & 2.6 & 3.1 \\
1660-1760 & Featureless & 1.3 & 1.7 \\
1900-2000 & Featureless & 1.6 & 1.0 \\
\end{tabular}
\tablecomments{The integrated root squared flux difference in a given wavelength region (Column 1) that includes the spectral features discussed in the text (Column 2). Column 3 gives the  root square flux difference of increasing the age of a 0.2~Z$_\odot$ \starburst\ model from 2~Myr to 8~Myr (gold line in \autoref{fig:df}) integrated over the region in Column 1. Column 4 gives the same value but for increasing the stellar metallicity from 0.05~Z$_\odot$ to 0.4~Z$_\odot$ for a 5~Myr population (blue line in \autoref{fig:df}).}
\label{tab:df}
\end{table}

Throughout this section we have compared the light-weighted stellar population properties (age and metallicity) to individual stellar features, but in all cases the inferred stellar properties were determined from the entire observed wavelength regime. In \autoref{fig:df} we plotted the change in flux density ($\Delta F_\lambda$) that occurs due to varying the stellar age (gold) or metallicity (blue). The largest $\Delta F_\lambda$ arises from the strong stellar wind and photospheric lines that we emphasized above. We demonstrated this by integrating the $\sqrt{\Delta F_\lambda^2}$ within specific wavelength regions to determine the total flux change attributable to individual spectral regions (\autoref{tab:df}). The \nv\ profile changes almost exclusively with age and is weakly dependent on \zs. Similarly, the \siiv region largely depends on \zs\ and hardly depends on stellar age (compare Column 3 and 4 in \autoref{tab:df}). While the notable stellar wind features dominate \autoref{fig:df}, small spectral features occur outside these regions that on aggregate differentiate between stellar properties. For instance, the region between 1420--1520\AA\ is devoid of prominent stellar wind features, but the shape of the stellar continuum noticeably changes over 100\AA\ due to the stellar population age and \zs. The 1420--1520\AA\ region changes the flux comparably to the broad \civ P-Cygni wind feature and the \ion{Si}{3}~1299\AA\ photospheric regions, and has more power to differentiate stellar properties than the \siiv wind region. This implies that "featureless" regions of stellar continuum hold significant power to determine stellar age and \zs. The power in these regions stems from the fact that they are not featureless, but crucially contain weak photospheric metal features like \ion{Fe}{3}, \ion{Fe}{4}, and \ion{Fe}{5} \citep{nemry91, demello, rix04}. In other words, the previous sections were not advocating to use a single feature to determine the stellar properties, rather the sections illustrated that the spectral features consistently and distinctly vary according to the inferred stellar population properties from the full stellar continuum.

Young stellar populations produce the most ionizing photons. Thus, it is important to have spectral features that distinguish immense sources of ionizing photons. There are spectral indices that are unique to very young (<7~Myr) stellar populations: (1) strong and broad \civ \textit{and}  \nv\ P-Cygni features (\autoref{fig:age}), (2) broad (>300~\kmsp) \heii\ emission (\autoref{fig:heii}), and (3) undetected \siiiip~1299\AA\ photospheric absorption features (equivalent widths <0.1\AA; \autoref{fig:photo}). These spectral signatures suggest that the stellar population is dominated by stars younger than 7~Myr. Older stellar populations appear to have combinations of both young and old stars, obscuring single trends in spectral features (see \autoref{cont}).  

\section{Discussion}
\subsection{Nebular gas and massive stars have similar metallicities}
\label{metal}
\begin{figure}
\includegraphics[width = 0.5\textwidth]{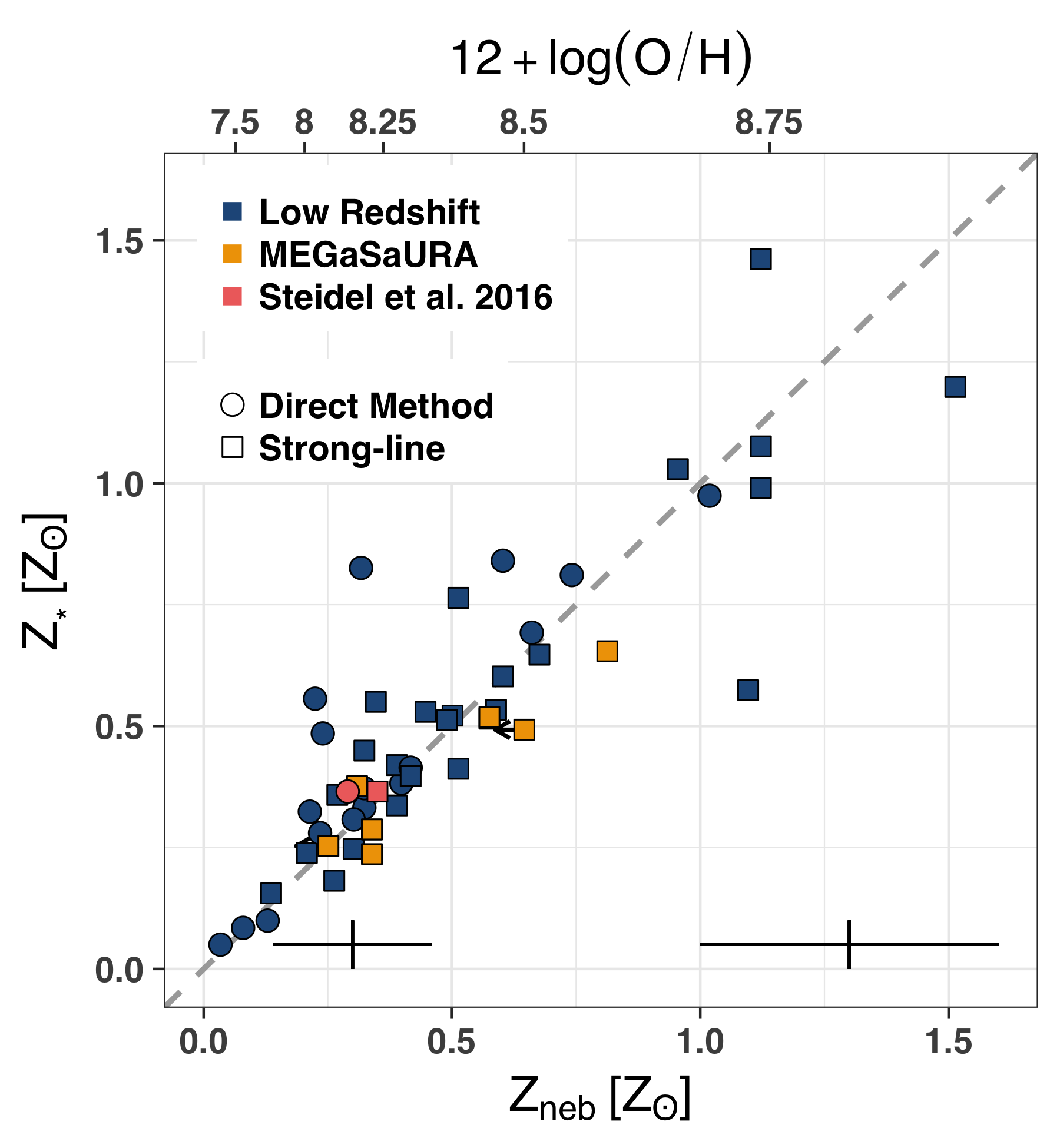}
\caption{Comparison of the light-weighted stellar metallicity (\zs) derived from the \starburst\ fitting to the gas-phase nebular metallicity (\zism). The points are color-coded according to the three samples: blue are low-redshift galaxies, orange are the individual $z \sim 2$ galaxies from the \megasaura\ sample, and the red point is the composite spectrum from \citet{steidel16}.We include both the direct and strong-line (O3N2) metallicity calculations from \citet{steidel16}. The symbol shapes correspond to whether the \zism\ was calculated using the direct method (circles) or a strong-line calibration (squares). Note, only 8 of the 19 \megasaura\ galaxies have a measured \zism. Two error bars, one at 0.3~Z$_\odot$ and one at 1.2~Z$_\odot$, are shown in the bottom of the plot using the median \zs\ error and the 0.4~dex 12+log(O/H) spread between \zism\ diagnostics suggested by \citet{kewley08}. A one-to-one line is shown in gray: \zs\ is largely equivalent to \zism.}
\label{fig:z_comp}
\end{figure}
\begin{figure*}
\includegraphics[width = \textwidth]{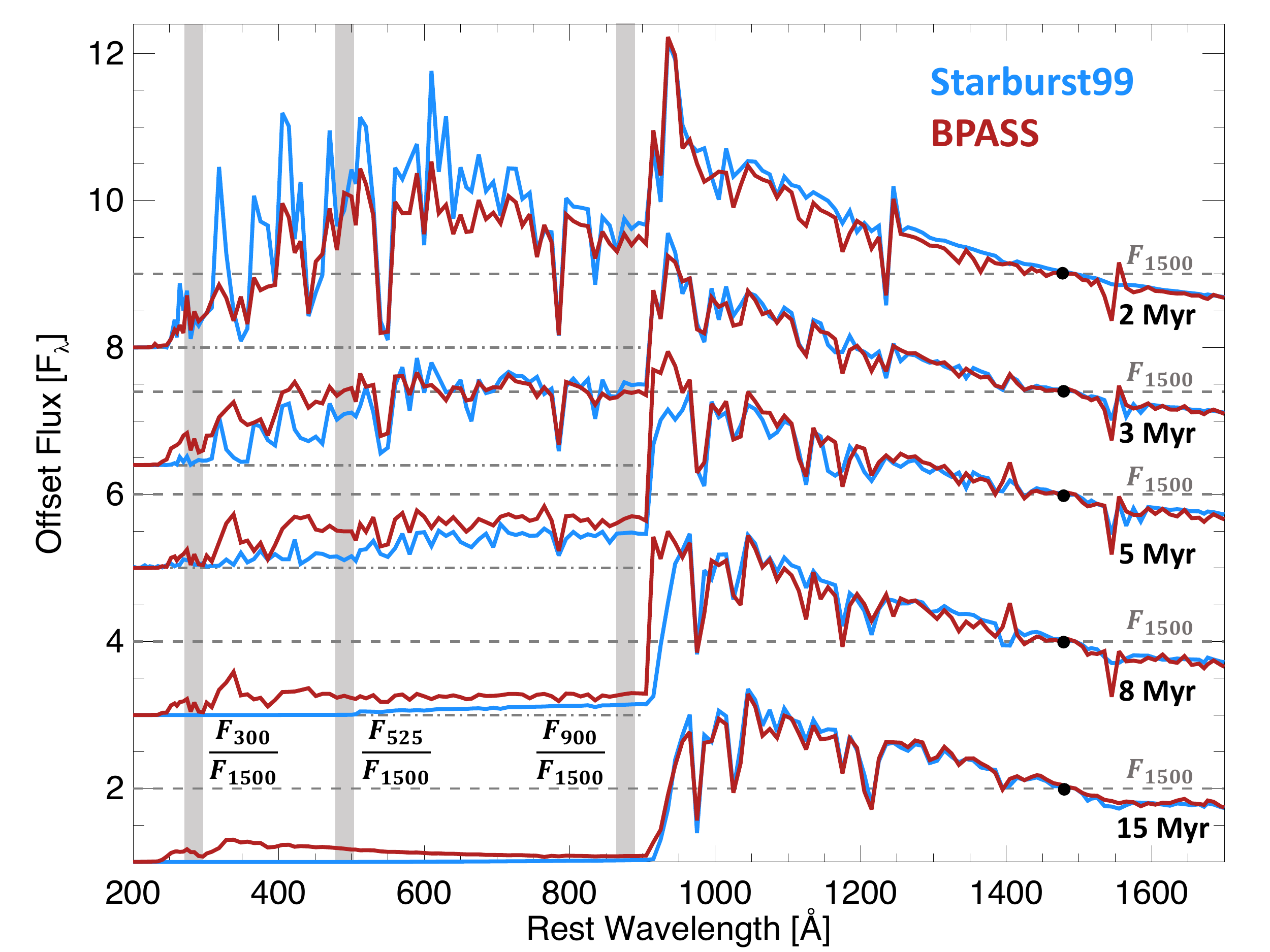}
\caption{The ionizing continua of five different single age, fully theoretical stellar models with a constant metallicity of 0.4~Z$_\odot$. We plot both \starburst\ single star models (blue lines) and \bpass\ models which include binary star evolution (red). The stellar populations are ordered by descending age, with the youngest (2~Myr) population at the top and the oldest (15~Myr) population at the bottom. Each age is offset by a constant for presentation purposes. Each population is normalized by the flux density at 1500\AA\ (F$_{1500}$; dashed line) such that zero flux is given by the dot-dashed line. The gray regions highlight the three regions that we measure the ratio of the ionizing to non-ionizing flux density to quantify the shape of the ionizing continuum: \fratt, \fratf, and \frat\ (the flux density at 300, 525, and 900\AA\ relative to the flux density at 1500\AA). \frat\ steadily decreases as the stellar population ages. }
\label{fig:ionizing_mods}
\end{figure*}

The stellar metallicity is a fundamental galaxy property that is traditionally challenging to measure, but crucially determines the production of ionizing photons. Previous observations have either used spectral indices  \citep[e.g.,][]{rix04, leitherer11, byler18} or full spectral synthesis \citep[e.g.,][]{pettini2000, kudritzki12, steidel16, steidel18, hernandez} to estimate the metallicities of young stellar populations. Here, we explore the relation between the light-weighted \zs\ derived from the full spectral synthesis to the observed gas-phase metallicities derived from rest-frame optical nebular emission lines (\zism).  

\autoref{fig:z_comp} shows that the light-weighted \zs\ and the \zism\ are correlated at the $8\sigma$ significance (with a Pearson's correlation coefficient of 0.87). The metallicities scatter about the one-to-one line, indicating that the young massive stars have a similar metallicity as the surrounding nebular gas. Both the low and high-redshift galaxies are near the one-to-one line; this relationship does not appear to evolve with redshift. There is a dearth of points above 1~Z$_\odot$, largely as a selection effect of both samples. While we have used direct metallicities whenever possible (see the discussion in \autoref{properties}), we have used different metallicity calibrations which can have a 0.4~dex scatter in their relative metallicity calibrations. The largest outlier from a one-to-one relationship is NGC~3256, in the lower right quadrant with \zs~=~$0.59\pm0.13$~Z$_\odot$ and \zism~=~$1.10\pm0.25$~Z$_\odot$ \citep[including calibration uncertainties;][]{engelbracht}. Consequently, the \zism\ of the largest outlier is still consistent with \zs\ at the 1.3$\sigma$ significance level. 

Further, we measure the residual standard error of the trend to be 0.16~Z$_\odot$, or 0.17~dex in 12+log(O/H) at the median \zism\ of 0.4~Z$_\odot$. This residual standard error is consistent with the full 0.4~dex spread found by \citet{kewley08} between the \citet{pettini04} and \citet{kobulnicky04} metallicity calibrations (the two most commonly used strong-line methods here). Thus, the observed dispersion in \autoref{fig:z_comp} is entirely consistent with the spread of the different 12+log(O/H) calibration methods.

\zs\ is surprisingly consistent for the lowest metallicity populations. The stellar mass-loss rate, and in turn the stellar wind profile, is only empirically constrained for \zs~>~0.2~Z$_\odot$ and the 0.05~Z$_\odot$ \starburst\ models instead rely on an extrapolation of the mass-loss rates to lower metallicities \citep{leitherer92, vink02}. The close metallicity correspondence below 0.2~Z$_\odot$ suggests that the stellar wind extrapolation adequately reproduces the observed stellar wind profiles and their mass-loss rates. \zism\ for 1~Zw~18 is lower than 0.05~Z$_\odot$, the lowest \citet{geneva94} atmospheric model, leading to a possible over-estimate of the light-weighted \zs\ for this galaxy.

The strong relationship between \zism\ and \zs\ in \autoref{fig:z_comp} suggests that \zs\  robustly determines the metallicities of galaxies. A robust metallicity indicator is need at high-redshifts as the crucial, yet extremely faint, metallicity sensitive lines, like [\ion{O}{3}]~4363\AA, are redshifted out of the optical \citep[][]{yuan, james14b, sanders}. Recent work has attempted to calibrate \zism\ using rest-frame UV emission lines \citep{perez, byler18}, but these calibrations depend on rather uncertain assumptions of the carbon-to-oxygen abundance that are not yet well calibrated. Fortunately, \autoref{fig:z_comp} implies that observations from the ground using upcoming large telescopes and the rest-frame FUV stellar continuum may constrain the chemical enrichment of galaxies to redshifts of $z\sim6$. 

\begin{figure*}
\includegraphics[width = \textwidth]{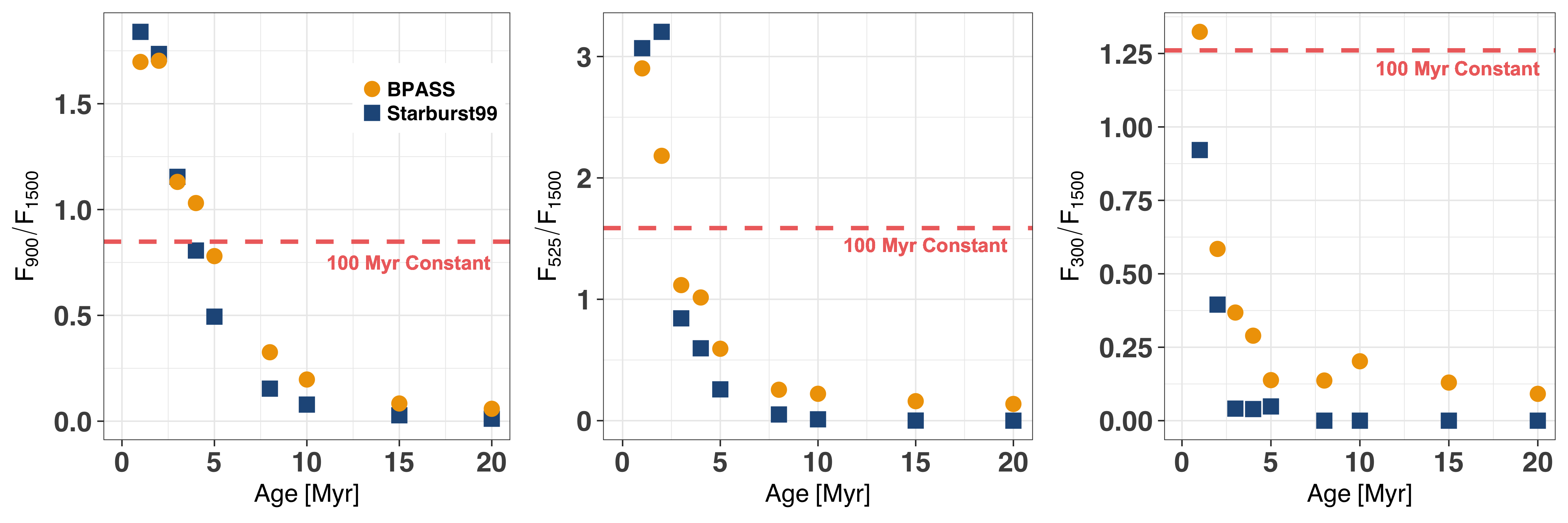}
\caption{The temporal variation of the ratio of the ionizing to non-ionizing flux density at three ionizing wavelengths (900, 525, and 300\AA\ from left to right) of a theoretical, 0.4~Z$_\odot$, single burst stellar population. We include both \starburst\ (blue squares) and \bpass\ models (gold circles). Young stellar populations emit more ionizing photons than non-ionizing photons. The flux ratio of a \bpass\ model with a constant star formation law and a 0.05~Z$_\odot$ metallicity is included as a dashed red line.}
\label{fig:fluxrat_age}
\end{figure*}

\begin{figure*}
\includegraphics[width = \textwidth]{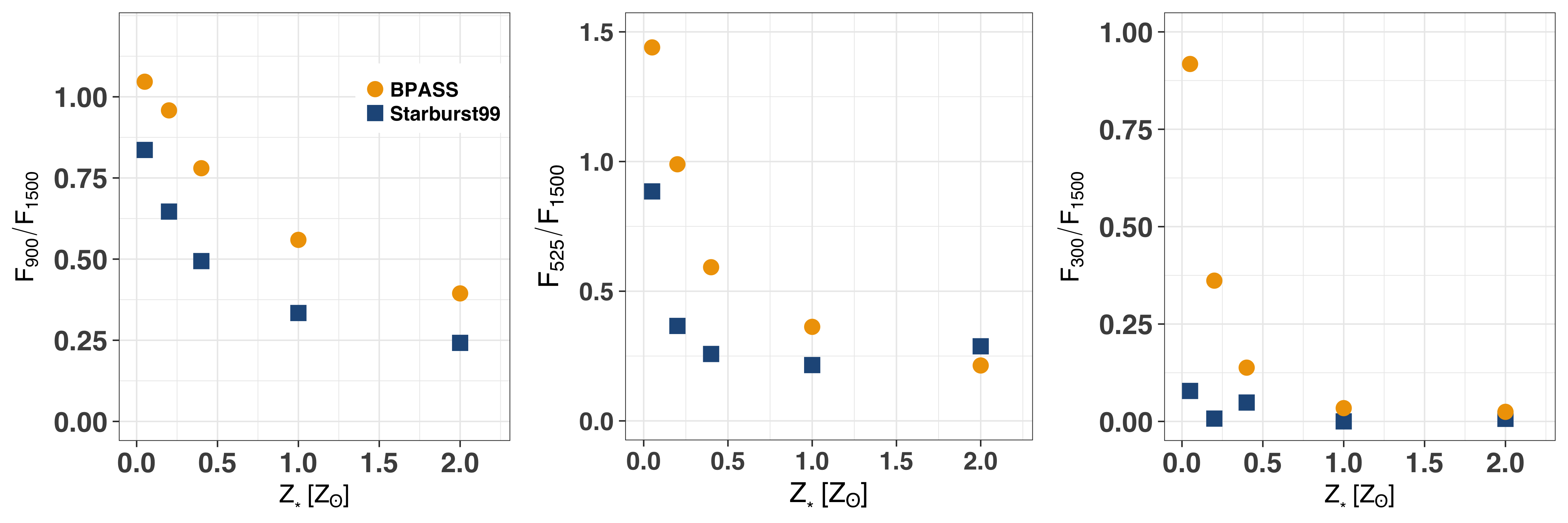}
\caption{The variation of the ionizing to non-ionizing flux density ratios with stellar metallicity (\zs). The three panels show these ratios at three wavelengths (900, 525, and 300\AA\ from left to right) for a theoretical 5~Myr single burst stellar population.  We include both \starburst\ (blue squares) and \bpass\ models (gold circles). Low metallicity \bpass\ models produce significantly more ionizing photons, especially at 300\AA, due to the effects of binary star evolution (\autoref{bpass_mods}).}
\label{fig:fluxrat_z}
\end{figure*}

The similarity of \zism\ and \zs\ indicates that the gas surrounding high-mass stars is not instantaneously metal-enriched by massive stars. In other words, it takes longer than the inferred lifetimes of the massive star populations to increase the metallicity of adjacent ISM gas. Newly synthesized metals are ejected from star-forming galaxies as very hot supernovae ejecta and galaxies must take longer than the 10$^{7}$~yr timescales observed here to fully mix these newly synthesized metals \citep{kobulnicky}. In fact, we did not find a statistical correlation between \zism\ and the stellar age, implying that gaseous enrichment either occurs on longer timescales than observed here, or that the enrichment of a single stellar population is within the scatter of \autoref{fig:z_comp}.

\subsection{The ionizing continua of massive stars}
\label{ion}

The non-ionizing FUV continua of massive stars are rich with features that diagnose the stellar population age and metallicity. The primary goal of this paper is to use these inferred light-weighted metallicities and ages to constrain the ionizing continua of massive stars. The same massive stars that produce the FUV spectral features also produce the unseen ionizing continua; these spectral features are the most direct probe of the ionizing continua because they only depend on the massive star properties (unlike nebular emission lines which also depend on the gaseous properties). We therefore use the theoretical stellar models to develop simple prescriptions for how the ionizing continuum varies with stellar properties (\autoref{ion_vary}) and then infer the ionizing continua from the FUV observations using the stellar fits (\autoref{ion_fits}). We refer to the ionizing continua determined from the models as the inferred ionizing continua to emphasize that we do not directly observe the ionizing continua. Throughout this sub-section we only focus on the \starburst\ fits; we return to the \bpass\ fits in \autoref{comp}. 

\subsubsection{Stellar population properties determine the ionizing continua of massive stars}
\label{ion_vary}

The full ionizing continuum will likely never be observed because foreground neutral hydrogen efficiently absorbs these photons. Observations are typically fortunate to observe the ionizing continuum at a single wavelength, which is most feasible at 900\AA. This ionizing flux density is then normalized by the observed non-ionizing continuum to control for stellar mass and star formation contributions. Thus, the literature typically quantifies the ionizing continuum through the ratio of the flux at 900\AA\ (ionizing) to the flux at 1500\AA\ (non-ionizing), or \frat\ \citep{steidel01}. In this subsection, we use the \starburst\ fully theoretical models to explore how the stellar properties determine this ratio as well as flux density ratios at other ionizing to non-ionizing wavelengths. 

\autoref{fig:ionizing_mods} shows the theoretical ionizing continua of single burst, 0.4~Z$_\odot$, \starburst\ stellar populations at five different ages. At 2~Myr the intrinsic ionizing flux density at 900\AA\ is actually 1.7 times more luminous than the non-ionizing flux density at 1500\AA. This is because the blackbody spectrum of a 43,000~K object, or a 40~M$_\odot$ star, peaks at 670\AA.  \frat\ steadily decreases as the stellar population ages, from 1 at 3~Myr to 0 at 15~Myr. Clearly, the stellar age crucially determines the stellar ionizing flux density of single bursts. 

We use three different ionizing to non-ionizing flux density ratios to quantify the shape of the ionizing continuum (\frat, \fratf, \fratt).  As hinted in \autoref{fig:ionizing_mods}, the left panel of \autoref{fig:fluxrat_age} shows a smooth temporal trend of \frat\ with age at a fixed metallicity. Higher energy photons (probed by the flux density ratio of \fratf) have a similarly strong temporal evolution (middle panel of \autoref{fig:fluxrat_age}), but shifted to earlier ages, such that only stars younger than 5~Myr are sufficiently hot to emit appreciably at 525\AA. The ionizing continua of the youngest stars peak near 525\AA\ where F$_{525}$ is three times more luminous than F$_{1500}$. Finally, the highest energy photons (probed by the flux density ratio of \fratt) are exclusively produced by stellar populations with ages less than 2~Myr (right panel of \autoref{fig:age_mods}). The flux density ratios of theoretical \starburst\ models show that the stellar age correlates with the shape and strength of the ionizing continua.  

The stellar metallicity (\zs) also impacts the shape of the ionizing continuum as measured by the flux density ratios \citep[\autoref{fig:fluxrat_z};][]{smith}. At a constant stellar age of 5~Myr, the \frat\ increases by a factor of 3 as \zs\ decreases by a factor of 40 (left panel of \autoref{fig:fluxrat_z}). Similarly, \fratf\ and \fratt\ decrease by a factor of three and twelve from \zs~=~0.05 to 2~Z$_\odot$, respectively. The ionizing to non-ionizing flux density ratios also quantify the impact of \zs\ on the shape and strength of the stellar ionizing continua.

\begin{figure}
\includegraphics[width = 0.5\textwidth]{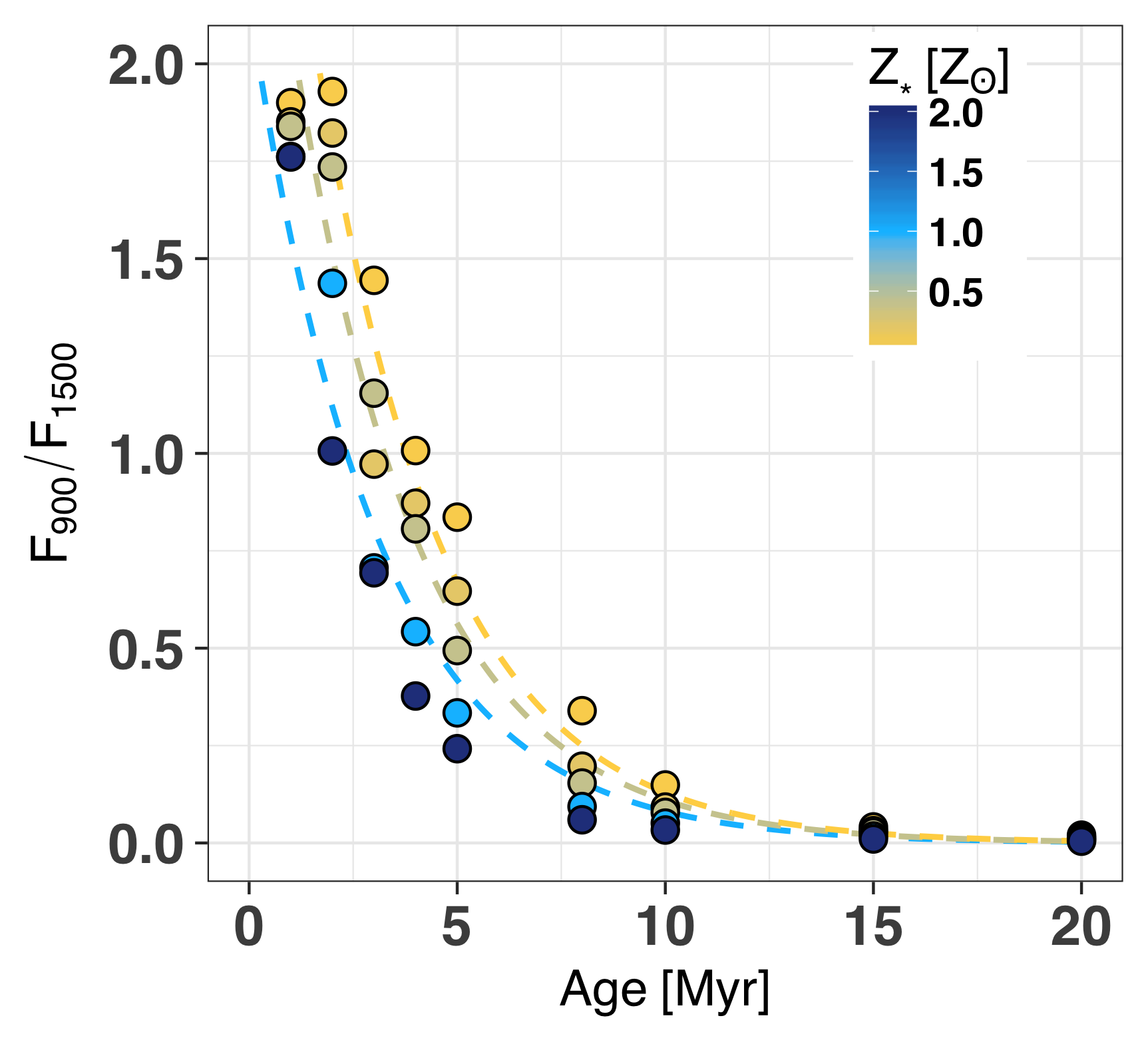}
\caption{The theoretical evolution of the ionizing to non-ionizing flux density ratio (\frat) with stellar age of a single burst \starburst\ model. All five \starburst\ metallicities are included (as indicated by the color-bar). After 10~Myr a single burst produces an insignificant amount of ionizing photons. The colored curves are single metallicity trends, using \autoref{eq:f9}, for three metallicities color-coded the same as the points (light-blue, tan, and gold are 1, 0.4, and 0.05~Z$_\odot$, respectively). }
\label{fig:f9_rat}
\end{figure}

\autoref{fig:f9_rat} illustrates that both the stellar population age and metallicity impact the ionizing continuum. A  multivariate, robust M-estimator linear regression \citep{huber} finds a relationship between these three variables for the single burst \starburst\ models (only including ages $<20$~Myr) as 
\begin{equation}
    \frac{\text{F}_{900}}{\text{F}_{1500}} = \left(3.5 \pm 0.3 \right) \times 10^{-(0.14 \pm 0.01)  \left(\frac{\text{Age}}{\text{1 Myr}}\right) -(0.21 \pm 0.02) \left(\frac{Z_\ast}{1Z_\odot}\right)} .
    \label{eq:f9}
\end{equation}
This relation determines the ratio of the intrinsic ionizing flux density at 900\AA\ to the non-ionizing flux density at 1500\AA\ for a single age burst of star formation given the stellar population age and metallicity.

In \autoref{fig:f9_rat}, we use \autoref{eq:f9} to overplot three curves of constant \zs\ onto the full \starburst\ grid. The curves generally approximate the variations in the models, but there still exists real variations at a given age. \frat\ evolves significantly at the youngest ages and small measurement errors lead to large \frat\ errors. The median error of the estimated age and \zs\ is 13\% and 11\% of the estimated values, respectively. For an estimated age of 3~Myr and metallicity of 0.4~Z$_\odot$, with these median uncertainties, there is a 15\% uncertainty on \frat, including calibration errors. The median \megasaura\ SNR, 21, is high for restframe FUV observations, but extremely high-quality observations are \textit{required} to determine \frat,  even with a 15\% error.

The \frat\ scales strongly with both stellar age and metallicity because both impact the stellar temperature. As stellar populations age, their stellar temperatures decrease which increases the fraction of neutral hydrogen within the stellar atmospheres. This is observed in the optical where Balmer absorption lines increase in older B and A-stars. Thus, the rapidly evolving \frat\ ratio with stellar age and \zs\ in \autoref{fig:f9_rat} probes the increasing X(H$^{0}$)/X(H$^{+}$) ionization fraction within the stellar atmospheres with decreasing stellar temperature. By the time the stellar population ages to 15~Myr, the temperature has dropped below 20,000~K and there is sufficient neutral hydrogen within the stellar atmospheres to absorb all of the ionizing photons. Thus, \frat\ traces the hydrogen ionization fraction within the stellar atmospheres. 

\subsubsection{Inferred ionizing continua from observed galaxies}
\label{ion_fits}
\begin{figure}
\includegraphics[width = 0.5\textwidth]{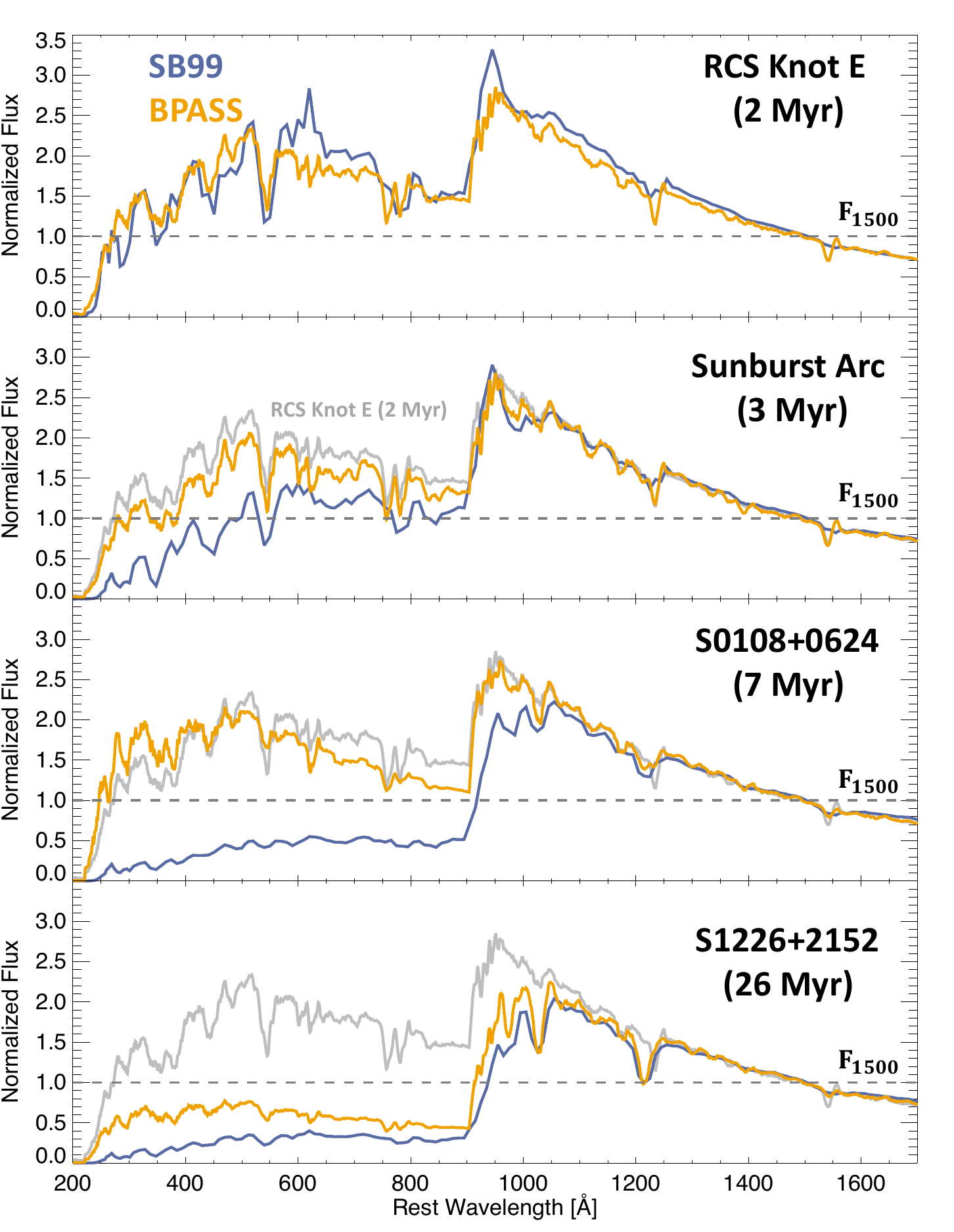}
\caption{The inferred ionizing continua, normalized at 1500\AA, for four galaxies: RCS Knot E (with a \starburst\ light-weighted age of 2~Myr; top panel), the Sunburst Arc (3~Myr; second panel), S0108+0624 (7~Myr; third panel), and S1226+2152 (26~Myr; bottom panel). Both the \starburst\ (blue line) and \bpass\ models (gold line) are plotted to illustrate the large differences between the inferred populations at older light-weighted ages. The spectra are ordered by increasing light-weighted age.  The RCS~Knot~E inferred \bpass\ continuum is included in gray in all lower panels to illustrate how the ionizing continuum changes as the stellar population ages. 
}
\label{fig:ion}
\end{figure}

We inferred the ionizing continua of the observed massive star populations by extending the stellar fits blueward beyond the observations.  The fits are linear combinations of multiple single age models,  and therefore the inferred ionizing continua may be more complicated than the single burst relations presented in \autoref{ion_vary}. How much do the inferred continua depart from a single burst model?

In \autoref{fig:ion}, we show the inferred ionizing continua of four galaxies that span the full age range of the sample. Focusing on only the \starburst\ fits (blue lines), the general trend found in \autoref{fig:ionizing_mods} is reproduced: the inferred ionizing continua at 900\AA\ decreases with fitted stellar age from RCS~Knot~E (top panel; 2~Myr), to the Sunburst Arc (top-middle panel; 3~Myr), to S0108+0624 (bottom-middle panel; 7~Myr), and finally S1226+2152 (bottom panel; 26~Myr). The galaxies with the strongest inferred ionizing continua are the youngest galaxies with the strongest stellar wind profiles (\autoref{features}). 

At wavelengths blueward of 900\AA, the \starburst\ inferred ionizing continua have substantially different morphologies. The ionizing continuum of RCS~Knot~E increases from 900 to 525\AA,  the Sunburst Arc stays relatively flat from 900 to 525\AA, while the ionizing continua of S0108+0624 and S1226+2152 both decline with decreasing wavelength. The inferred shapes of the ionizing continua constrain the total number of ionizing photons produced by the stellar population and the hardness of the nebular emission spectra (see \autoref{esc}). In this parlance, RCS~Knot~E has a particularly hard ionizing spectrum, while S1226+2152 has a relatively soft spectrum. 

The inferred \frat\ qualitatively follows the general trends outlined in \autoref{ion_vary}. RCS~Knot~E is the youngest population and has a \frat\ that is 5 times larger than the oldest galaxy, S1226+2152. However, there are large departures in the inferred \frat\ values from the single burst models: S1226+2152 has a light-weighted age of 26~Myr, yet it still has a \frat~$= 0.3$. S1226+2152 has an old stellar population, the \civ profile is nearly flat (\autoref{fig:age_mods}), yet the inferred ionizing continuum is 500 times stronger than a single burst model with the same age and metallicity (\frat~=~0.0006). Similarly, S1527$+$0652 and the Cosmic Eye have light-weighted ages of 21 and 29~Myr but inferred \frat\ of 0.2 and 0.1, respectively. These values are 50 and 400 times larger than predicted by the single burst models. This over-production of ionizing photons extends to all populations older than 9~Myr.

\begin{figure}
\includegraphics[width = 0.5\textwidth]{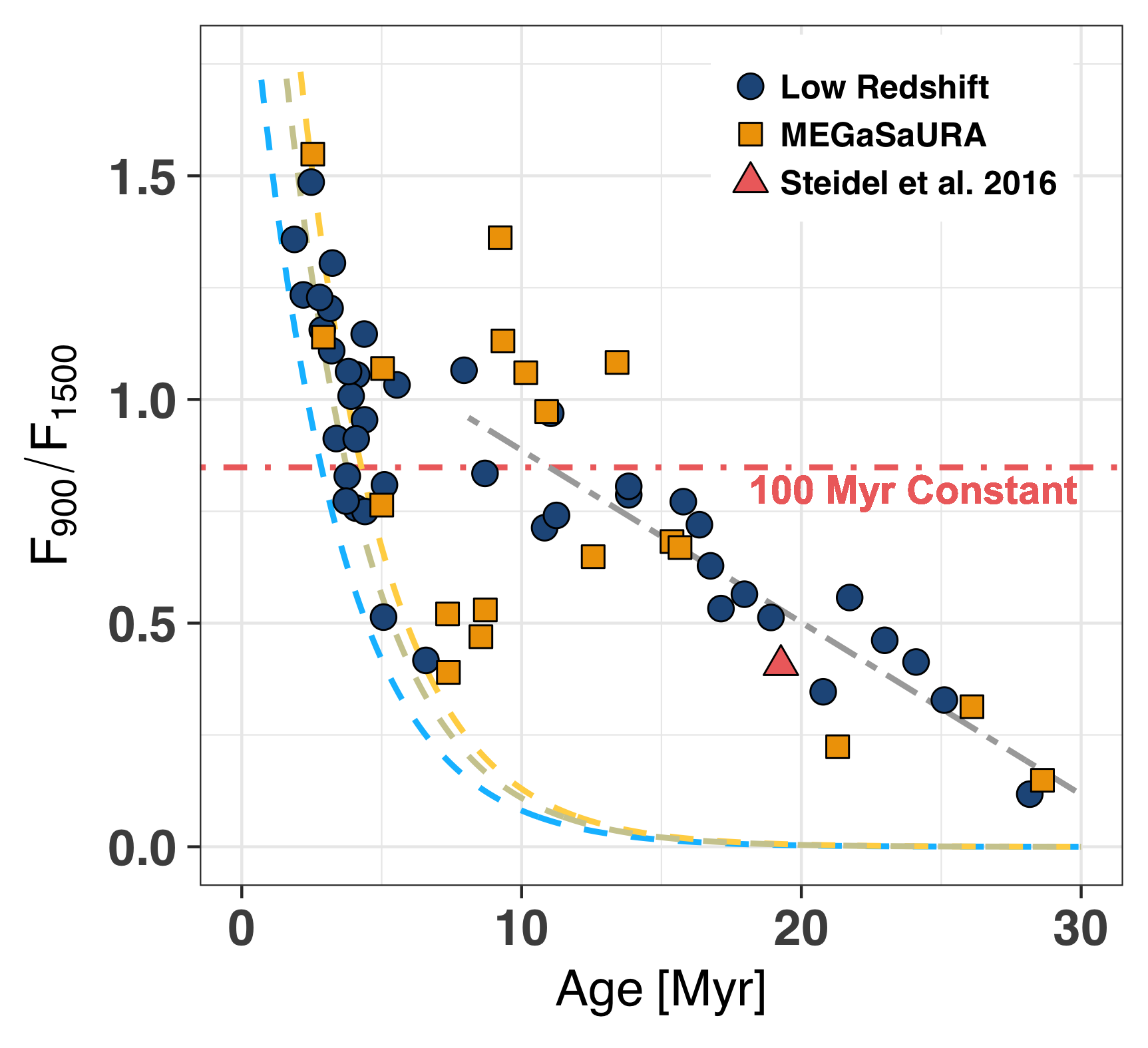}
\caption{The inferred ratio of the ionizing to non-ionizing flux density (\frat) versus the inferred age from the multiple age \starburst\ stellar continuum fits to the \megasaura\ (gold squares) and low-redshift (blue circles) samples. The curves are the single burst \starburst\ models at three metallicities from \autoref{fig:f9_rat} (blue is 1~Z$_\odot$, tan is 0.4~Z$_\odot$, and gold is 0.05~Z$_\odot$). The observed points are described by two star formation histories: a single burst of star formation that aligns with the single burst models at young ages and a mixed age population. The gray line is the fit to the mixed age population (\autoref{eq:mixedq}). The red dashed line is the \frat\ value of a 0.05~Z$_\odot$ constant star formation \bpass\ model and the red triangle is the inferred value from the \starburst\ fit to the \citet{steidel16} spectra. }
\label{fig:f9_rat_obs}
\end{figure}

Whether the inferred ionizing continua agree with the single burst models splits the observed stellar populations into two star formation histories: a single burst and a mixed age population (\autoref{fig:f9_rat_obs}). Bursts have stellar ages less than 8.6~Myr, while mixed age populations have ages greater than 8.6~Myr. This divides the full sample into 31 burst-dominated stellar populations and 30 mixed age populations. There is not a statistical redshift dependence as both the low-redshift sample and the \megasaura\ sample are nearly split evenly.

We overlay the single population model tracks of \autoref{fig:f9_rat} onto the inferred \frat\ in \autoref{fig:f9_rat_obs} to show that young populations have single burst-dominated ionizing spectra. 
All of the burst-dominated non-ionizing spectra have the spectral properties outlined in \autoref{young} for young stellar populations: strong \nv\ and \civ P-Cygni features, broad \heii\ emission (when the transition is within the wavelength coverage), and non-detected \siiii photospheric features. 

The spectra of the second population are a mixture of old and young stellar populations. The stellar continua of these stellar populations are complex, with contributions from young stars shaping the stellar winds and B-stars contributing photospheric absorption features.
This creates an averaged ionizing continuum that is non-zero due to contributions from massive O-stars, but diluted by contributions from older populations. 

The spectral differences between the two populations are seen by comparing the \civ features of the burst-dominated RCS~Knot~E and the mixed age S1226$+$2152 in \autoref{fig:age_mods}. RCS~Knot~E has very strong and pronounced \civ absorption and emission profiles, similar to the single age \starburst\ model. Meanwhile, single burst models do not fit S1226$+$2152: the 20~Myr single age \civ profile under-estimates the \civ absorption and emission, indicating a weak O-star population. Besides this weak O-star contribution, the pronounced photospheric absorption clearly indicates an older B-star population  (\autoref{fig:photo}). The observed spectral features demonstrate that as the inferred light-weighted age increases, a single stellar population ceases to dominate the stellar continuum, rather the stellar light becomes a mix of young and old spectral features. 

In \autoref{fig:f9_rat_obs}, the \frat\ of the mixed age population strongly (5.5$\sigma$, Pearson's correlation coefficient of $-0.70$) and linearly decreases with light-weighted age between 8-30~Myr as
\begin{equation}
    \frac{\text{F}_{900}}{\text{F}_{1500}} = (1.35 \pm 0.09) - (0.042 \pm 0.005) \times \left(\frac{\text{Age}}{\text{1~Myr}}\right) .
    \label{eq:f9older}
\end{equation}
This relationship is overplotted on the observations in \autoref{fig:f9_rat_obs} as a gray dot-dashed line. There is not a  statistically significant trend between \frat\ and \zs, such that the statistical significance of the trend decreases if we introduce a multivariate fit. \autoref{eq:f9older} analytically quantifies the evolution of the strength of the ionizing continua of older mixed aged stellar populations. 

We investigated the origin of mixed age populations by envisioning that the observations of individual star-forming regions sample a random mixture of stellar populations at various ages. This can be analytically tested using a random assortment of the theoretical stellar models. The model grid contains more young stars because their spectral features vary on shorter timescales, consequently, for modeling purposes, we only included \starburst\ models with ages less than 4~Myr or greater than 10~Myr. We then simulated whether a burst occurred using a binomial distribution with a uniform probability of success for each model. We chose the probability for success to be 15\% such that the expected value of the number of bursts matched the number of models typically included in the fits ($\sim6$). The success probability was tested for a range between 1-50\%, but the results were not sensitive to the chosen probability. If the random binomial for a given model returned a success (P($M_i$) = 1), we assigned a random light-fraction drawn from a Gaussian distribution. This process was repeated for all possible stellar models and the sum of all light-fractions was normalized to total one. We multiplied the stellar models by the randomized light-fractions and summed the synthetic stellar populations according to \autoref{eq:flux}. We inferred the light-weighted age and \frat\ using the same method as for the observations. We then repeated this process one million times to create a statistical sample of random mixtures of stellar populations. This sample of synthetic observations is hereafter referred to as the MCMC sample. We found a strong correlation (Pearson's correlation coefficient of $-0.81$) between the age and the \frat\ of this random mixed age population to be 
\begin{equation}
    \frac{F_{900}}{F_{1500}} = 1.36 -0.046 \times \left(\frac{\text{Age}}{\text{1~Myr}}\right) .
    \label{eq:sim} 
\end{equation}
This relationship is statistically similar to the relationship inferred from the mixed age population in \autoref{eq:f9older}. This suggests that older inferred light-weighted ages are composites of multiple epochs of stellar populations within a single aperture. A single stellar population does not dominant the FUV light of mixed age populations, rather the FUV light is broadly distributed distributed over many ages. The resultant composite is a detailed, light-weighted mixture of the different epochs of star formation.

The relative proportions of this age mixture are determined by the light-fraction of each model at a given age, or $f_{\rm age}$ (where $f_{\rm age}$ is the light-fraction of each model age; see \autoref{eq:flux}). As a simplified example, assume that the light at 1270\AA\ of a 0.4~Z$_\odot$ stellar population comes 70\% from a 2~Myr population ($f_2 = 0.7$) and 30\% from a 25~Myr population ($f_{25} = 0.3$).   The light-weighted age from \autoref{eq:age} is 8.9~Myr and the light-weighted \frat\ from \autoref{eq:f9} is 1.05. These inferred values are on the age-\frat\ mixed age relationship of \autoref{eq:f9older}. The 8.9~Myr stellar continuum has a deep \nv\ profile from the 2~Myr population, but weak photospheric lines from the 25~Myr population. This hypothetical stellar population resembles RCS~Knot~G in \autoref{fig:age_mods} which is dominated by light-fractions of $f_1$, $f_3$, and $f_{40}$ of 0.30, 0.43, and 0.20, respectively. If the assumed light-fractions of the toy model are reversed, the light-weighted age increases to 18~Myr and \frat\ decreases to 0.45. The younger population contributes less to the integrated stellar light and the stellar wind lines of this older hypothetical population are now flatter, resembling S1429-1202 in \autoref{fig:nv}. The observed stellar populations are more complicated than these toy examples, but they illustrate the evolution of the ionizing continua of mixed age populations. 

The ionizing continuum of a mixed age population looks very different than that of a single burst. A single burst 8~Myr, 0.4~Z$_\odot$, \starburst\ population has \frat~$=0.2$ and \fratt~$=0.05$, but the mixed age population envisioned above has \frat~$=1.1$ and \fratt~$= 0.3$. This \lq\lq{}{}old\rq{}\rq{} mixed age stellar population produces \textit{six} times more photons per F$_{1500}$ than a single burst of the same age. Moreover, mixed age populations produce a large number of extremely hard ionizing photons due to their out-sized presence of very young stellar populations (in this example $f_2 = 0.7$). Similarly, the 18~Myr population theorized above has an \fratt\ that is still \textit{three} times larger than a single burst 3~Myr stellar population, among the youngest in our sample, even though the stellar age inferred from the non-ionizing continuum is five times older. Mixed age stellar populations produce dramatically more ionizing photons than suggested by their light-weighted ages.

The fitted light fractions illustrate the difference between a population dominated by a single burst and a mixture of bursts (see \autoref{fig:light_frac}). RCS~Knot~E has a burst-dominated stellar spectra with a light-weighted age of $2.5\pm0.1$~Myr. 100\% of the FUV light comes from populations with ages less than 4~Myr (and $f_{2} = 61$\%; left panel of \autoref{fig:light_frac}). Alternatively, S1527+0652 has a mixed age stellar spectra with a light-weighted age of $21\pm3$~Myr. Young stellar populations (<5~Myr) within S1225+2152 contribute 19\% of the FUV light, while older stellar populations supply 81\% of the FUV light (where the dominate ones are $f_{10}$ and $f_{40}$ with 31 and 44\%, respectively). The age distribution of burst populations is relatively simple and clustered, but the light from mixed age populations is broadly distributed over many  stellar ages.

A principle worry is whether the derived continuum fits, and the associated mixed age populations, are robust. 
The errors on the stellar properties were derived by varying the observed flux of the stellar continuum by a random Gaussian with a width equal to the observed flux uncertainty and then refitting the stellar continuum. This process was repeated 100 times and the standard deviation of the ensemble was taken as the stellar property uncertainty. We find that the SNR of the parameter scales with the spectral SNR at the 3$\sigma$ significance (p-value of 0.0006, Kendall's $\tau$ value of 0.65). As a general guide, we find that the median SNR of the inferred stellar population age is 0.44 times the observed spectral SNR per resolution element. Thus, a spectral SNR per resolution element of 22 (7) is required to determine the stellar age at the 10$\sigma$ (3$\sigma$) significance level. In \autoref{ion} we also noted that a spectral SNR of 21 estimates the \frat\ with a 15\% uncertainty. Thus, a SNR~$\approx 20$ is a general guideline to robustly determine the stellar population properties and to infer the ionizing continua. 

If the age mixtures were arbitrary mixtures that were not constrained by spectral properties, then we would measure a different fitted age for each iteration and the inferred age and metallicity uncertainty would be largely unconstrained and uncorrelated with the spectral SNR. We conclude that the properties of both single bursts and mixed age populations are relatively robust and constrained by spectral signatures.

\subsection{Mixed age versus continuous star formation histories}
\label{cont}

In \autoref{features} we emphasized that the O and B-star spectral features reflect changes in the inferred stellar population age and metallicity. In \autoref{fig:age}, we showed that the stellar wind lines are sensitive to the age of the stellar population. Young populations show strong \civ P-Cygni emission, while older stellar populations have more moderate \civ features. The \nv\ wind feature only depends on age (\autoref{fig:age}) and we see strong galaxy-to-galaxy \nv\ variations (\autoref{fig:nv}). Short-lived Wolf-Rayet stars create a broad \heii\ feature that is only seen in the youngest (<5~Myr) spectra (\autoref{fig:heii}). Similarly, older stellar populations have weak, but distinct, photospheric absorption features that are absent in the youngest stellar populations (\autoref{fig:photo}). We concluded that there is real and quantifiable age variation within the observed stellar continua. 

The stellar continua vary from galaxy-to-galaxy, but not always as predicted from a single stellar population. The stellar wind profiles from populations with older light-weighted ages require a mixture of ages (see S1226+2152 in \autoref{fig:age_mods}). This mixture extends to the ionizing continua, where \autoref{fig:f9_rat_obs} demonstrated that the inferred ionizing spectra are divided into two star formation histories: single bursts and mixed age populations. A mixed age stellar population contains some assortment of young (<5~Myr) and old (>8 Myr) stellar populations that varies from galaxy-to-galaxy.


Constant star formation is the canonical alternative to a single burst star formation history. A constant star formation history assumes that every year a fixed number of stars are formed by consistently populating the IMF. When averaged over a sufficiently long temporal baseline ($>$100~Myr, similar to the ages of older B-stars), continuous star formation is a good physical description of the star formation history of evolved star-forming galaxies, such as normal spirals. However, it is unlikely that stars with lifetimes 1-10\% of the dynamical time of galaxies will form at a constant rate. However, in principle, a constant star formation is a singular, simplified mixed age solution that can be recovered with the proper light-fractions.

A constant star formation history has a fixed number of O-stars which produce a fixed number of ionizing photons. The red lines in \autoref{fig:fluxrat_age} show the ionizing flux density ratios of a 0.05~Z$_\odot$ \bpass\ constant star formation model. This constant star formation law is similar to a mixed age population with an age of 12~Myr (\autoref{fig:f9_rat_obs}), but a continuous star formation history under-predicts the ionizing continuum of a 8~Myr mixed age population by a factor of 2, and over-predicts the ionizing continuum of a 25~Myr mixed age population by a factor of 3 (\autoref{fig:f9_rat_obs}). The number of ionizing photons produced by a constant star formation history only matches the inferred ionizing continua of the mixed age fits near 12~Myr, and a constant star formation history does not match the ionizing continua at other stellar ages. 

Only \zs\ can change the stellar continuum of a constant star formation history. Predominately age sensitive tracers, such as \nv\ and the ionizing continuum, have fixed strengths in a constantly star-forming model. A constant star formation model may fit part of the \civ absorption profile well, but over (or under) estimate the \nv\ profile because the age has been fixed (see \autoref{fig:age}). The constant star formation model fails to simultaneously fit the different stellar wind profiles of very young populations (e.g. the Sunburst Arc in \autoref{fig:nv}) or older populations that lack stellar wind signatures (e.g. S1527+0652 in \autoref{fig:nv}). 

Unlike continuous star formation histories, FUV spectral features determine the mixed age star formation histories and observationally constrain the relative contributions of young and old stars. By accounting for the dominance of young stellar populations to certain ionizing continua, mixed age star formation histories generate significantly more ionizing photons, at higher energies, than continuous star formation histories. The different star formation histories alter the ionizing continua by factors of 2--3 and must be determined to accurately infer the number of ionizing photons produced by  stellar populations.

\subsection{Impact of binary evolution on the stellar continua}
\label{comp} 
Most massive stars form in binaries \citep{sana12}, and mass can be transferred between the two stars if they are sufficiently close. The star that receives the mass is rejuvenated and its main-sequence lifetime is extended \citep{eldridge09, gotberg17}. Mass transfer typically occurs when one star is in an evolved stage and expands to fill the Roche lobe. Consequently, binary star evolution extends the duration that stars spend in late evolutionary phases such as the Wolf-Rayet phase \citep{vanbeveren, georgy}. By increasing the pathways to form evolved stars, populations with binary evolution emit more ionizing photons and have harder ionizing spectra than populations with only single star evolution \citep{stanway16, gotberg17, gotberg18}. Binary evolution establishes a significant evolved population at lower metallicities because lower metallicity stars are hotter and must expand more in their later evolutionary phases. On the other side of mass transfer, the star that donates its hydrogen envelope is hypothesized to remain as an extremely hot helium core with a black body that peaks blueward of 912\AA, possibly as an X-ray binary \citep[][]{vanbever, mirabel, fragos, gotberg17}. These processes add an additional source of extreme ionizing radiation. 

Binary star evolution drastically increases the number of ionizing photons produced by a stellar population. Can the rest-frame FUV observationally constrain the importance of binary evolution? Here we discuss the observational differences between \bpass\ and \starburst\ models in their non-ionizing spectra (\autoref{bpass_mods}), the differences between the derived stellar population properties (\autoref{bpass_props}), and then compare the ionizing continua of the two evolution models (\autoref{bpass_fits}). We conclude that the non-ionizing continua alone cannot discern the impact of binary evolution, and explore alternative observational methods to distinguish between binary and single star models (\autoref{heii_comp}). 
 
\subsubsection{Does binary evolution change the observed non-ionizing stellar continuum?}
\label{bpass_mods}
\starburst\ and \bpass\ predict comparable non-ionizing continua (see \autoref{fig:mods}). The two use similar O-star libraries and produce largely identical non-ionizing spectral features. The stellar wind shapes (e.g. \nv\ and \civp) slightly differ, but this reflects the subtle differences in the O-star models rather than the impact of binary evolution \citep{claus2010, conroy14, eldridge17}. The SNRs of individual galaxies are typically too low to distinguish between the two stellar models (compare the blue and gold lines in \autoref{fig:age_mods}--\ref{fig:nv}). However, the \starburst\ models fit the \nv\ and \civ wind features of the high SNR (SNR~=~103 per resolution element) \megasaura\ stack marginally better than the \bpass\ models \citep{rigbyb}. 

\bpass\ and \starburst\ have significant differences in their older populations. The \ciii\ and \siiiip~1299\AA\ photospheric features, which we identified in \autoref{photo} as strong B-star diagnostics \citep{demello}, are very weak in the \bpass\ models (\autoref{fig:mods}). This is likely due to the fact that \bpass\ models below 20,000~K have lower spectral resolution at wavelengths less than 1500\AA\ and the weak absorption lines are nonphyiscally blended into the continuum  (see the red 15~Myr model in \autoref{fig:mods}). The absence of \ciii\ and \siiii absorption features is likely due to the inadequate spectral resolution, not binary star evolution.

The broad stellar \heii\ emission is more prominent in 5-8~Myr \bpass\ models than \starburst\ models \citep[\autoref{fig:mods};][]{steidel16}.  The \bpass\ models do fit the \heii\ region substantially better than the \starburst\ models, but the overall fit is still occasionally poor at young ages (\autoref{fig:heii}). This is likely because binary evolution provides an additional pathway to create and rejuvenate WR-stars such that the \bpass\ models contain more evolved stars, but either the WR evolution or atmosphere models still need refinement. The improved fit of the \heii\ region may indicate that binary stars contribute to the FUV light, but even the \bpass\ fits do not completely describe the \heii\ region.

Similarly, the increased population of evolved massive stars produces \siiv P-Cygni that are stronger than the \civ features at ages of 5--10~Myr (\autoref{fig:mods}). Differences between \siiv stellar wind profiles are a key prediction of models that do and do not include binary evolution. \siiv is not prominently observed as a stellar P-Cygni profile at low \zs, but \autoref{fig:siiv} shows that the fitted \siiv profiles are statistically similar whether \bpass\ or \starburst\ models are used. Even though the single age \bpass\ models show broad \siiv P-Cygni profiles, the multiple age fits do not indicate \siiv P-Cygni features. This is because the population is a mixed age population with young (<5~Myr) and old (>10~Myr) populations rather than a single age 8~Myr population. Thus, the non-ionizing stellar continua do not disentangle the impact of binary star evolution on the ionizing continua.

\subsubsection{Does binary evolution impact the inferred light-weighted stellar ages and metallicities?}
\label{bpass_props}

\begin{figure}
\includegraphics[width = 0.5\textwidth]{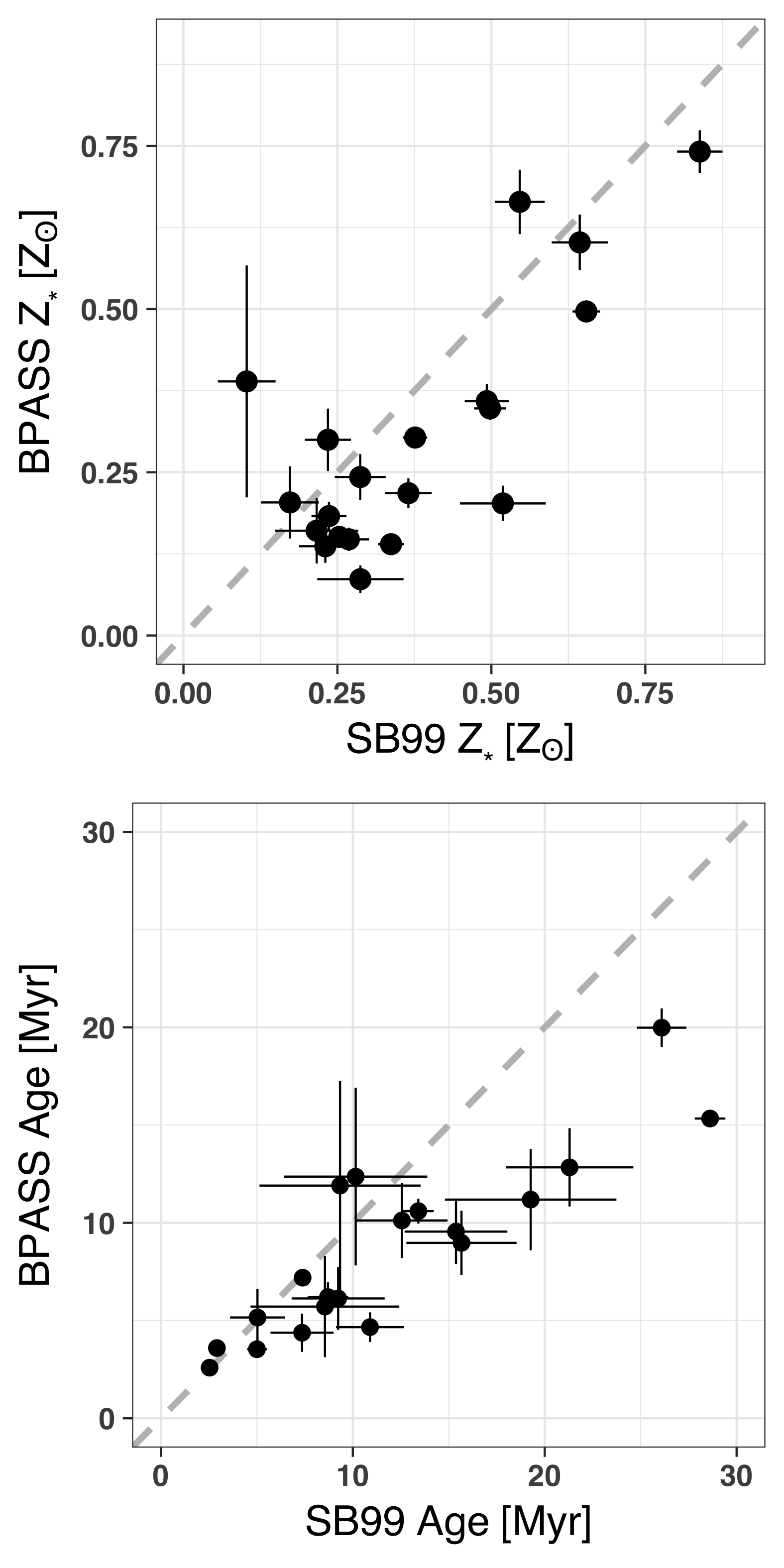}
\caption{Comparison of the derived stellar population properties of the \megasaura\ sample computed using a linear combination of either \starburst\ (SB99; x-axis) or \bpass\ (y-axis) stellar models. \textit{Top Panel:}  Comparison of the light-weighted stellar metallicities (\zs). \zs\ generally follows the one-to-one trend (gray dashed line). \textit{Bottom Panel:}  Comparison of the light-weighted stellar age.  There is a slope change near 15~Myr largely because high spectral resolution \bpass\ stellar templates do not exist at all wavelengths past this age (\autoref{fig:mods}). Consequently, the oldest inferred \bpass\ ages are younger than the inferred \starburst\ ages. Even still, there is a qualitative agreement such that old inferred \bpass\ populations are also old \starburst\ populations.}
\label{fig:mod_comp}
\end{figure}

The light-weighted metallicities and ages derived with \bpass\ and \starburst\ in \autoref{fig:mod_comp} are similar. However, \bpass\ models have a lower spectral resolution at wavelengths $<1500$\AA\ for stellar temperatures $<25$,000~K. This means that the \bpass\ models do not have a single spectral resolution over the bandpass. This biases the COS fits either toward younger stellar populations where the spectral resolution matches the data or towards older populations where multiple individual features are combined into single unresolved features. These resultant \bpass\ fits to the COS data often times do not match the \nv\ stellar wind profiles. We tested that the reduced wavelength coverage causes the poor fitting of the COS data by fitting the \megasaura\ data with only rest-frame wavelengths $<1500$\AA. We found that the \bpass\ fits to the \megasaura\ data with only wavelengths <1500\AA\ were similarly poorly fit by the \bpass\ models, but the \starburst\ fits were not impacted by using the restricted wavelength regime. Consequently, we only compare the \bpass\ fits for the \megasaura\ data, which have a complete wavelength coverage.

The inferred light-weighted stellar metallicities (at the $3\sigma$ significance) and ages ($5\sigma$ significance) scale along the one-to-one relationship in \autoref{fig:mod_comp}. The ages are similar for young populations (<15~Myr), while \bpass\ models systematically estimate younger ages when \starburst\ ages are >15~Myr. This split happens at the same age as the degradation of the \bpass\ spectral resolution discussed above. Thus, we conclude that, while the poor \bpass\ spectral resolution inhibits a firm conclusion, the ages and \zs\ derived using the different stellar models are similar. In the next section, we compare the \bpass\ ionizing continua, but we note that the inferred \bpass\ ages are up to 30\% younger for the oldest populations.

\subsubsection{Binary evolution strongly impacts the ionizing continua}
\label{bpass_fits}
The \civ (\autoref{fig:age_mods} \& \ref{fig:age_mods_z}) and \nv\ (\autoref{fig:nv}) regions cannot discriminate between binary and single star populations. While \bpass\ models fit the \heii\ region better than \starburst\ models (\autoref{fig:heii}), neither successfully reproduces the full \heii\ spectral shape. We concluded that the fitted non-ionizing continua cannot determine the contribution of binary star evolution to the stellar continuum.

Conversely, the \bpass\ modeled ionizing continua drastically differ from the \starburst\ continua. This is most extreme for older and lower metallicity stellar populations (\autoref{fig:fluxrat_age} $\&$ \ref{fig:fluxrat_z}). At young ages (<3~Myr), the production of ionizing photons is similar regardless of the binary evolution model because young stars have not yet evolved off the main-sequence. Meanwhile at older ages, such as  8~Myr, there is a factor of two difference in \frat\ between \bpass\ and \starburst\ models. This difference is similar to the difference between a burst and a 8~Myr mixed age population (\autoref{ion_fits}).

The modeled ionizing continua differ most drastically at higher energies (right two panels of \autoref{fig:fluxrat_age}). A single 10~Myr \starburst\ burst has negligible \fratt~$= 3\times10^{-8}$, but a population that includes binary evolution produces a staggering \fratt~$= 0.20$ (the furthest right circle in the right panel of \autoref{fig:ionizing_mods}). Populations with binary evolution have substantially harder ionizing spectra due to the enhanced contribution of evolved stars. 

Thus, the inferred \bpass\ ionizing continua are typically stronger and harder than those inferred from \starburst\ (blue versus gold curves in \autoref{fig:ion}). While the \bpass\ models exhibit the same qualitative trend of decreasing \frat\ with inferred stellar age, there are important differences between the ionizing continua inferred with \bpass\ and \starburst. RCS~Knot~E (top panel) has the youngest \megasaura\ population (2.5~Myr) and the inferred ionizing continuum is similar for both \starburst\ and \bpass\ models. Conversely, the Sunburst Arc, which is only 0.5~Myr older, has similar \frat\ values, but the \fratf\ and \fratt\ are 1.4 and 4.9 times larger for the \bpass\ models than the \starburst\ models, respectively. The oldest galaxy, S1226$+$2152, again has the lowest \frat, but \frat, \fratf, and \fratt\ are 1.4, 2.0, and 7.2 times larger for the \bpass\ models than the \starburst\ models.

 S0108$+$0624 has the most pronounced difference in the \bpass\ and \starburst\ ionizing continua. The inferred \frat, \fratf, and \fratt\ are 2, 4, and 12 times larger for the \bpass\ fits than \starburst. 
 Surprisingly, the \bpass\ models in \autoref{fig:ion} predict  larger \fratt\ for the older population of S0108$+$0624 than the youngest stellar populations in the sample, such as RCS~Knot~E (1.4 times larger) and the Sunburst Arc (1.7 times larger). 

The inferred ionizing continua of stellar populations that include binary evolution can produce up to 12 times more hard ionizing photons per FUV luminosity than single star populations with the same FUV spectral features. This presents an uncomfortable conclusion: there is up to an order of magnitude difference in the number of ionizing photons predicted from populations with binary evolution that cannot be observationally constrained by the non-ionizing stellar continuum. Additional observations must constrain the impact of binary star evolution.  

\subsubsection{How to observationally distinguish binary and single star populations}
\label{heii_comp}

\begin{figure}
\includegraphics[width = 0.5\textwidth]{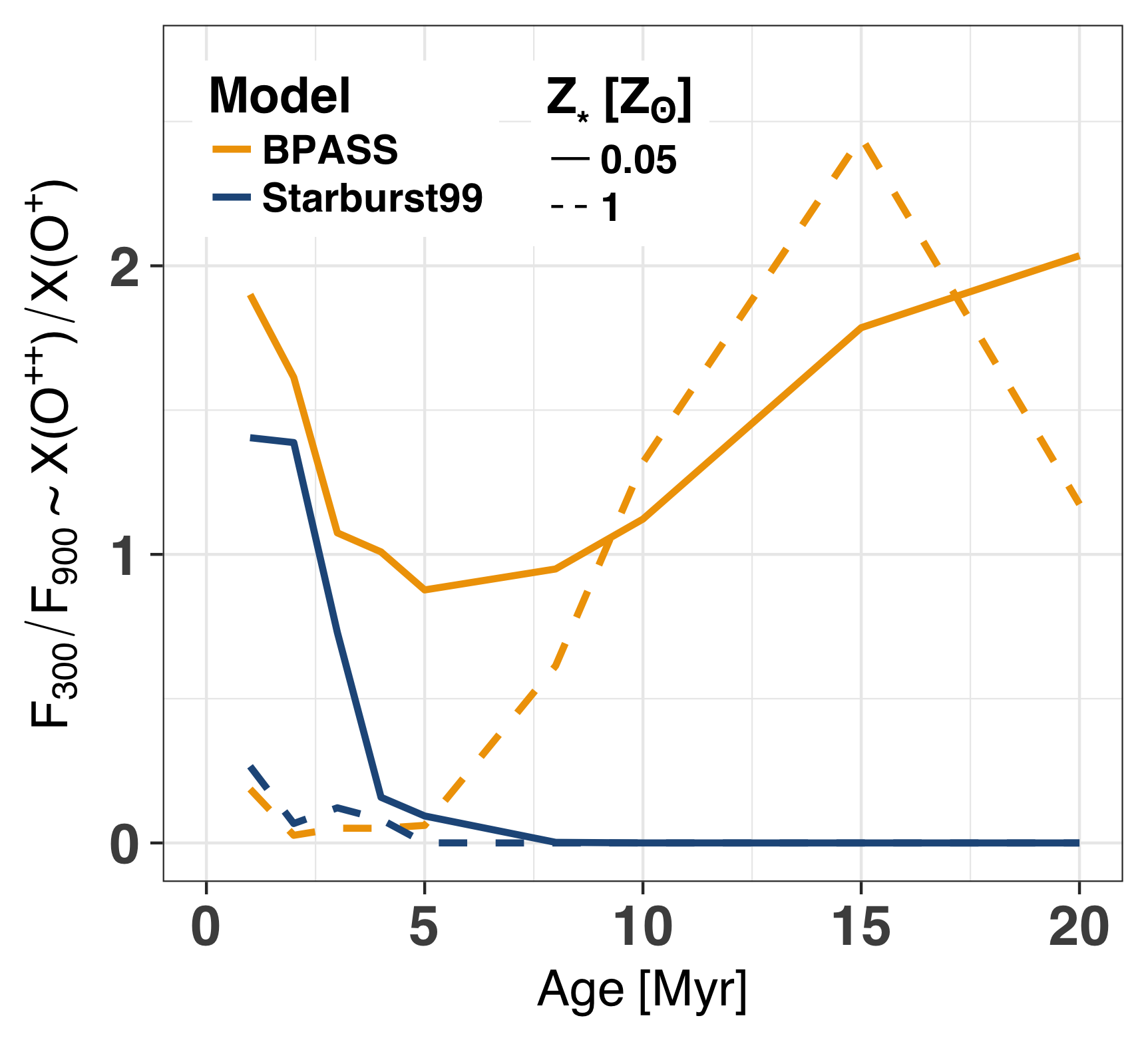}
\caption{The age variation of the theoretical stellar continuum flux density ratio at 300\AA\ (F$_{300}$) and 900\AA\ (F$_{900}$) for \starburst\ (gold lines) and \bpass\ (blue line) single burst models. Solid lines are 0.05~Z$_\odot$ stellar metallicities and dashed lines correspond to 1~Z$_\odot$ stellar metallicities. Photons at 300\AA\ create O$^{++}$ and photons at 900\AA\ create O$^{+}$, thus the F$_{300}$/F$_{900}$ ratio probes the ionization fraction of X(O$^{++}$)/X(O$^{+}$). Full photoionization modeling is required to explore the actual nebular emission line ratios ([\ion{O}{3}]/[\ion{O}{2}]). \starburst\ models only produce extreme amounts of hard ionizing photons within the first 5~Myr and at low metallicities. \bpass\ models produce hard ionizing photons at later ages and all metallicities. Older, higher metallicity \bpass\ models actually produce relatively more high-energy photons than younger metal poor populations.  }
\label{fig:o32}
\end{figure}


The previous sub-sections demonstrated that the non-ionizing stellar continua cannot quantify the impact of binary star evolution on the integrated stellar spectra, but there are extreme differences in their respective ionizing continua. While observations suggest that massive stars typically form in binaries, the exact binary parameters--mass ratios, period distributions, and remnant masses--are largely unknown and must be assumed to create the \bpass\ models \citep{eldridge17}. Therefore, the impact of binary evolution on the stellar continuum is observationally unconstrained. Observational tests must break this degeneracy.

Direct observations of the extreme-UV continuum (EUV; wavelengths <912\AA) are a definitive method to break this degeneracy. With up to a factor of 12 flux density difference between the \bpass\ and \starburst\ modeled ionizing continua, direct observations would indisputably determine the impact of binary star evolution on the ionizing continuum. Further, there are strong stellar wind lines in the EUV (e.g. \ion{N}{3}~763\AA) which are more prevalent in binary populations than single star populations (\autoref{fig:ion}). The intrinsic LyC break (near 912\AA) is also different for 5--10~Myr populations due to the higher stellar temperatures of binary populations (\autoref{fig:ionizing_mods}). However, the EUV of an O-star dominated stellar population has never been observed.

\citet{steidel16} broke this degeneracy using nebular emission lines. For an ionization bounded nebula (or where the gas absorbs all of the ionizing photons), the relative strength of nebular emission lines at different ionization states (e.g., [\ion{O}{3}]~5007\AA\ versus [\ion{O}{2}]~3727\AA) traces the intrinsic hardness of the stellar spectrum, and \textsc{H} and \textsc{He} recombination lines trace the intensity (normalization) of the ionizing continuum (modulo the escape of ionizing photons and the effects of dust). Constraining whether the inferred \bpass\ or \starburst\ ionizing continua better reproduce the observed nebular emission line structure will illuminate the importance of binary evolution. 

Comparing [\ion{O}{3}] and [\ion{O}{2}] emission lines may resolve the different ionizing continua of the two evolutionary models.  \bpass\ models create more more O$^{++}$ gas relative to low ionization, O$^{+}$, gas than single star models \citep{stanway14}. We illustrate this by comparing the F$_{300}$/F$_{900}$ ratio of the \starburst\ (blue lines in \autoref{fig:o32}) and \bpass\ models (gold lines in \autoref{fig:o32}). Photons at 300\AA\ produce O$^{++}$ gas, while photons at 900\AA\ only generate O$^{+}$; F$_{300}$/F$_{900}$ probes the formation of O$^{++}$ and O$^{+}$ gas which emit [\ion{O}{3}] and [\ion{O}{2}], respectively. This is meant to illustrate the possibility of observing differences in the [\ion{O}{3}]/[\ion{O}{2}] ratio with stellar age, but full photoionization modeling is required to compare to the observed emission lines. Models without binary evolution produce O$^{++}$ at low metallicities and young ages, while \bpass\ models produce large X(O$^{++}$)/X(O$^{+}$) ratios at all metallicities and ages. In fact, high \zs\ \bpass\ models produce larger X(O$^{++}$)/X(O$^{+}$) ratios than the youngest low-metallicity single star models. Correlations between X(O$^{++}$)/X(O$^{+}$) and stellar population properties may constrain the importance of binary evolution.  

Individual emission lines may also break the degeneracy between binary and single-star models. Nebular \heii\ emission requires extremely high-energy photons ($\lambda < 227$\AA). Stellar continua are typically considered too weak and too soft to reproduce the observed \heii~1640\AA\ and \heii~4686\AA\ nebular emission \citep{kehrig, berg18, nanayakkara, stanway19, schaerer19}. S0108$+$0624 is the only spectrum in the \megasaura\ sample with a weak and narrow \heii\ emission line (\autoref{fig:heii}; equivalent width of $-0.3\pm0.1$\AA). \autoref{fig:ion} shows that the \bpass\ fit to S0108$+$0624 has a harder spectrum than the youngest stellar populations (compare the gold and gray lines in the third panel), possibly suggesting that evolved, mixed age binary star populations are required to produce the observed nebular \heii.

\subsection{Fits to stacked \megasaura\ spectra}
\label{stack}

Stacking, or averaging many spectra together, increases the SNR of the composite spectrum. Often it is assumed that a stacked spectrum probes the average properties of the underlying population. The combination of the high-quality individual \megasaura\ spectra and the stacked \megasaura\ spectrum allows us to test this hypothesis.

By weighting the individually measured stellar ages and metallicities by the SNR of their respective \megasaura\ spectrum at 1500\AA, we estimated the stellar age and metallicity of the \megasaura\ ensemble to be $14.0\pm1.3$~Myr and $0.47\pm0.03$~Z$_\odot$ ($9.7\pm1.0$~Myr and $0.35\pm0.02$~Z$_\odot$ for \bpass). We then fit the stacked \megasaura\ spectrum and estimated a light-weighted age and metallicity of $12.0\pm0.7$~Myr and $0.42\pm0.01$~Z$_\odot$  ($9.9\pm0.6$~Myr and $0.27\pm0.01$~Z$_\odot$ for \bpass). Thus the inferred stellar properties of the stack are statistically similar to the ensemble averages.

The SNR weighted average \frat\ of the individual fits is $0.68\pm0.42$ and the inferred \frat\ from the stack is 0.47. Meanwhile, if we input the fitted light-weighted age of the stellar population into the \frat\ relationship (\autoref{eq:f9older}), we expect \frat~$= 0.77$. We find that the fitted stellar properties of the \megasaura\ stacked spectra reasonably match the underlying population averages. 

\subsection{Comparison to previous work}
\label{prev} 

The most directly comparable work to our stellar population synthesis fitting is \citet{steidel16}. Those authors used a stack of 30 galaxies at $z \sim 2.4$ with rest-frame FUV and optical observations (for the optical emission lines) to determine the stellar population and nebular emission properties. They fit a variety of \zs, IMFs, and stellar models (\bpass\ versus \starburst) to the observed stellar continuum, but always used a continuous star formation history. Low metallicity ($\sim0.05$~Z$_\odot$) continuous star formation models fit the stellar continuum best, even though the measured nebular metallicities were \zism~=~0.29--0.49~Z$_\odot$. 

The various continuous star formation stellar continuum models were then used as the ionizing source within \textsc{cloudy} photoionization models to predict the nebular emission lines and compare these predictions to the observed values. Continuous star formation \starburst\ models produced an insufficient number of ionizing photons to match the observed classical \citet{baldwin} diagnostics such as [\ion{O}{3}]~5007\AA/H$\beta$ and [\ion{N}{2}]~6585\AA/H$\alpha$. Similarly, \bpass\ models with \zs~=~\zism\ failed to reproduce these nebular emission-line ratios. 

However, \citet{steidel16} could reproduce the emission lines if \zs\ was five times smaller than \zism. \citet{steidel16} argued that rather than a lower \zs, the $\alpha$/\textsc{F}e relative abundance ratio of $z\sim2$ galaxies must be super-solar. This would lead to a measured \zs\ that is smaller than the actual metallicity, and the stars would produce more ionizing photons because \textsc{F}e is the chief opacity source in the atmospheres of massive stars. Thus, they concluded that the non-solar $\alpha$/\textsc{F}e relative abundances produced an illusion that the stars had five times fewer metals than the gas. 
\begin{figure}
\includegraphics[width = 0.5\textwidth]{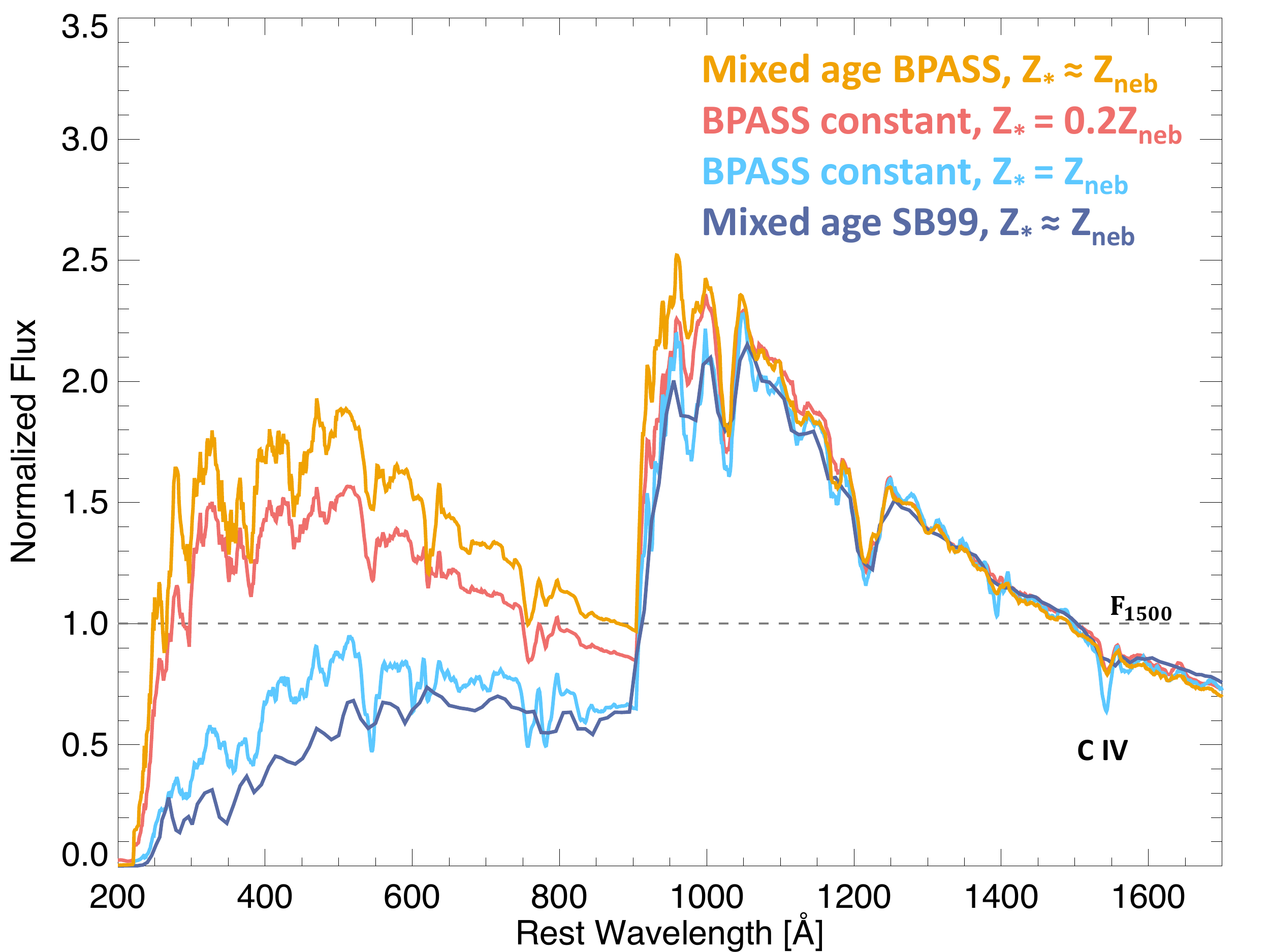}
\caption{Comparison of the inferred stellar continua of the stacked spectrum from \citet{steidel16} using a linear combination of single age \bpass\ (in orange with a light-weighted age of 10~Myr and 0.22~Z$_\odot$) and \starburst\ (dark blue with a light-weighted age and metallicity of 13~Myr and 0.37~Z$_\odot$). The 0.05~Z$_\odot$ (1/5$\times$\zism) constant star formation model that reproduces the nebular emission structure is shown in red, while the constant star formation model with \zs~=~\zism, which does not reproduce the nebular emission lines, is in light blue. The constant and mixed age \bpass\ models have similar shapes, even though their metallicities differ by a factor of 4. The light-weighted \zs\ of the mixed age fits are similar to the measured nebular metallicity of 0.29--0.49~Z$_\odot$ \citep{steidel16}.  Mixed age populations produce a strong and hard ionizing continuum, while having a similar metallicity as the nebular gas.}
\label{fig:chuck}
\end{figure}

We fit the \citet{steidel16} stack with the methods outlined above, and derived a light-weighted \zs~=~0.37$\pm0.04$~Z$_\odot$  using \starburst\ models ($0.22\pm0.02$~Z$_\odot$ using \bpass). The inferred \starburst\ stellar population age of the \citet{steidel16} stack is 12.6~Myr, implying that it is a mixed age population. These metallicities are similar to the nebular metallicities of \zism~=~0.29--0.49 calculated using the direct and strong-line methods, implying that the stars and gas have similar metallicities (we plot both the direct and strong-line \zism\ calculations in \autoref{fig:z_comp}).

While the inferred light-weighted \zs\ is meant to convey a total metal content, the strong stellar wind lines, which are only observed through $\alpha$-element transitions in the FUV, could hide discrepant $\alpha$/\textsc{F}e abundance ratios \citep{steidel16}. Therefore, \citet{steidel16} suggested using regions devoid of strong wind features, but containing many weak photospheric lines (the so-called \lq{}\lq{}featureless\rq{}\rq{} regions in \autoref{tab:df}) to test the stellar {F}e abundance. Using \lq{}\lq{}Mask 2\rq{}\rq{} from \citet{steidel16}, we refit their composite spectra using only wavelengths without strong stellar wind lines and find a very similar light-weighted \zs\ as when we fit the entire spectrum (0.43 and 0.37~Z$_\odot$ for \starburst\ and \bpass, respectively). Thus, even in stellar spectral regions dominated by \textsc{F}e absorption features, we still infer that \zs\ and \zism\ are similar.

The only difference between the inferred \zs\ from this work and \citet{steidel16} is the assumed star formation history. As emphasized in \autoref{cont}, a constant star formation history predetermines a fixed number of young and older stars, each with their own distinct spectral signatures. Since young populations have \ion{Fe}{4} and \ion{Fe}{5} photospheric lines (\autoref{fig:mods}) and slightly older populations generally have \ion{Fe}{3} lines \citep[fig.~5 in ][]{demello}, a continuously star-forming population is always expected to have \ion{Fe}{3}, \ion{Fe}{4}, and \ion{Fe}{5} absorption.  The assumption of constant star formation may force the \citet{steidel16} fits toward lower metallicities in order to balance the strength of the \textsc{F}e absorption across different ionization states. By contrast, a mixed age stellar population can reduce the \textsc{F}e absorption lines by either reducing the metallicity or by selecting the proper combination of stellar population ages to match the observed \textsc{F}e lines (i.e. an older population without \ion{Fe}{4} and \ion{Fe}{5} or a younger population without \ion{Fe}{3}). Thus, we emphasize that the a priori assumed star formation history (constant versus mixed age) has a profound impact on the inferred stellar metallicity.

Even though the light-weighted \zs\ derived from our mixed age fits to the stellar continuum are four times larger, the \civp~1550\AA\ stellar wind P-Cygni profile appears similar to the lower metallicity, 0.05~Z$_\odot$ continuous star formation model (orange and red lines \autoref{fig:chuck}) and weaker than the 0.3~Z$_\odot$ constant star formation model (light blue line). As discussed in \autoref{cont}, continuous star formation models remove possible age variations, leaving \zs\ as the only lever to influence the \civ P-Cygni shape. The mixed age fits to the stacked data have light-weighted ages of $12.6\pm2.4$~Myr using \starburst\ ($10.1\pm1.9$~Myr using \bpass) such that a moderate metallicity, moderate age population has relatively weak \civ absorption.

\autoref{fig:chuck} shows the inferred ionizing continuum from the fit to the \citet{steidel16} stack. This figure strongly emphasizes that mixed age populations produce substantially more ionizing photons, even at higher metallicities, than continuous star formation models. The \starburst\ fit (dark blue line) has a weaker ionizing continuum than the \bpass\ fit (orange line), and the \starburst\ fit is more similar to the \bpass\ \zs~=~\zism\ continuous star formation model (light-blue), which failed to reproduce the nebular emission. As suggested by \citet{steidel16}, the \starburst\ fits are unlikely to generate sufficient ionizing photons to match the nebular emission. 

The \bpass\ mixed age fit to the stack (orange curve in \autoref{fig:chuck}) produces more ionizing photons, with a similar spectral shape, as the low metallicity continuous star formation model which reproduced the nebular emission structure (red curve).  The youngest stellar populations produce the most ionizing photons (\autoref{fig:f9_rat}), thus, allowing younger populations to have larger light-fractions increases the total number of ionizing photons produced by a stellar population (see \autoref{esc}). Consequently, a mixed age stellar population produces \textit{more} ionizing photons than a continuous star forming model, while having a \textit{similar} stellar and nebular metallicity.  

\subsection{The production efficiency of ionizing photons by massive star populations}
\label{esc}

\begin{figure}
\includegraphics[width = 0.5\textwidth]{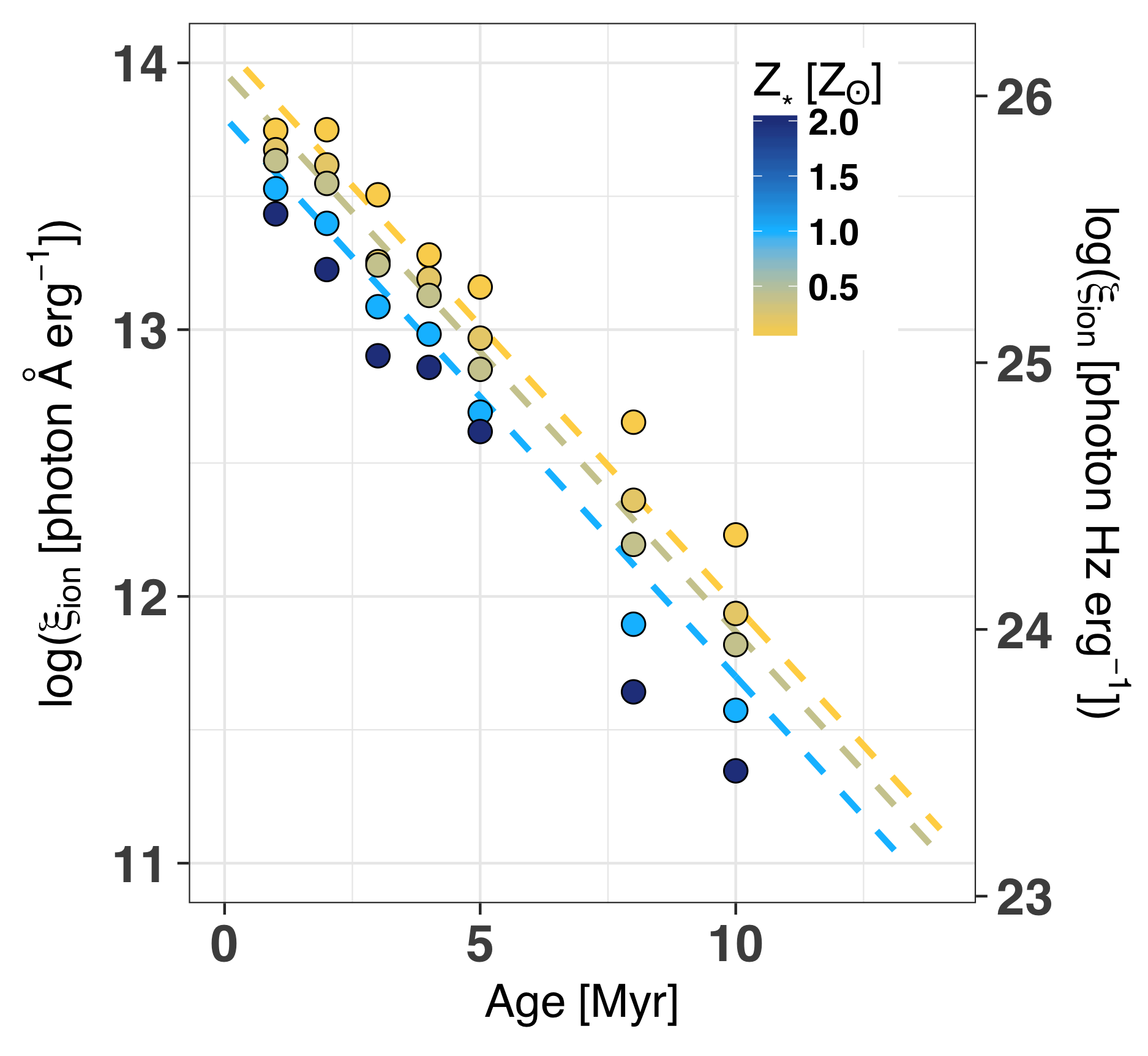}
\caption{The scaling of the ionizing photon production efficiency ($\xi_{\rm ion}$; the number of ionizing photons per FUV luminosity) for a single burst \starburst\ stellar population with age (x-axis) and metallicity (\zs; color-bar). Younger stellar populations produce an order of magnitude more ionizing photons per FUV luminosity than older stellar populations. The blue, tan, and gold curves are single metallicity relations for $\xi_{\rm ion}$ using \autoref{eq:xi_age_aa} with \zs\ of 1, 0.4, and 0.05~Z$_\odot$, respectively.}
\label{fig:xi_age}
\end{figure}

Throughout this paper we have described the ionizing continua using \frat, a measure of the ionizing continuum at a single wavelength. However, the principle goal of this paper is to use the FUV stellar continua to determine a more physically important parameter: the total number of ionizing photons produced by massive star populations ($Q$). $Q$ scales linearly with the star formation rate, such that populations forming more stars generate more ionizing photons \citep{madau98, kennicutt98, kennicutt2012, madau14}. Therefore, the ionizing photon production efficiency, $\xi_{\rm ion}$, compares the capacity of different stellar populations to produce ionizing photons at a given FUV luminosity. The ionizing photon production efficiency is defined from the stellar models as
\begin{equation}
     \xi_{\rm ion} [\text{photons}~\text{\AA}~\text{erg}^{-1}] = \frac{Q[\text{photons s}^{-1}]}{\text{L}_{1500} [\text{erg~\AA}^{-1}~\text{s}^{-1}] },
\end{equation}
or in the more common units as
\begin{equation}
     \xi_{\rm ion} [\text{photon}~\text{Hz}~\text{erg}^{-1}] = \frac{Q[\text{photon s}^{-1}]}{\text{L}_{1500} [\text{erg~Hz}^{-1}~\text{s}^{-1}] }.
\end{equation}
When multiplied by the reddening corrected FUV luminosity, L$_{1500}$, $\xi_{\rm ion}$ determines the total number of ionizing photons generated by a stellar population. 

\autoref{fig:xi_age} shows that the stellar population parameters, particular stellar age, determine $\xi_{\rm ion}$. A 2~Myr, 0.4~Z$_\odot$ \starburst\ stellar population produces 50 times more ionizing photons per L$_{1500}$ than a similar 10~Myr population. Similarly, a 0.05~Z$_\odot$, 2~Myr population produces 3 times more ionizing photons than a 2~Z$_\odot$ population of the same age. These numbers stress the relative importance of age for the production of ionizing photons. The multivariate relationship between $\xi_{\rm ion}$ and the \starburst\ stellar population properties is
\begin{equation}
    \xi_{\rm ion}[\text{photon~\AA~erg}^{-1}] = \left(1.2\times10^{14} \right)10^{-0.21 
    \frac{\text{Age}}{\text{1~Myr}}-0.28\frac{\text{Z}_\ast}{\text{1~Z$_\odot$}}} ,
    \label{eq:xi_age_aa}
\end{equation}
or 
\begin{equation}
    \xi_{\rm ion}[\text{photon~Hz~erg}^{-1}] = \left(1.6\times10^{26}\right) 10^{-0.21 
    \frac{\text{Age}}{\text{1~Myr}}-0.28\frac{\text{Z}_\ast}{\text{1~Z$_\odot$}}} .
        \label{eq:xi_age_hz}
\end{equation}
This relation powerfully illustrates that the total number of ionizing photons, $Q$, can be determined from the stellar population age, metallicity, and the extinction-corrected L$_{1500}$.  As with the \frat\ relations, the models vary from the fitted relation at a given age (the curves in \autoref{fig:xi_age} are single metallicity relations), but \autoref{eq:xi_age_aa} and \autoref{eq:xi_age_hz} characterize the overall evolution of the efficiency of ionizing photon production with an observed stellar age and metallicity. 

\begin{figure}
\includegraphics[width = 0.5\textwidth]{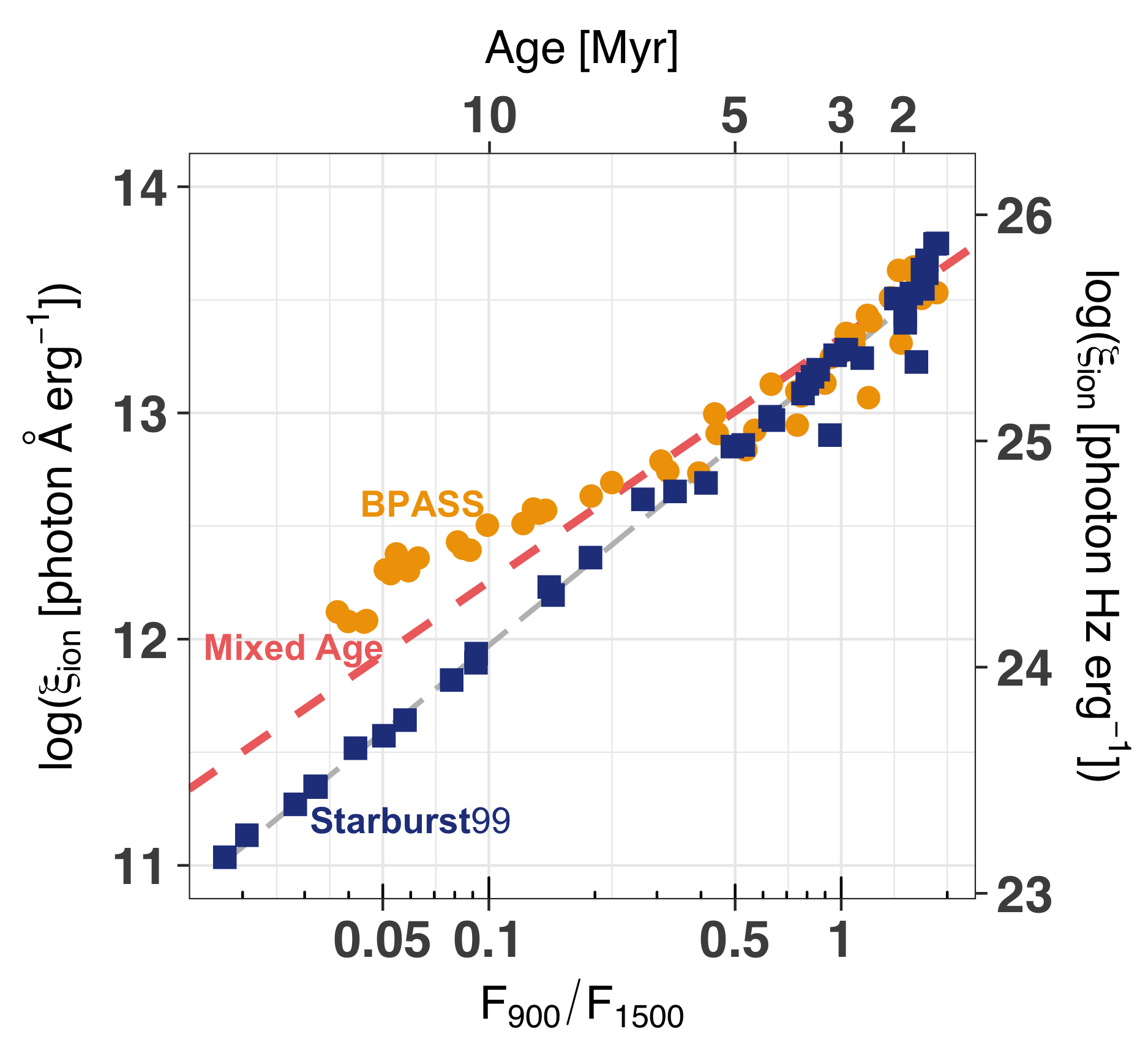}
\caption{The intrinsic number of ionizing photons per FUV luminosity generated by a theoretical stellar population ($\xi_{\rm ion}$) versus the ratio of the flux density at 900\AA\ to the flux density at 1500\AA\ (F$_{900}$/F$_{1500}$). The blue points are single burst \starburst\ populations and the gray line is the best-fit relation to these points (\autoref{eq:q}). The relation found from the MCMC \starburst\ mixed age population is shown by the red line (\autoref{eq:mixedq}). Models that include binary evolution are gold circles. The upper x-axis shows the stellar age of a 0.4~Z$_\odot$ model: at young ages (<5~Myr) all models produce a similar number of ionizing photons per FUV luminosity.}
\label{fig:q}
\end{figure}

\autoref{eq:xi_age_aa} \& \ref{eq:xi_age_hz} have exceptionally similar scalings with age and \zs\ as \frat\ does in \autoref{fig:f9_rat}. In fact, \frat\ strongly scales with  $\xi_{\rm ion}$ (\autoref{fig:q}) as
\begin{equation}
    \text{log}\left(\xi_{\rm ion}[\text{\AA}~\text{erg}^{-1}]\right) =13.23 + 1.23\times 
    \text{log}\left(\frac{\text{F}_{900}}{\text{F}_{1500}}\right) .
    \label{eq:q}
\end{equation}
The strong relationship between \frat\ and $\xi_{\rm ion}$ may come as a surprise, but it is simply related by changes in stellar temperature. As the stellar temperature decreases with increasing stellar age, the hydrogen within the stellar atmospheres becomes more neutral and the stellar Lyman break increases (compare the 2~Myr to 3 and 5~Myr models in \autoref{fig:ionizing_mods}). Thus, \frat\ effectively measures the hydrogen ionization fraction within the stellar atmosphere, which is related to the stellar temperature by both the stellar age and metallicity.  Since stars are to first order complicated black bodies, it follows that the integrated ionizing energy emitted by massive stars, $\xi_{\rm ion}$, must strongly scale with the stellar temperature (or proxies thereof, such as \frat) through the Stephan-Boltzmann law. Thus, the analysis of the variation of \frat\ with age and \zs\ from \autoref{ion_vary} directly extends to $\xi_{\rm ion}$. 

Populations with binary evolution produce a flatter relationship between $\xi_{\rm ion}$ and \frat\ (orange circles in \autoref{fig:q}), because binary populations produce more evolved stars, which in turn have higher temperatures at older ages. $\xi_{\rm ion}$ from \bpass\ models scales as
\begin{equation}
    \text{log}\left(\xi_{\rm ion}[\text{\AA}~\text{erg}^{-1}]\right) = 13.27 + 0.82 \times\text{log} \left(\frac{\text{F}_{900}}{\text{F}_{1500}}\right) .
    \label{eq:qbp}
\end{equation}
At \frat~=~$0.1$ (or an age of $\sim$10~Myr), populations with binary evolution produce three times more total ionizing photons than populations without binary evolution. However, at ages $<5$~Myr both models produce a similar number of ionizing photons. Thus, binary evolution increases the total number of ionizing photons produced by older stellar populations, but has minimal impact on the youngest, 2-4~Myr stellar populations. 

Similarly, mixed age stellar populations produce more ionizing photons per L$_{1500}$ at older light-weighted ages than a single burst star formation history (red line in \autoref{fig:q}), due to the outsized contribution of younger populations (\autoref{cont}). Using the same MCMC sample as \autoref{ion_fits}, there is a strong correlation between $\xi_{\rm ion}$ and F$_{900}$/F$_{1500}$ for the mixed age \starburst\ population:
\begin{equation}
        \text{log}\left(\xi_{\rm ion}[\text{\AA}~\text{erg}^{-1}]\right) = 13.33 + 1.08\times \text{log}\left(\frac{\text{F}_{900}}{\text{F}_{1500}}\right)  .
        \label{eq:mixedq}
\end{equation}
These three cases illustrate that at older light-weighted ages $\xi_{\rm ion}$ strongly depends on both the star formation history and impact of binary evolution.

\begin{figure}
\includegraphics[width = 0.5\textwidth]{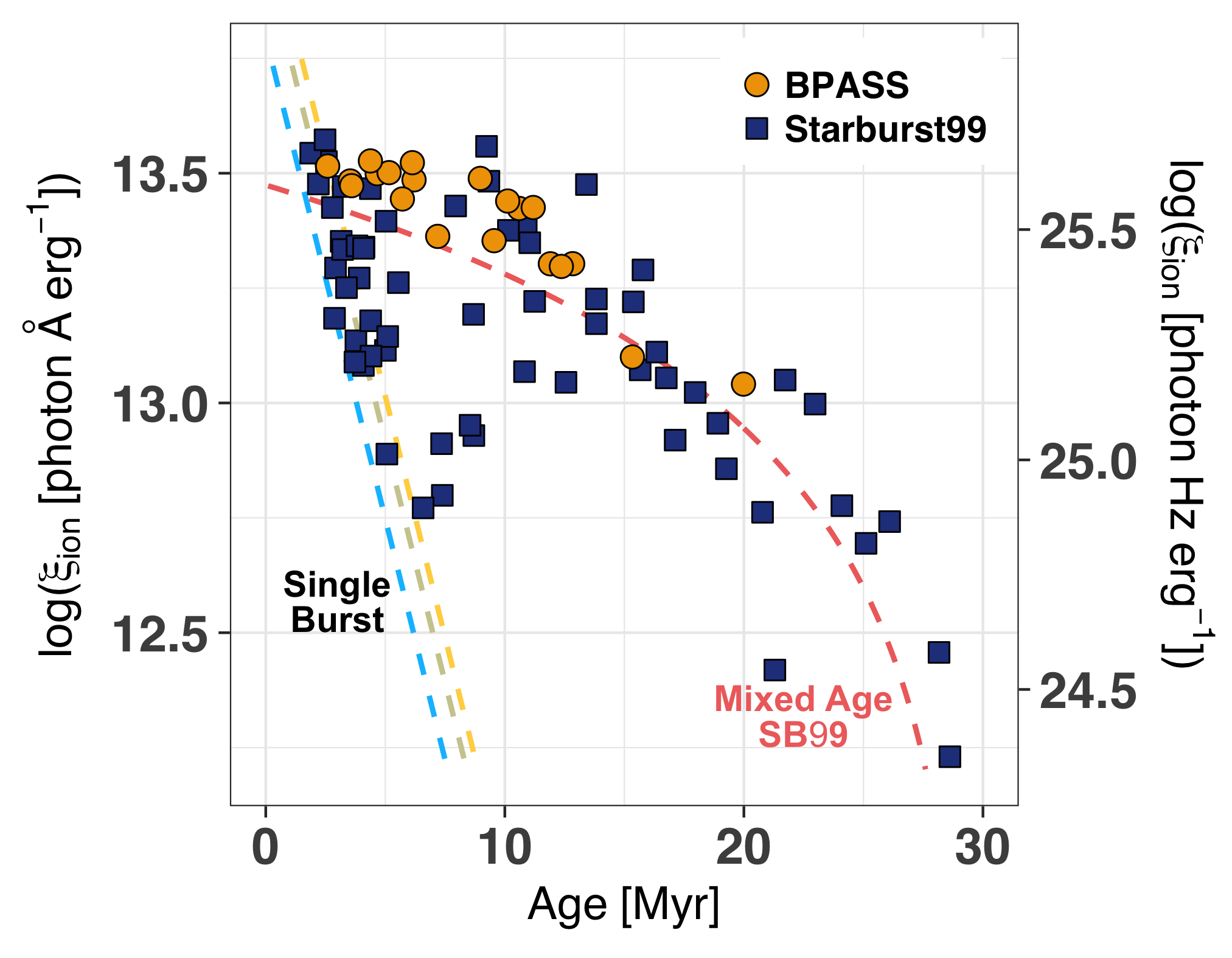}
\caption{The inferred ionizing photon production efficiency ($\xi_{\rm ion}$) from integrated the ionizing continuum of the stellar continuum fits versus the inferred light-weighted stellar age using the \starburst\ (blue squares) and \bpass\ models (orange circles). Only the \megasaura\ spectra were fit with \bpass\ models due to the resolution discrepancy. The blue, tan, and gold curves are single burst \zs\ \starburst\ curves (\autoref{eq:xi_age_aa}) with  \zs\ of 1, 0.4, and 0.05~Z$_\odot$, respectively.  The red dashed line is the relationship from the MCMC \starburst\ mixed age models. }
\label{fig:xi_obs}
\end{figure}

Finally, we compare the ionizing photon production efficiencies of the observed stellar spectra in \autoref{fig:xi_obs}. The $\xi_{\rm ion}$ values are determined by integrating the entire ionizing continuum, divided by the photon energy, of the FUV stellar continuum fits. Galaxies that we defined as having a single burst star formation history in \autoref{ion_fits} closely track the single burst $\xi_{\rm ion}$ relations of \autoref{eq:xi_age_aa} (blue, tan, and yellow curves), while mixed age populations follow the curve defined by the synthetic MCMC population (red curve). 

The median log($\xi_{\rm ion}$[\AA~erg$^{-1}$]) of the observed sample and the \starburst\ models is $13.2\pm0.2$ ($25.3\pm0.2$ for log($\xi_{\rm ion}$[Hz~erg$^{-1}$])) and the full log($\xi_{\rm ion}$[\AA~erg$^{-1}$])  range is 12.2--13.6 (24.4--25.7 for log($\xi_{\rm ion}$[Hz~erg$^{-1}$])). The galaxies within our sample have over an order of magnitude range in the number of ionizing photons that high-mass stars produce per FUV luminosity. $\xi_{\rm ion}$ strongly varies from galaxy-to-galaxy; accurate stellar continuum modeling must determine how efficiently ionizing photons are produced. Perhaps most importantly, the range of inferred $\xi_{\rm ion}$ shows that the number of ionizing photons \textit{cannot} be estimated from the observed FUV luminosity to better than an \textit{order of magnitude} without estimating the stellar population age and metallicity.

The measured $\xi_{\rm ion}$ values are largely consistent with values inferred in the literature. \citet{kennicutt2012} used a constant star formation history, 1~Z$_\odot$ \starburst\ model to derive expressions for the star formation rate of normal local galaxies. These models have log($\xi_{\rm ion}$[\AA~erg$^{-1}$])~=~13.0 (log($\xi_{\rm ion}$[Hz~erg$^{-1}$])~=~25.1), which is statistically consistent with the median value of our sample. While this assumed $\xi_{\rm ion}$ value agrees with the median of our sample, it will \textit{not} agree on a galaxy-by-galaxy basis, to which a detailed analysis of the massive star properties is required.  At higher redshifts, it is typically presumed that more extreme objects, with  log($\xi_{\rm ion}$[Hz~erg$^{-1}$])$~> 25.2-25.3$, are required to reionize the early universe if 20\% of their ionizing photons escape the interstellar medium \citep{robertson13, bouwens16, finkelstein19}. Half of our sample, 31 of 61 (51\%), have $\xi_{\rm ion}$ greater than the value typically assumed for cosmic reionization (log($\xi_{\rm ion}$[Hz~erg$^{-1}$])$~>25.3$). This is not surprising because the galaxies were selected to be bright in the rest-frame FUV, such that the stars must be relatively young. Finally, local galaxies that have been observed to emit ionizing photons also have very young stellar populations \citep{izotov18b} and correspondingly extreme values of $\xi_{\rm ion}$[Hz~erg$^{-1}$]~=~25.6 measured from their \ion{C}{3}]~1909\AA\ emission lines \citep{schaerer18}. These $\xi_{\rm ion}$ values are consistent with the youngest 2--3~Myr stellar populations in both of our samples.

The \bpass\ fits to the \megasaura\ spectra indicate that binary evolution increases the \textit{duration} of the peak $\xi_{\rm ion}$, but not the peak $\xi_{\rm ion}$ itself (orange circles versus blue squares in \autoref{fig:xi_obs}). At the youngest ages, the inferred $\xi_{\rm ion}$ from  the \starburst\ and \bpass\ fits are very similar; binary evolution does not increase the maximum inferred $\xi_{\rm ion}$. Consequently, the youngest stellar populations, which produce the majority of the ionizing photons, have consistent $\xi_{\rm ion}$ values regardless of the chosen binary evolution model. Thus, binary evolution does not increase the maximum $\xi_{\rm ion}$, rather it elongates the period which stellar populations produce their maximum $\xi_{\rm ion}$. If binary evolution is important at high redshifts, it would increase the total number of ionizing photons that high-mass stellar populations produced over their entire lifetimes, and make it easier for stars to reionize the universe.

A surprising prediction from the \bpass\ models in \autoref{fig:xi_obs} is that the $\xi_{\rm ion}$ values of 4-7~Myr populations do \textit{not} decrease along the single burst tracks of \autoref{fig:xi_age}. Rather, the WR-stars in populations with binary evolution keep the $\xi_{\rm ion}$ elevated, and nearly constant at the maximum $\xi_{\rm ion}$ value, over the 10~Myr lifetime of O-stars. Comparing the inferred ionizing continua of RCS~Knot~E and S0108+0624 in \autoref{fig:ion} demonstrates this behavior: the two galaxies have substantially different light-weighted ages but similar \bpass\ ionizing continua. Thus, the number of ionizing photons produced by a stellar population with binary evolution does not vary until the stellar population is older than 10~Myr. As mentioned above, this could substantially increase the total number of ionizing photons generated over the integrated lifetime of stellar populations within the epoch of reionization. This is also a testable observation of the impact of binary evolution using the nebular emission lines.

Determining the total number of ionizing photons intrinsically produced by a given stellar population is a key strength of FUV spectral synthesis. This analysis infers the total number of ionizing photons, as well as their spectral distribution, using observable spectral features from the same massive stars that emit the ionizing photons. In theory, spectral synthesis promises to be an extremely powerful tool for a wide assortment of extragalactic applications, from understanding the generation of nebular emission lines to uncovering the sources of cosmic reionization. In practice, determining the total number of ionizing photons requires precise constraints on the stellar population age, metallicity, star formation history, and the impact of binary star evolution (\autoref{fig:xi_age} and \autoref{fig:q}). We demonstrated throughout this paper that massive star spectral features constrain the age, metallicity, and star formation history. However, we also emphasized that the non-ionizing FUV stellar continuum alone cannot constrain the impact of binary evolution. Binary evolution does not dramatically impact the number of ionizing photons produced per FUV luminosity of the youngest stellar populations (<4~Myr), but it significantly increases the number of ionizing photons produced by older populations. Without constraining the impact of binary evolution, there is up to a factor of 7 uncertainty in the total number of ionizing photons produced by older stellar populations.

\section{Summary}
We have studied the massive star populations in a sample of 61
star-forming galaxies comprised of 42 at  $z < 0.3$ from the HST archive and 19 at $z\sim2$ from the \megasaura\ sample. We fit a simple linear combination of single age, fully theoretical stellar models to the far-ultraviolet stellar continuum ($\lambda_{\rm r} = 1220-2000$\AA). From these fits we derived light-weighted ages and metallicities of the stellar populations, and extrapolated the fits to infer their ionizing continua ($\lambda_{\rm r} < 912$\AA). 

Individual, stellar spectral features agree with the inferred light-weighted properties. The \civ stellar wind line varies with the inferred stellar age and metallicity in different and distinguishable ways (\autoref{fig:age}), such that the \civ emission depends on stellar age (\autoref{fig:age_mods}) and the absorption depends on metallicity (\autoref{fig:age_mods_z}). The \nv\ line strongly depends on age and is nearly independent of metallicity (\autoref{fig:nv}). Very young stellar populations have broad ($\sim400$~\kmsp) \heii\ emission from Wolf-Rayet stars, while older populations have negligible \heii\ features (\autoref{fig:heii}). Finally, old populations have \ion{C}{3} and \ion{Si}{3}  photospheric absorption features, while these features are undetected in younger populations (\autoref{fig:photo}). 

The light-weighted stellar metallicity is strongly correlated with ($8\sigma$), and consistent with, the nebular metallicity (\autoref{fig:z_comp}). The metallicity relationship does not depend on redshift, as we find $z \sim 2$ stars and gas to have similar metallicities. Stellar metallicities may be the most reliable, yet time-consuming, method to infer metallicities at high redshift (\autoref{metal}).

We used the stellar continuum models to demonstrate how the stellar ionizing continuum depends on stellar population properties (\autoref{ion}). Specifically, we characterized the ionizing continuum using three ratios of the ionizing flux density at different wavelengths to the non-ionizing flux density at 1500\AA. We focused on how the flux density ratio at 900\AA\ to 1500\AA, or \frat, depends on stellar population properties, because 900\AA\ is the most common wavelength to observe the ionizing continuum. Theoretical stellar models demonstrate that the strength and spectral shape of the ionizing continuum of a single burst strongly depends on the stellar age (\autoref{fig:ionizing_mods} and \autoref{fig:fluxrat_age}) and metallicity (\autoref{fig:fluxrat_z}). A multivariate relationship between \frat\ and both the age and metallicity (\autoref{eq:f9}) describes most of the variation in \frat\ for a single burst of star formation (\autoref{fig:f9_rat}). 

We then inferred the ionizing continuum of the observed stellar populations by extrapolating the fits to the observed stellar continua. Stellar populations with older light-weighted ages have weaker inferred ionizing continua (\autoref{fig:ion}). However, only half (31 of 61) of the sample follows the single burst \frat\ relations (\autoref{fig:f9_rat_obs}). The other half of the sample has a mixture of stellar populations at multiple ages. The mixed age populations have stronger inferred ionizing continua at older ages than prescribed by a single burst model. The inferred \frat\ of these galaxies is strongly linearly anti-correlated (5.5$\sigma$) with the light-weighted age (\autoref{eq:f9older}). These two populations are split at a light-weighted age of 8.6~Myr. 


Mixed age populations have stronger ionizing continua than single bursts of a  similar age due to the presence of very young stellar populations (\autoref{fig:f9_rat_obs}). We emphasized the differences between  mixed age populations and the typically assumed continuous star formation history (\autoref{cont}). Chiefly, the mixed age continua depend on the relative mixture of age \textit{and} metallicity. Thus, age sensitive spectral features  (\civp, \nv, and \heii) can vary, and the ionizing continuum can correspondingly increase or decrease. Mixed age stellar populations produce significantly more ionizing photons than both a single burst with the same light-weighted age and a continuous star formation history, while having the same metallicity as the nebular gas (\autoref{fig:chuck}). This is in stark contrast with previous work that assume a constant star formation history and we emphasized that the only difference is the assumed star formation history. 

Our fiducial stellar models are single star \starburst\ models mainly because these provided sufficient spectral resolution at all wavelengths (see the discussion in \autoref{bpass}). We also fit the \megasaura\ sample with stellar models that include binary evolution (\bpass\ models), but do not find significant differences between the \starburst\ and \bpass\ fits to the FUV features (compare the blue and gold lines in \autoref{fig:age}--\ref{fig:nv}). Similarly, the light-weighted metallicities and ages are similar, as long as we only consider young ages for which the \bpass\ spectral resolution is adequate (\autoref{fig:mod_comp}). The most significant difference is that \bpass\ models produce up to 12 times more high-energy ionizing photons than \starburst\ models at older ages and lower metallicities (\autoref{fig:f9_rat} and \autoref{fig:ion}). Consequently, the number of ionizing photons produced by a stellar population depends on the observationally \textit{un}constrained impact of binary evolution. 

We provide scaling relations for the total number of ionizing photons produced per FUV luminosity ($\xi_{\rm ion}$). $\xi_{\rm ion}$ strongly depends on the age, metallicity, star formation history, and presence of binary stars (\autoref{fig:xi_age} and \autoref{fig:q}). The inferred $\xi_{\rm ion}$ values from our observations have an order of magnitude scatter that strongly depends on stellar population properties (\autoref{fig:xi_obs}). We emphasize that $\xi_{\rm ion}$ times the extinction corrected FUV luminosity determines the total number of ionizing photons produced by a stellar population. While fitting the FUV stellar continuum promises to be a powerful tool in determining the total number of ionizing photons produced by high-mass stars, without constraining the impact of binary evolution there is a factor of 7 uncertainty in the total number of ionizing photons produced by older stellar populations.

The rest-frame FUV spectral features of massive stars mirror the production of ionizing photons. Stellar population synthesis can estimate many properties required to infer the number of ionizing photons produced by stars and determine the source of cosmic reionization. This type of modeling is currently infeasible for galaxies within the epoch of reioniziation, but observations of the stellar continuum will be possible with future 30~m class telescopes and the \textit{James Webb Space Telescope}. Stellar population synthesis of the most massive stars, as presented here, will be the most direct method to constrain the number of ionizing photons produced by the most distant star-forming galaxies.

\acknowledgements
We thank the anonymous referee for careful reading the original manuscript and their insightful suggestions. We thank Chuck Steidel for kindly providing his stacked spectra for our comparison. JC appreciates Christy Tremonti's early stellar continuum fitting guidance and discussions throughout this project. JC thanks Joe Cassinelli for early discussions that provided the fundamental stellar physics that inspired a substantial portion of this project. JC is grateful for helpful discussions and clarifications from Claus Leitherer. We thank X. Prochaska for insightful comments and clarifications that greatly improved the scope and clarity of the paper. JC thanks the Ohio State University for their hospitality while writing portions of this paper. JR is grateful for helpful discussions with George Sonneborn and Sally Heap. JR and DB acknowledge discussion and ideas presented at the Carnegie Symposium in honor of Leonard Searle, on the topic of \lq{}\lq{}Understanding Nebular Emission in High-Redshift Galaxies,\rq{}\rq{} held at the Carnegie Observatories in Pasadena in July 2015. 

This paper includes data gathered with the 6.5 meter Magellan Telescopes located at Las Campanas Observatory, Chile. We thank the staff of Las Campanas for their dedicated service, which has made possible these observations. We thank the telescope allocation committees of the Carnegie Observatories, The University of Chicago, The University of Michigan, Massachusetts Institute of Technology, and Harvard University, for supporting the \megasaura\ project over several years of observing. This paper includes data from observations made with the Nordic Optical Telescope, operated by the Nordic Optical Telescope Scientific Association at the Observatorio del Roque de los Muchachos, La Palma, Spain, of the Instituto de Astrofisica de Canarias.


\appendix

Here we give the tables of the derived stellar parameters for the \megasaura\ (\autoref{tab:megasaura}) and low-redshift samples (\autoref{tab:cos}). We then give the table of the ionization production efficiency ($\xi_{\rm ion}$) for both samples in \autoref{tab:xi_meg} and \autoref{tab:cos_xi}. In the electronic version of the paper we also include the models used to make the fits, the light fractions of each fit, and the fitted stellar continua for each galaxy. 
\clearpage

{
\begin{centering}
\begin{table*}
\caption{Stellar continuum properties of the \megasaura\ spectra}
\begin{tabular}{lcccccccccc}
(1) & (2) & (3) & (4) & (5) & (6) & (7) & (8) & (9) & (10) & (11)\\
Galaxy name&$z$&E(B-V)&E(B-V)&Z$_\text{neb}$&Z$_\ast$&
Z$_\ast$&Age&Age&F$_{900}$/F$_{1500}$ & Spectral\\
& & SB99 & BPASS & & SB99 & BPASS & SB99 & BPASS & SB99 & Reference \\
&&[mag]&[mag]&[Z$_\odot$]& [Z$_\odot$] &[Z$_\odot$]&[Myr]&[Myr]&\\
\hline

RCS-0327-1326~Knot~E&              1.7034&                0.31&                0.29&0.34$^{\rm c}$&                0.24$\pm$                0.03&                0.18$\pm$                0.02&                2.54$\pm$                0.14&                2.59$\pm$                0.14&                1.55$\pm$                0.08
 & 3, 13, 10\\
Sunburst~Arc~Region~5&              2.3709&                0.15&                0.15&--&                0.55$\pm$                0.04&                0.66$\pm$                0.05&                2.92$\pm$                0.08&                3.59$\pm$                0.09&                1.14$\pm$                0.05 &
4, 11\\
S003341.5+024217&              2.3887&                0.33&                0.32&--&                0.84$\pm$                0.04&                0.74$\pm$                0.03&                5.00$\pm$                0.51&                3.53$\pm$                0.36&                0.76$\pm$                0.08 & 10
\\
RCS-0327-1326~Knot~U&              1.7039&                0.27&                0.28&0.34$^{\rm c}$&                0.29$\pm$                0.07&                0.09$\pm$                0.02&                5.03$\pm$                1.44&                5.16$\pm$                1.47&                1.07$\pm$                0.31 &3, 10, 13
\\
S095738.7+050929&              1.8204&                0.17&                0.17&--&                0.17$\pm$                0.05&                0.20$\pm$                0.06&                7.36$\pm$                1.64&                4.38$\pm$                0.98&                0.52$\pm$                0.20 &  1, 10 \\
S090003.3+223408&              2.0326&                0.10&                0.09&0.31$^{\rm a}$&                0.38$\pm$                0.02&                0.30$\pm$                0.02&                7.39$\pm$                0.41&                7.19$\pm$                0.39&                0.39$\pm$                0.03 & 5, 10
\\
SPT0356 &  2.362 &0.14 & 	0.14 & -- & 0.22$\pm$0.07 & 0.16$\pm$0.05 & 8.55$\pm$3.88 & 5.71$\pm$2.59 & 0.47$\pm$0.33 & -- \\
S010842.2+062444&              1.9102&                0.18&                0.20&--&                0.50$\pm$                0.03&                0.35$\pm$                0.02&                8.71$\pm$                1.06&                6.20$\pm$                0.75&                0.53$\pm$                0.09 & 9, 10
\\
SPT2325&              1.5799&                0.17&                0.17&--&                0.27$\pm$                0.03&                0.15$\pm$                0.02&                9.24$\pm$                2.42&                6.13$\pm$                1.61&                1.36$\pm$                0.54 & --
\\

Cosmic~Horseshoe&              2.3812&                0.11&                0.09&0.57$^{\rm b}$&                0.52$\pm$                0.07&                0.20$\pm$                0.03&                9.33$\pm$                4.20&               11.91$\pm$                5.35&                1.13$\pm$                0.76 & 2, 10
\\

SPT0142 & 2.674 &  0.19 &	0.17 & --&	0.23$\pm$0.04 & 0.14$\pm$0.03 & 10.15$\pm$3.73 & 12.36$\pm$4.54 & 1.06$\pm$0.65  & -- \\
RCS-0327-1326~Knot~G&              1.7039&                0.24&                0.24&0.34$^{\rm c}$&                0.29$\pm$                0.04&                0.24$\pm$                0.04&               10.90$\pm$                1.77&                4.66$\pm$                0.76&                0.97$\pm$                0.34
&3, 10, 13\\
S000451.7-010321&              1.6811&                0.33&                0.30&0.25$^{\rm d}$&                0.25$\pm$                0.02&                0.15$\pm$                0.01&               13.41$\pm$                0.80&               10.60$\pm$                0.63&                1.08$\pm$                0.22 & 9, 10
\\
S142954.9+120239&              2.8245&                0.18&                0.20&--&                0.64$\pm$                0.05&                0.60$\pm$                0.04&               15.38$\pm$                2.68&                9.55$\pm$                1.66&                0.68$\pm$                0.30 & 8, 10
\\

SPT0310&              1.9960&                0.27&                0.29&--&                0.23$\pm$                0.04&                0.30$\pm$                0.05&               15.66$\pm$                2.87&                8.97$\pm$                1.65&                0.67$\pm$                0.37 & --
\\
PSZ0441&              1.8008&                0.30&                0.30&--&                0.10$\pm$                0.05&                0.39$\pm$                0.18&               19.27$\pm$                4.47&               11.19$\pm$                2.59&                0.41$\pm$                0.57 & --
\\
S152745.1+065219&              2.7623&                0.33&                0.32&<0.65$^{\rm \rm f}$&                0.49$\pm$                0.04&                0.36$\pm$                0.03&               21.30$\pm$                3.32&               12.84$\pm$                2.00&                0.22$\pm$                0.12 & 6, 10
\\
S122651.3+215220&              2.9260&                0.13&                0.13&--&                0.34$\pm$                0.02&                0.14$\pm$                0.01&               26.10$\pm$                1.29&               19.98$\pm$                0.99&                0.31$\pm$                0.09 & 7, 10
\\
Cosmic~Eye&              3.0735&                0.38&                0.36&0.81$^{\rm e}$&                0.65$\pm$                0.02&                0.50$\pm$                0.02&               28.62$\pm$                0.80&               15.34$\pm$                0.43&                0.15$\pm$                0.03   & 10, 12\\
\end{tabular}
\tablecomments{Column 1 gives the galaxy names ordered by the inferred \starburst\ stellar age (Column 8). Most names are given here by their full right ascension and declination but throughout the text we shorten the names to the first four digits. Column 2 gives the redshift. Column 3 and 4 give the fitted attenuation parameter (E(B-V)) measured using \starburst\ and \bpass\ models, respectively. Column 5 gives the nebular metallicities (\zism) from the literature. Column 6 and 7 give the light-weighted stellar metallicities (\zs) using \starburst\ and \bpass, respectively. Column 8 and 9 gives the light-weighted stellar age using \starburst\ and \bpass, respectively. Column 10 gives the inferred flux density ratio at 900\AA\ and 1500\AA. Column 11 gives references for the galaxies and spectra. All spectra will be fully introduced in Rigby et al.\ (in preparation). \\
Metallicity references: (a) \citet{bian10}, (b) \citet{hainline09}, (c) \citet{rigby11}, (d) \citet{rigbya}, (e) \citet{stark08}, (f) \citet{wuyts}.\\ Spectral references: (1) \citet{bayliss11}, (2) \citet{belokurov07}, (3) \citet{bordoloi16}, (4) \citet{dahle}, (5) \citet{diehl09}, (6) \citet{hennawi08}, (7) \citet{koester10}, (8) \citet{marques17}, (9) \citet{rigby14}, (10) \citet{rigbya}, (11) \citet{rivera-thorsen17, rivera19}, (12) \citet{smail07}, (13) \citet{wuyts}. }
\label{tab:megasaura}
\end{table*}
\end{centering}

}
{ 
\begin{table*}
\caption{Stellar continuum properties of the low-redshift HST/COS Spectra}
\begin{tabular}{lcccccccccccccc}
(1) & (2) & (3) & (4) & (5) & (6) & (7) & (8) & (9) & (10) & (11)\\
Galaxy name&$z$&E(B-V)&Z$_\text{neb}$&Z$_\ast$&Age& \frat &PID&
Z$_\text{neb}$ & Metallicity &Spectra\\
& &[mag] & [Z$_\odot$] & [Z$_\odot$] & [Myr] & & & Ref & Calibration & Ref \\
\hline
J1416+1223& 0.1231&0.14&0.60&0.60$\pm$0.05& 1.88$\pm$ 0.35&1.36$\pm$0.28&13017&\citet{heckman15} &PP04&1~6 \\
J0150+1260& 0.1467&0.12&0.50&0.52$\pm$0.13& 2.20$\pm$ 2.17&1.23$\pm$1.25&11727&\citet{heckman15} &PP04&1~6 \\
J0824+2806 & 0.0472 & 0.28 & 0.35 &	0.55$\pm$0.08 & 2.48$\pm$3.12 &	1.49$\pm$0.88 &	13017	& \citet{heckman15} & PP04 & 1~6 \\
J0907+5327& 0.0299&0.08&0.22&0.56$\pm$0.19& 2.79$\pm$ 1.31&1.23$\pm$0.72&12583&\citet{Ostlin14} &Direct&5~9~10\\
NGC~7552& 0.0054&0.71&1.12&0.99$\pm$0.05& 2.88$\pm$ 0.04&1.16$\pm$0.07&12173&\citet{moustakas} &KK04&8\\
J0055-0021& 0.1674&0.25&0.39&0.42$\pm$0.15& 3.16$\pm$ 2.27&1.20$\pm$0.97&11727&\citet{heckman15} &PP04&1~6\\
J1429+0643& 0.1736&0.20&0.27&0.36$\pm$0.09& 3.21$\pm$ 0.36&1.11$\pm$0.30&13017&\citet{heckman15} &PP04&1~6\\
J0926+4427& 0.1807&0.00&0.14&0.16$\pm$0.06& 3.24$\pm$ 0.37&1.31$\pm$0.49&11727&\citet{heckman15} &PP04&1~6\\
J0808+3948 & 0.0912 & 0.15 & 1.12 &	1.46$\pm$0.04 & 3.38$\pm$0.07 &0.91$\pm$0.02	& 11727 & \citet{heckman15} & PP04 & 1~6	\\
J0921+4509 &	0.2350 &	0.19 &	0.95 &	1.03$\pm$0.15 &3.73$\pm$0.62 &	0.77$\pm$0.13 &	11727 &	\citet{heckman15} & PP04&	1~2~6 \\
NGC~6090& 0.0293&0.22&0.51&0.76$\pm$0.03& 3.77$\pm$ 0.11&0.83$\pm$0.04&12173&\citet{Cortijo-Ferrero} &M13&8\\
J0021+0052 & 0.0984 &0.10 &	0.32 &	0.45$\pm$0.13 &	3.82$\pm$1.28 &	1.06$\pm$0.30 &13017 &	\citet{heckman15} & PP04 & 1~6	\\
J1415+0540& 0.0819&0.13&0.39&0.34$\pm$0.10& 3.91$\pm$ 2.40&1.01$\pm$0.69&13017&\citet{heckman15} &PP04&1~6\\
NGC~3256& 0.0094&0.49&1.10&0.57$\pm$0.12& 4.08$\pm$ 1.45&0.76$\pm$0.31&12173&\citet{engelbracht} &PP04&8\\
J0213+1259 &	0.2190 & 0.40 &	1.12 & 1.08$\pm$0.22 & 4.09$\pm$1.65 & 0.91$\pm$0.33 & 13017 & \citet{heckman15} & PP04 &1~6\\
KISSR~1578& 0.0279&0.11&0.24&0.48$\pm$0.06& 4.11$\pm$ 2.03&1.05$\pm$0.54&11522&\citet{Ostlin14} & Direct & 11\\
J1112+5503& 0.1315&0.25&0.68&0.65$\pm$0.09& 4.38$\pm$ 3.54&1.15$\pm$0.94&13017&\citet{heckman15} &PP04&1~6\\
NGC~4214& 0.0010&0.30&0.32&0.33$\pm$0.06& 4.39$\pm$ 0.30&0.95$\pm$0.20&11579&\citet{kobulnicky96} &Direct&7\\
J1315+6207& 0.0308&0.43&1.02&0.97$\pm$0.17& 4.40$\pm$ 2.23&0.75$\pm$0.40&12583& This work &Direct&5~9~10\\
NGC~7714& 0.0093&0.42&0.60&0.84$\pm$0.02& 5.07$\pm$ 0.06&0.51$\pm$0.01&12604&\citet{gonzalez} &Direct&3\\
SBS~1415+437& 0.0020&0.14&0.08&0.08$\pm$0.02& 5.10$\pm$ 0.21&0.81$\pm$0.19&11579&\citet{thuan99} &Direct&7\\
KISSR~218& 0.0209&0.46&1.51&1.20$\pm$0.19& 5.55$\pm$ 3.73&1.03$\pm$0.71&11522&\citet{salzer05} &EP84&11\\
J1429+1653& 0.1816&0.19&0.42&0.40$\pm$0.07& 6.58$\pm$ 2.73&0.42$\pm$0.19&13017&\citet{heckman15} &PP04&1~6\\
J0938+5428& 0.1021&0.11&0.21&0.24$\pm$0.08& 7.94$\pm$ 3.34&1.07$\pm$0.58&11727&\citet{heckman15} &PP04&1~6\\
J1025+3622& 0.1265&0.11&0.26&0.18$\pm$0.05& 8.69$\pm$ 3.64&0.83$\pm$0.43&13017&\citet{heckman15} &PP04&1~6\\
KISSR~182& 0.0224&0.32&0.45&0.53$\pm$0.17&10.82$\pm$ 4.58&0.71$\pm$0.38&12027&\citet{salzer05} &EP84&11\\
J1525+0757& 0.0757&0.14&0.59&0.53$\pm$0.09&11.04$\pm$ 5.18&0.97$\pm$0.48&13017&\citet{heckman15} &PP04&1~6\\
Shoc~22& 0.0179&0.10&0.23&0.28$\pm$0.05&11.25$\pm$ 2.97&0.74$\pm$0.23&15099& This work &Direct& --\\
J1250+0734& 0.0382&0.19&0.66&0.69$\pm$0.17&13.83$\pm$ 6.80&0.81$\pm$0.44&12583&\citet{Ostlin14} &Direct&5~9~10\\
MRK~1486& 0.0338&0.12&0.13&0.10$\pm$0.05&13.83$\pm$ 2.58&0.79$\pm$0.46&12583&\citet{Ostlin14} &Direct&5~9~10\\
Haro~11& 0.0206&0.09&0.40&0.38$\pm$0.02&15.78$\pm$ 1.48&0.77$\pm$0.08&13017&\citet{james13} &Direct&1~6\\
SBS~0926+606A& 0.0137&0.14&0.21&0.32$\pm$0.08&16.36$\pm$ 4.83&0.72$\pm$0.28&15099&\citet{berg16} &Direct&--\\
J1307+5427& 0.0325&0.37&0.30&0.83$\pm$0.14&16.75$\pm$ 4.87&0.63$\pm$0.21&12583& This work &Direct&5~9~10\\
NGC~5253\_1& 0.0014&0.34&0.30&0.31$\pm$0.04&17.13$\pm$ 2.43&0.53$\pm$0.10&11579&\citet{walsh89} &Direct&7\\
KISSR~108& 0.0236&0.19&0.30&0.25$\pm$0.14&17.96$\pm$ 5.12&0.56$\pm$0.36&12027&\citet{salzer05} &EP84&11\\
J1144+4012& 0.1270&0.26&0.51&0.41$\pm$0.10&18.91$\pm$ 5.71&0.51$\pm$0.20&13017&\citet{heckman15} &PP04&1~6\\
1~Zw~18& 0.0025&0.12&0.03&0.05$\pm$0.01&20.78$\pm$ 0.71&0.35$\pm$0.07&11579&\citet{Izotov97} &Direct&7\\
KISSR~242& 0.0378&0.22&0.49&0.51$\pm$0.13&21.73$\pm$ 5.16&0.56$\pm$0.20&11522&\citet{salzer05} &EP84&4~11\\
IRAS~08339& 0.0191&0.23&0.58&0.52$\pm$0.03&22.98$\pm$ 1.22&0.46$\pm$0.04&12173&\citet{lopez} &D02&8\\
NGC~4670& 0.0036&0.26&0.32&0.37$\pm$0.04&24.10$\pm$ 1.49&0.41$\pm$0.05&11579&\citet{heckman98} &Direct&7\\
NGC~4449& 0.0007&0.37&0.42&0.41$\pm$0.03&25.11$\pm$ 1.47&0.33$\pm$0.03&11579&\citet{marble10} &Direct&7\\
NGC~3690& 0.0102&0.21&0.74&0.81$\pm$0.10&28.16$\pm$ 1.07&0.12$\pm$0.01&11579&\citet{heckman98} &Direct&7\\
\end{tabular}
\tablecomments{The fitted properties for the 42 low-redshift galaxies with COS/HST observations. Column 1 gives the galaxy name. Column 2 gives the galaxy redshift ($z$). Column 3 gives the fitted stellar attenuation from the \starburst\ fits to the stellar continua (E(B-V)). Column 4 gives the literature nebular metallicity of each galaxy (\zism). Column 5 gives the inferred stellar metallicity from the \starburst\ fits to the stellar continua (\zs). Column 6 gives the inferred stellar age from the \starburst\ fits to the stellar continua. Column 7 is the inferred ratio of the flux at 900\AA\ to the flux at 1500\AA. Column 8 is the HST project ID for each spectra. Column 9 is the literature reference for each \zism. Column 10 is the calibration method used to determine \zism. Codes for metallicity calibration are: PP04 \citep{pettini04}, KK04 \citep{kobulnicky04}, EP84 \citep{edmunds84}, D02 \citep{denicolo}, and M13 \citep{marino13}.  Column 11 gives literature references for each spectra.    \\Spectral references are coded as: (1) \citet{alexandroff}, (2) \citet{borthakur}, (3) \citet{fox2013}, (4) \citet{france2010}, (5) \citet{hayes14}, (6) \citet{heckman15}, (7)  \citet{james14}, (8) \citet{leitherer14}, (9) \citet{Ostlin14}, (10) \citet{rivera15}, (11) \citet{wofford2013}. } 
\label{tab:cos}
\end{table*}
}

\begin{table}
\caption{$\xi_{\rm ion}$ values for the \megasaura\ sample. }
\begin{tabular}{lcc}
(1) & (2) & (3)\\
Galaxy name&  log($\xi_{\rm ion}$) SB99 &  log($\xi_{\rm ion}$) BPASS \\
 & log([photon~\AA~erg$^{-1}$])  & log([photon~\AA~erg$^{-1}$])  \\
\hline
RCS-0327-1326~Knot~E &	13.53$\pm$0.03 & 13.52$\pm$0.03\\
Sunburst~Arc~Region~5 &	13.29$\pm$0.02 & 13.47$\pm$0.02 \\
S003341.5+024217 &	13.11$\pm$0.05 &	13.48$\pm$0.05 \\\
RCS-0327-1326~Knot~U &	13.40$\pm$0.16 & 13.50$\pm$0.16 \\
S095738.7+050929 &	12.91$\pm$0.20 & 13.53$\pm$0.21 \\
S090003.3+223408 &	12.80$\pm$0.05	& 13.36$\pm$0.05\\
SPT0356 &12.95$\pm$0.37 &13.44$\pm$0.38 \\
S010842.2+062444 &12.93$\pm$0.09 &13.49$\pm$0.10\\
SPT2325 & 13.56$\pm$0.21 &	13.52$\pm$	0.21\\
Cosmic~Horseshoe &		13.48$\pm$	0.36 &	13.30$\pm$	0.35\\
SPT0142 &		13.38$\pm$	0.33 &	13.30$\pm$	0.33\\
RCS-0327-1326~Knot~G &		13.38$\pm$	0.19 &	13.50$\pm$	0.19\\
S000451.7-010321 &		13.48$\pm$	0.11 &	13.42$\pm$	0.11\\
S142954.9+120239 &		13.22$\pm$	0.23 &	13.35$\pm$	0.24\\
SPT0310 &		13.07$\pm$	0.29 &	13.49$\pm$	0.30\\
PSZ0441 &		12.86$\pm$	0.75 &	13.43$\pm$	0.79\\
S152745.1+065219 &		12.42$\pm$	0.29 &	13.30$\pm$	0.31\\
S122651.3+215220 &		12.74$\pm$	0.16 &	13.04$\pm$	0.16\\
Cosmic~Eye &		12.23$\pm$	0.10 &	13.10$\pm$	0.11\\
\end{tabular}
\label{tab:xi_meg}
\tablecomments{The photon production efficiency ($\xi_{\rm ion}$) for the \megasaura\ sample. Column 2 gives the values inferred using the \starburst\ models and Column 3 gives the values for the \bpass\ models. The galaxies are listed in descending light-weighted age inferred using the \starburst\ models. }
\end{table}

{\centering
\begin{table}
\caption{$\xi_{\rm ion}$ values for the low-redshift HST/COS sample}
\begin{tabular}{lc}
(1) & (2) \\
Galaxy name& log($\xi_{\rm ion}$)\\
& log([photon \AA\ erg$^{-1}$]) \\
\hline
J1416+1223&13.54$\pm$0.11\\
J0150+1260&13.48$\pm$0.54\\
J0824+2806&13.57$\pm$0.32\\
J0907+5327&13.43$\pm$0.31\\
NGC~7552&13.18$\pm$0.03\\
J0055-0021&13.35$\pm$0.43\\
J1429+0643&13.33$\pm$0.14\\
J0926+4427&13.47$\pm$0.20\\
J0808+3948&13.25$\pm$0.01\\
J0921+4509&13.09$\pm$0.09\\
NGC~6090&13.14$\pm$0.03\\
J0021+0052&13.34$\pm$0.15\\
J1415+0540&13.27$\pm$0.36\\
NGC~3256&13.08$\pm$0.22\\
J0213+1259&13.34$\pm$0.19\\
KISSR~1578&13.34$\pm$0.27\\
J1112+5503&13.47$\pm$0.44\\
NGC~4214&13.18$\pm$0.11\\
J1315+6207&13.10$\pm$0.29\\
NGC~7714&12.89$\pm$0.01\\
SBS~1415+437&13.14$\pm$0.13\\
KISSR~218&13.26$\pm$0.37\\
J1429+1653&12.77$\pm$0.24\\
J0938+5428&13.43$\pm$0.29\\
J1025+3622&13.19$\pm$0.27\\
KISSR~182&13.07$\pm$0.28\\
J1525+0757&13.35$\pm$0.27\\
Shoc~22&13.22$\pm$0.17\\
J1250+0734&13.23$\pm$0.29\\
MRK~1486&13.17$\pm$0.31\\
Haro~11&13.29$\pm$0.06\\
SBS~0926+606A&13.11$\pm$0.21\\
J1307+5427&13.05$\pm$0.18\\
NGC~5253\_1&12.92$\pm$0.10\\
KISSR~108&13.02$\pm$0.34\\
J1144+4012&12.96$\pm$0.20\\
1~Zw~18&12.76$\pm$0.11\\
KISSR~242&13.05$\pm$0.19\\
IRAS~08339&13.00$\pm$0.05\\
NGC~4670&12.78$\pm$0.07\\
NGC~4449&12.69$\pm$0.05\\
NGC~3690&12.46$\pm$0.07\\
\end{tabular}
\tablecomments{Values of the ionizing photon production efficiency ($\xi_{\rm ion}$) for the low-redshift HST/COS galaxies calculated using the \starburst\ models. The tables is in descending age order. } 
\label{tab:cos_xi}
\end{table}
}

\clearpage
\bibliographystyle{mnras}
\bibliography{hotstars}
\clearpage
\end{document}